\documentclass{aa}
\usepackage{txfonts}
\usepackage{natbib}
\usepackage{graphicx}
\bibpunct{(}{)}{;}{a}{}{,} 
\usepackage{xspace}
\usepackage{longtable}
\usepackage{pdflscape}

\def \bea {\begin{eqnarray}}
\def \ena {\end{eqnarray}}                  
\def \bee {\begin{equation}}
\def \ene {\end{equation}}
\def    \simlt  {\lower.5ex\hbox{$\; \buildrel < \over \sim \;$}}
\def    \simgt  {\lower.5ex\hbox{$\; \buildrel > \over \sim \;$}}

\def	\xhat		{\hat{\bf x}}
\def	\yhat		{\hat{\bf y}}
\def	\zhat		{\hat{\bf z}}
\def	\ahat		{\hat{\bf a}}

\def    \ali     	{{\rm ali}}

\newcommand     \mum    {\,\mu{\rm m}}  

\begin{document}
\title{Spectropolarimetry of Galactic stars with anomalous extinction sightlines}

\author{Aleksandar Cikota \inst{1}
\and Thiem Hoang \inst{2,3}
\and Stefan Taubenberger \inst{1,4}
\and Ferdinando Patat \inst{1}
\and Paola Mazzei \inst{5}
\and Nick L.J. Cox \inst{6}
\and Paula Zelaya \inst{7,8}
\and Stefan Cikota \inst{9,10}
\and Lina Tomasella \inst{5}
\and Stefano Benetti \inst{5}
\and Gabriele Rodeghiero \inst{11,5}} 

\institute{European Southern Observatory, Karl-Schwarzschild-Str. 2, 85748 Garching b. M\"{u}nchen, Germany, \email{acikota@eso.org}
\and Korea Astronomy and Space Science Institute, Daejeon 34055, Korea, \email{thiemhoang@kasi.re.kr}
\and Korea University of Science and Technology, 217 Gajungro, Yuseong-gu, Daejeon, 34113, Korea
\and Max-Planck-Institut f\"{u}r Astrophysik, Karl-Schwarzschild-Str. 1, 85741 Garching b. M\"{u}nchen, Germany
\and INAF-Osservatorio Astronomico di Padova, Vicolo dell'Osservatorio 5, I-35122 Padova, Italy
\and Anton Pannekoek Institute for Astronomy, University of Amsterdam, NL-1090 GE Amsterdam, The Netherlands
\and MAS—Millennium Institute of Astrophysics, Casilla 36-D,7591245, Santiago, Chile
\and Instituto de Astrof\'{i}sica, Pontificia Universidad Cat\'{o}lica de Chile, Casilla 306, Santiago 22, Chile
\and University of Zagreb, Faculty of Electrical Engineering and Computing, Department of Applied Physics, Unska 3, 10000 Zagreb, Croatia
\and Ru{\dj}er Bo\v{s}kovi\'{c} Institute, Bijeni\v{c}ka cesta 54, 10000 Zagreb, Croatia
\and Max-Planck-Institut f\"{u}r Astronomie, K\"{o}nigstuhl 17, D-69117 Heidelberg, Germany
}

\abstract{Highly reddened type Ia Supernovae (SNe Ia) with low total-to-selective visual extinction ratio values, $R_V$, also show peculiar linear polarization wavelength dependencies with peak polarizations at short wavelengths ($\lambda_{\rm max} \lesssim 0.4 {\rm \mu m}$). It is not clear why sightlines to SNe Ia display such different continuum polarization profiles from interstellar sightlines in the Milky Way with similar $R_V$ values. We investigate polarization profiles of a sample of Galactic stars with low $R_V$ values, along anomalous extinction sightlines, with the aim to find similarities to the polarization profiles that we observe in SN Ia sightlines. We undertook spectropolarimetry of 14 stars, and used archival data for three additional stars, and run dust extinction and polarization simulations (by adopting the picket-fence alignment model) to infer a simple dust model (size distribution, alignment) that can reproduce the observed extinction and polarization curves. Our sample of Galactic stars with low $R_V$ values and anomalous extinction sightlines displays normal polarization profiles with an average $\lambda_{\rm max} \sim 0.53 {\rm \mu m}$, and is consistent within 3$\sigma$ to a larger coherent sample of Galactic stars from literature. Despite the low $R_V$ values of dust towards the stars in our sample, the polarization curves do not show any similarity to the continuum polarization curves observed towards SNe Ia with low $R_V$ values. There is a correlation between the best-fit Serkowski parameters $K$ and $\lambda_{\rm max}$, but we did not find any significant correlation between $R_V$ and $\lambda_{\rm max}$. Our simulations show that the $K$--$\lambda_{\rm max}$ relationship is an intrinsic property of polarization. Furthermore, we have shown that in order to reproduce polarization curves with normal $\lambda_{max}$ and low $R_V$ values, a population of large (a $\geq 0.1 \mu m$) interstellar silicate grains must be contained in the dust's composition.}


\maketitle


\section{Introduction}

The motivation to study anomalous sightlines towards highly reddened Galactic stars derives from Type Ia Supernova (SN) observations which show peculiar extinction curves with very low $R_V$ values, as well as peculiar polarization wavelength dependencies (polarization curves).

Past studies that include large samples of SNe Ia show that the total-to-selective visual extinction ratio, $R_V$, of dust in type Ia SN host galaxies ranges from 1 to 3.5, and is in most cases lower than the average value of Milky Way dust, $R_V\sim$3.1 \citep{1996ApJ...473..588R,1999AJ....118.1766P,2004MNRAS.349.1344A,2005ApJ...624..532R,2007ApJ...664L..13C,Wang2006ApJ...645..488W,2008ApJ...686L.103G,Nobili2008A&A...487...19N,2009ApJS..185...32K,2009ApJ...700.1097H,2010AJ....139..120F,2010ApJ...722..566L,2011ApJ...731..120M,2016ApJ...819..152C}. 

Observations of individual highly reddened SNe Ia also reveal host galaxy dust with low $R_V$ values, e.g. $R_V$=2.57$^{+0.23}_{-0.21}$ for the line of sight of SN 1986G \citep{2013ApJ...779...38P}, $R_V \sim$ 1.48 for SN 2006X \citep{WangSN2006X2008ApJ...677.1060W}, $R_V$=1.20 $^{+0.26}_{-0.14}$ for SN 2008fp \citep{2013ApJ...779...38P} and $R^{obs}_V$=1.64$\pm$0.16 for SN 2014J \citep{2014MNRAS.443.2887F}.

Linear (spectro)polarimetric observations of these four SNe also display anomalous interstellar polarization curves \citep{2015A&A...577A..53P}, steeply rising towards blue wavelengths. \citet{2015A&A...577A..53P} fitted their observations with a Serkowski curve \citep{1975ApJ...196..261S} to characterize the polarization curves.

The Serkowski curve is an empirical wavelength dependence of interstellar linear polarization:
\begin{equation}
\label{eq_serkowski}
\frac{P(\lambda)}{P_{\mathrm{\rm max}}} = \exp \left[-K \mathrm{ln}^2 \left( \frac{\lambda_{\mathrm{\rm max}}}{\lambda} \right) \right] 
\end{equation}
The wavelength of peak polarization, $\lambda_{\rm max}$, depends on the dust grain size distribution. For an enhanced abundance of small dust grains, $\lambda_{\rm max}$ moves to shorter wavelengths, and for an enhanced abundance of large dust grains to longer wavelengths. 
Thus, linear spectropolarimetry probes the alignment of dust grains and the size distribution of the aligned dust grains.

\citet{2015A&A...577A..53P} found that the polarization curve of all four SNe display an anomalous behavior, with $\lambda_{\rm max}\sim 0.43 {\rm \mu m}$ for SN 1986G, and $\lambda_{\rm max}\lesssim 0.4 {\rm \mu m}$ for SN 2006X, SN 2008fp and SN 2014J. Because SNe Ia have a negligible intrinsic continuum polarization \citep{2008ARA&A..46..433W}, the anomalous polarization curves likely have to be associated with the properties of host galaxies dust. \citet{Zelaya2017arxiv} expanded the sample of four SNe Ia investigated in \citet{2015A&A...577A..53P}, and present a study of 19 Type Ia SNe. They group the SNe in the "sodium-sample", consisting of 12 SNe which show higher continuum polarization values and interstellar Na I D lines at the redshift of their host galaxies, and the "non-sodium-sample" with no rest-frame Na I D lines and smaller peak polarization. Eight sodium-sample SNe have $\lambda_{\rm max} \lesssim 0.4 {\rm \mu m}$ and their polarization angles are aligned with their host galaxies spiral arms, which is evidence that the polarizing dust is likely located in their host galaxies, and aligned due to the host galaxies magnetic fields. The non-sodium-sample SNe are less polarized, with $P_{\rm max} \lesssim 0.5 \%$, have $\lambda_{\rm max}$ values similar to the common Galactic dust (with $\lambda_{\rm max} \sim 0.55 \mu m$), and their polarization angles do not align with host-galaxy features, which might be interpreted as the continuum polarization being produced by the Galactic foreground dust.

It is not understood why these reddened SN Ia sightlines show such a different polarization profile compared to the typical Milky Way dust. A natural explanation is that the composition of dust in the SN Ia host galaxies is different from that in the Galaxy. 

However, there are alternative explanations. Scattering might explain the low $R_V$ values, as well as the peculiar polarization profiles. As illustrated by \citet{2015A&A...577A..53P} (see their Fig. 6), the polarization profile of SN 2006X may, besides the Serkowski component, also have a component induced by Rayleigh scattering. However, in case a light echo propagates through local dust, we expect to observe variability in $R_V$ and polarization \citep{2005ApJ...635L..33W}, which is usually not the case (see Fig. 4 in \citealt{Zelaya2017arxiv}). \citet{2017ApJ...834...60Y} used HST observations to map the interstellar medium (ISM) around SN 2014J through light echoes. These authors observed two echo components: a diffuse ring and a luminous arc, produced through dust scattering of different grain sizes. From the wavelength dependence of the scattering optical depth, the arc dust favors a small $R_V$ value of $\sim$ 1.4, which is consistent with the $R_V$ measured along the direct line of sight, while the ring is consistent with a common Milky Way $R_V \sim 3$ value.

Another interesting explanation for the peculiar SNe Ia sightlines is given by \citet{2015arXiv151001822H} who simultaneously fits a two-component (interstellar and circumstellar) extinction and polarization model to photometric and (spectro)polarimetric observations of SNe 1986G, 2006X, 2008fp and 2014J, to investigate the grain size distribution and alignment functions of dust along those lines of sights. \citet{2015arXiv151001822H} could reproduce the observational data of SN 1986G and SN 2006X by assuming an enhanced abundance of small silicate grains in the interstellar dust only, while in case of SN 2014J, a contribution of circumstellar (CS) dust must be accounted for. In case of SN 2008fp, \citet{2015arXiv151001822H} found that the alignment of small dust grains must be as efficient as that of big grains, but the existence of CS dust is uncertain. \citet{2015arXiv151001822H} suggests that the enhanced abundance of small silicate grains might be produced by cloud collisions driven by the SN radiation pressure. Strong SN radiation might also induce efficient alignment of small grains via the radiative torque mechanism. However, in case of alignment via the radiative torque mechanism, the polarization angle alignment with host-galaxy features remains unexplained. \\

The aim of this work is to investigate Galactic stars with low $R_V$ values with spectropolarimetry, in order to possibly find similarities to the polarization curves observed towards SNe Ia. Numerical simulations will be used to infer general properties of interstellar dust towards these stars by simultaneously fitting to extinction curves with low $R_V$ values and normal polarization curves.\\

The paper is structured as follows: in Sect.~\ref{sec:stars} we describe our sample of stars, in Sect.~\ref{sec:datareduction} the instruments and observing strategies, in Sect.~\ref{sec:results} we present the data processing and results, in Sect.~\ref{sec:data_analysis} the analysis of the observations, in Sect.~\ref{sec:simulations} we run simulations in order to interpret the observed data, in Sect.~\ref{sec:discussion} we discuss the results, and finally we summarize and conclude in Sect.~\ref{sec:summary}.

\section{Target sample}
\label{sec:stars}

We selected our targets from the samples presented by \citet{2008MNRAS.390..706M,2011A&A...527A..34M}. \citet{2011A&A...527A..34M} obtained 785 extinction curves for sightlines with $E(B-V)$ $\geqslant$ 0.2 mag \citep{1985ApJS...59..397S}, observed with the Astronomical Netherlands Satellite (ANS) in five UV bands (1/$\lambda$ = 6.46, 5.56, 4.55, 4.01 and 3.04 ${\rm \mu m^{-1}}$) \citep{1982A&AS...49..427W}. They combined the UV observations with Two-Micron All-Sky Survey (2MASS) observations in the near-infrared $J$, $H$ and $K$ bands, applied a least square fit of the standard \citet*{1989ApJ...345..245C} extinction curve (CCM) with different $R_V$ values, and determined the residual differences between the observed values and best-fit CCM curve at five UV wavelengths. 
The curves were classified as anomalous if at least one UV wavelength deviated by more than 2$\sigma$ from the best-fit standard CCM curve.
Twenty curves with weaker UV bumps and steeper far-UV slopes (type A); or with stronger bumps and smoother far-UV rises (type B) compared to their best-fit CCM curve, were analyzed in \citet{2008MNRAS.390..706M}. 
\citet{2011A&A...527A..34M} focus on 64 lines of sight for which the corresponding best-fit CCM curve is always well below ($\geqslant 2 \sigma$) or well above the observed data (type C curves), with some exception at 1/$\lambda$ = 3.01 for five curves (see bottom panel of Fig. 1 in \citealt{2011A&A...527A..34M}). They conclude that the sightlines characterized by anomalous type C extinction curves, require lower dust abundances than environments characterized by normal CCM extinction curves.

From those 64 anomalous lines of sight, we selected 14 lines of sight with the lowest $R_V$ values and observed them with the FOcal Reducer and low dispersion Spectrograph (FORS2), the Asiago Faint Object Spectrograph and Camera (AFOSC), and the Calar Alto Faint Object Spectrograph (CAFOS). The observed targets are listed in Table~\ref{tab1}. Additionally we use archival HPOL data for 3 stars.

\begin{table*}
{\tiny
\centering
\caption{List of observed stars.}
\label{tab1}
\begin{tabular}{lllrllclll}
\hline\hline
Name		& RA 				& DEC 				& V	  & B-V		& Spec. &      &          & No. of   & \\
		& (J2000)			& (J2000)			&(mag) &(mag)	&type	& Type & Telescope &  Epochs & Comment\\
\hline
BD +23 3762 & 19 45 42.31 & +23 59 04.0 & 9.34 & 0.62  & B0.5III    & SCI & CAHA   & 1 & HD 344880, Star in Association \\
BD +45 3341 & 20 57 02.68 & +46 32 44.7 & 8.73 & 0.38  & B1II       & SCI & CAHA   & 2 & \\
HD 1337     & 00 17 43.06 & +51 25 59.1 & 6.14 & -0.13 & O9.2II+O8V & SCI & CAHA   & 3 & W UMa type \\
			&			  &				&	   &       &			& & Asiago & 2 & \\
HD 137569   & 15 26 20.82 & +14 41 36.3 & 7.91 & -0.05 & B9Iab:p    & SCI & CAHA   & 2& Post-AGB Star (proto-PN) \\
			&			  &				&	   &       &			& & Asiago & 2 & \\
			&			  &				&	   &       &			& & VLT    & 2 free + 4 GG435 & \\
HD 144579   & 16 04 56.79 & +39 09 23.4 & 6.67 & 0.73  & G8V        & unPolStd & CAHA   & 2 & High proper motion \\
			&			  &				&	   &       &			& & Asiago & 1 & \\
HD 154445   & 17 05 32.26 & -00 53 31.5 & 5.61 & 0.12  & B1V        & PolStd & CAHA   & 2 & HR 6353 \\
HD 194092   & 20 22 05.44 & +40 59 08.2 & 8.28 & 0.09  & B0.5III    & SCI & CAHA   &1 & Star in Cluster \\
			&			  &				&	   &       &			& & Asiago & 1& \\
HD 28446    & 04 32 01.84 & +53 54 39.1 & 5.77 & 0.10  & B0III+B0IV/V & SCI & CAHA & 2& Triple star (DL Cam) \\
			&			  &				&	   &       &			& & Asiago & 1& \\
HD 43384    & 06 16 58.71 & +23 44 27.3 & 6.25 & 0.45  & B3Iab      & PolStd & CAHA   & 2 & Pulsating variable Star \\ 
			&			  &				&	   &       &			& & Asiago & 3& \\
HD 90508    & 10 28 03.88 & +48 47 05.7 & 6.43 & 0.60  & G0V        & unPolStd & CAHA   & 1 & Double star \\
			&			  &				&	   &       &			& & Asiago & 2& \\
HD 39587 	& 05 54 22.98 & +20 16 34.2 & 4.40 & 0.60  & G0VCH+M    & unPolStd & Asiago & 2& Variable of RS CVn type \\
HD 54439    & 07 08 23.20 & -11 51 08.6 & 7.68 & 0.05  & B2/3II     & SCI & Asiago & 1& \\
			&			  &				&	   &       &			& & VLT    & 1 free + 1 GG435 & \\
HD 14357    & 02 21 10.44 & +56 51 56.4 & 8.52 & 0.31  & B2III      & SCI & Asiago & 1& Star in cluster \\	
HD 21291    & 03 29 04.13 & +59 56 25.2 & 4.22 & 0.41  & B9Ia       & PolStd & Asiago & 1& Pulsating variable Star \\
HD 73420    & 08 36 37.12 & -44 04 48.2 & 8.85 & 0.07  & B2III/III  & SCI & VLT    & 1 free + 1 GG435& \\
HD 78785    & 09 08 24.09 & -46 15 13.3 & 8.60 & 0.51  & B2III      & SCI & VLT    & 1 free + 2 GG435& \\
HD 96042    & 11 03 40.56 & -59 25 59.1 & 8.23 & 0.18  & B1(V)ne    & SCI & VLT    & 2 free + 2 GG435 &  Emission-line Star \\
HD 141318   & 15 51 06.80 & -55 03 19.9 & 5.77 & -0.01 & B2III      & SCI & VLT    & 2 free + 2 GG435 & Pulsating variable Star \\
HD 152245   & 16 54 00.48 & -40 31 58.2 & 8.37 & 0.13  & B0Ib       & SCI & VLT    & 1 free + 2 GG435 & Star in Cluster \\ 
HD 152853   & 16 58 07.93 & -45 58 56.5 & 7.94 & 0.11  & B2III      & SCI & VLT    & 1 free + 1 GG435 & Star in Cluster \\ 
\hline
\end{tabular}
\tablefoot{The coordinates, brightness and spectral type were taken from the SIMBAD Astronomical Database. Type indicates if the star is a polarized standard star (PolStd), unpolarized standard star (unPolStd) or one of our science targets (SCI). No. of Epochs is the number of epochs observed with a particular instrument: FORS2 (VLT), CAFOS (CAHA) or AFOSC (Asiago).}
}
\end{table*}

\section{Instruments and methods}
\label{sec:datareduction}

We observed our targets using three different instruments and telescopes: the FOcal Reducer and low dispersion Spectrograph (FORS2) in spectropolarimetric mode (PMOS) mounted on the UT1 Cassegrain focus of the Very Large Telescope (VLT) in Chile; the Asiago Faint Object Spectrograph and Camera (AFOSC) mounted at the 1.82 m Copernico telescope at the Asiago Observatory in northern Italy; and the Calar Alto Faint Object Spectrograph (CAFOS) mounted at the Calar Alto 2.2 m telescope in Andalusia, Spain. 

The characteristics of the instruments and corresponding differences in the data reduction are described in the following subsections.

\subsection{FORS2 at the VLT}
\label{subsection_FORS2}

FORS2 in PMOS mode is a dual-beam polarimeter. The spectrum produced by the grism is split by the Wollaston prism into two beams with orthogonal directions of polarization: ordinary (o) and extraordinary (e) beam. The data used in this work were obtained with the 300V grism, with and without the GG435 filter, and with the half-wave retarder plate positioned at angles of 0$^{\circ}$, 22.5$^{\circ}$, 45$^{\circ}$, and 67.5$^{\circ}$ (Program ID: 094.C-0686). The half-wave retarder plate angle is measured between the acceptance axis of the ordinary beam of the Wollaston prism (which is aligned to the north-south direction) and the fast axis of the retarder plate.

The data were reduced using standard procedures in IRAF. Wavelength calibration was achieved using He-Ne-Ar arc lamp exposures. The typical RMS accuracy is $\sim$ 0.3 \AA. The data have been bias subtracted, but not flat field corrected. However, the effects of improper correction were minimized by taking advantage of the redundant number of half-wave positions \citep[see][]{2006PASP..118..146P}.

Ordinary and extra-ordinary beams were extracted in an unsupervised way using the PyRAF apextract.apall procedure, with a fixed aperture size of 10 pixels.
The synthetic broad-band polarization degree was computed by integrating the total flux weighted with Bessel's BVRI passband filters. We binned the spectra in 50 $\AA$ bins, in order to obtain a larger signal-to-noise ratio, and calculated the Stokes parameters $Q$ and $U$, polarization degree $P$, and polarization angle $\theta_P$ as a function of wavelength.

The Stokes parameters $Q$ and $U$ were derived via Fourier transformation, as described in the FORS2 User Manual \citep{FORS2manual}:
\begin{equation}
\begin{array}{l}
Q = \frac{2}{N} \sum_{i=0}^{N-1} F(\theta_i)\cos(4\theta_i)  \\ 
U = \frac{2}{N} \sum_{i=0}^{N-1} F(\theta_i)\sin(4\theta_i)
\end{array}
\end{equation}
where $F(\theta_i)$ are the normalized flux differences between the ordinary ($f^o$) and extra-ordinary ($f^e$) beams:
\begin{equation}
\label{eqF}
F(\theta_i) = \frac{f^o (\theta_i) - f^e (\theta_i)}{f^o (\theta_i) + f^e (\theta_i)}
\end{equation}
at different half-wave retarder plate position angles $\theta_i = i * 22.5^{\circ}$.

Although FORS2 is equipped with a super-achromatic half wave plate, residual retardance chromatism is present. The wavelength dependent retardance offset ($\Delta\theta(\lambda)$) is tabulated in the FORS2 User Manual. The chromatism was corrected through the following rotation of the Stokes parameters:
\begin{equation}
\begin{array}{l}
Q_0 = Q \cos 2\Delta\theta(\lambda) - U \sin 2\Delta\theta(\lambda) \\
U_0 = Q \sin 2\Delta\theta(\lambda) + U \cos 2\Delta\theta(\lambda)
\end{array}
\end{equation}

Finally we calculated the polarization: 
\begin{equation}
\label{eqP}
P=\sqrt{Q^2+U^2}
\end{equation}
and the polarization angle:
\begin{equation}
\label{eqtheta}
\theta_0 = \frac{1}{2}\arctan(U_0/Q_0).
\end{equation}

The reliability of data obtained with FORS2 is demonstrated in \citet{2017MNRAS.464.4146C}. They used archival data of polarized and unpolarized stars to test the stability and capabilities of the spectropolarimetric mode (PMOS) of the FORS2 instrument, and found a good temporal stability since FORS2 was commissioned, and a good observational repeatability of total linear polarization measurements with an RMS $\lesssim$ 0.21$\%$.
\citet{2017MNRAS.464.4146C} also found a small ($\lesssim$ 0.1$\%$) instrumental polarization and fit linear functions to correct Stokes $Q$ and $U$, which we apply to the FORS2 data in this work.

\subsection{AFOSC at the 1.82 m Copernico telescope}
\label{sec:subsection_AFOSC}

Spectropolarimetry with AFOSC was obtained using a simple combination of two Wollaston prisms and two wedges, a grism and a slit mask of 2.5 arcsec wide and 20 arcsec long slitlets. This configuration permits measurements of the polarized flux at four polarimetric channels simultaneously, i.e. at angles 0, 45, 90 and 135 degrees, without the need of a half-wave retarder plate \citep{1997A&AS..123..589O}. 
For any given rotator adapter angle $\theta_i$ there are four fluxes that can be measured. We group them to two, which we call ordinary (O) and extraordinary (E). We indicate them as: $f_{O1,i}$, $f_{E1,i}$ and $f_{O2,i}$, $f_{E2,i}$. We will use them to indicate the generic four beams $f_0$, $f_{90}$ and $f_{45}$, $f_{135}$, respectively.

However, in order to remove possible instrumental problems (i.e. non perfect beam splitting, flat fielding, etc.) it is convenient to obtain at least two sets of data. This can be achieved by rotating the instrument by 90 degrees with respect to the sky, so that a pair-wise swap between the corresponding polarimeter channels, 0 to 90 and 45 to 135 degrees, is performed. 

The final $Q$ and $U$ were obtained via the Fourier approach:
\begin{equation}
\begin{array}{l}
Q = \frac{1}{N} \sum_{i=0}^{N-1} (F_{1,i}\cos(2\theta_i) - F_{2,i}\sin(2\theta_i))  \\ 
U = \frac{1}{N} \sum_{i=0}^{N-1} (F_{1,i}\sin(2\theta_i) + F_{2,i}\cos(2\theta_i))
\end{array}
\end{equation}
where N is the number of rotator adapter angles, $\theta_i=\frac{\pi}{4} i$, and $F_{1,i}$ and $F_{2,i}$ are normalized flux ratios:
\begin{equation}
\begin{array}{l}
F_{1,i} = \frac{f_{O1,i} - f_{E1,i}}{f_{O1,i} + f_{E1,i}}  \\
F_{2,i} = \frac{f_{O2,i} - f_{E2,i}}{f_{O2,i} + f_{E2,i}}
\end{array}
\end{equation}

Finally we calculated the polarization degree and angle as given in Eq. (\ref{eqP}) and (\ref{eqtheta}) respectively.

\subsection{CAFOS at the Calar Alto 2.2 m telescope}
\label{subsection_CAFOS}

CAFOS is a dual-beam polarimeter, similar to FORS2 in PMOS mode, composed of a half-wave retarder followed by a Wollaston prism which splits the incoming beam in an ordinary and extraordinary beam.
The data processing is as described for FORS2 in Sect.~\ref{subsection_FORS2}.

The CAFOS instrument was characterized in \citet{2011A&A...529A..57P}. They used polarized standard stars to quantify the HWP chromatism which causes a peak-to-peak oscillation of $\sim$ 11 degrees. From observations of unpolarized standard stars they found an instrumental polarization likely produced by the telescope optics which appears to be additive. The instrumental polarization is $\sim$ 0.3 $\%$ between 4000 $\AA$ and 8600 $\AA$, and  grows to $\sim$ 0.7 $\%$ below 4000 $\AA$. It can be removed by subtracting the instrumental components in the $Q$--$U$ Stokes plane. After correcting for the HWP chromatism and instrumental polarization, \citet{2011A&A...529A..57P} concluded that an accuracy of $\sim$ 0.1 $\%$ can be reached with four HWP angles and a sufficient signal-to-noise ratio.

\section{Data processing and results}
\label{sec:results}

The data were obtained with FORS2 during 8 different nights between 2014-10-10 and 2015-02-06 (Program ID: 094.C-0686), with CAFOS during the night of 2015-04-29, and with AFOSC during 5 nights at three observing runs starting on 2015-02-09, 2015-03-09, and 2016-08-02.

\subsection{Standard stars}

We investigate the accuracy and reliability of the instruments using unpolarized and polarized standard stars.

\citet{2017MNRAS.464.4146C} used archival data of 8 unpolarized standard stars observed at 40 epochs between 2009 and 2016 to test the stability and capabilities of the spectropolarimetric mode (PMOS) of the FORS2 instrument. They showed that the polarization degree and angle are stable at the level of $\lesssim$ 0.1$\%$ and $\lesssim$ 0.2 degrees, respectively. They found a small ($\lesssim$ 0.1$\%$) wavelength dependent instrumental polarization and derived linear functions for the Stokes $Q$ and $U$, which we apply to the observed Stokes parameters.

Thus, in this paper, we will focus on unpolarized and polarized standard stars observed with CAFOS and AFOSC only.

Two unpolarized standard stars (HD\,144579, and HD\,90508), and two polarized standard stars (HD\,154445 and HD\,43384) were observed with CAFOS. We did not find any significant instrumental polarization in the CAFOS observations, and the polarization values are consistent with the literature. The results are given in Sect.~\ref{sec:Standard stars with CAFOS}.

We used observations of 3 unpolarized standard stars to investigate possible instrumental polarization of AFOSC: HD\,90508, HD\,39587 and HD\,144579; and three polarized standard stars to test the reliability: HD\,43384, HD\,21291, and HD\,198478. We did not detect any significant instrumental polarization, however, the polarization degrees of polarized stars observed at different epochs vary by $\sim$0.3$\%$. The inconsistencies might be caused by diffraction of light from the edge of the slit \citep[see Keller C.U. in ][p. 303]{2002assp.book.....T}, or by an inaccuracy of the instruments rotation angle (see \citet{2017A&A...608A.146B}). The results are given in Sect.~\ref{sec:Standard stars with AFOSC}.

\subsection{FORS2 science data}

FORS2 is the most stable instrument used in this work and we are confident that the data gained with FORS2 are accurate and can be used as reference for comparison to other instruments (see \citealt{2017MNRAS.464.4146C}). Eight stars with anomalous extinction sightlines were observed with FORS2 (see Table~\ref{tab1}). HD 54439 was also observed with AFOSC, and HD\,137569 was observed with all three instruments, FORS2, CAFOS and AFOSC, which we will briefly discuss in Sect.~\ref{sub_HD_137569}. HD\,78785, HD\,141318, HD\,152853, HD\,152245, HD\,73420 and HD\,96042 were observed with FORS2 only.

\begin{figure}[h!]
\begin{center}
\includegraphics[trim=0mm 50mm 0mm 50mm, width=9.5cm, clip=true]{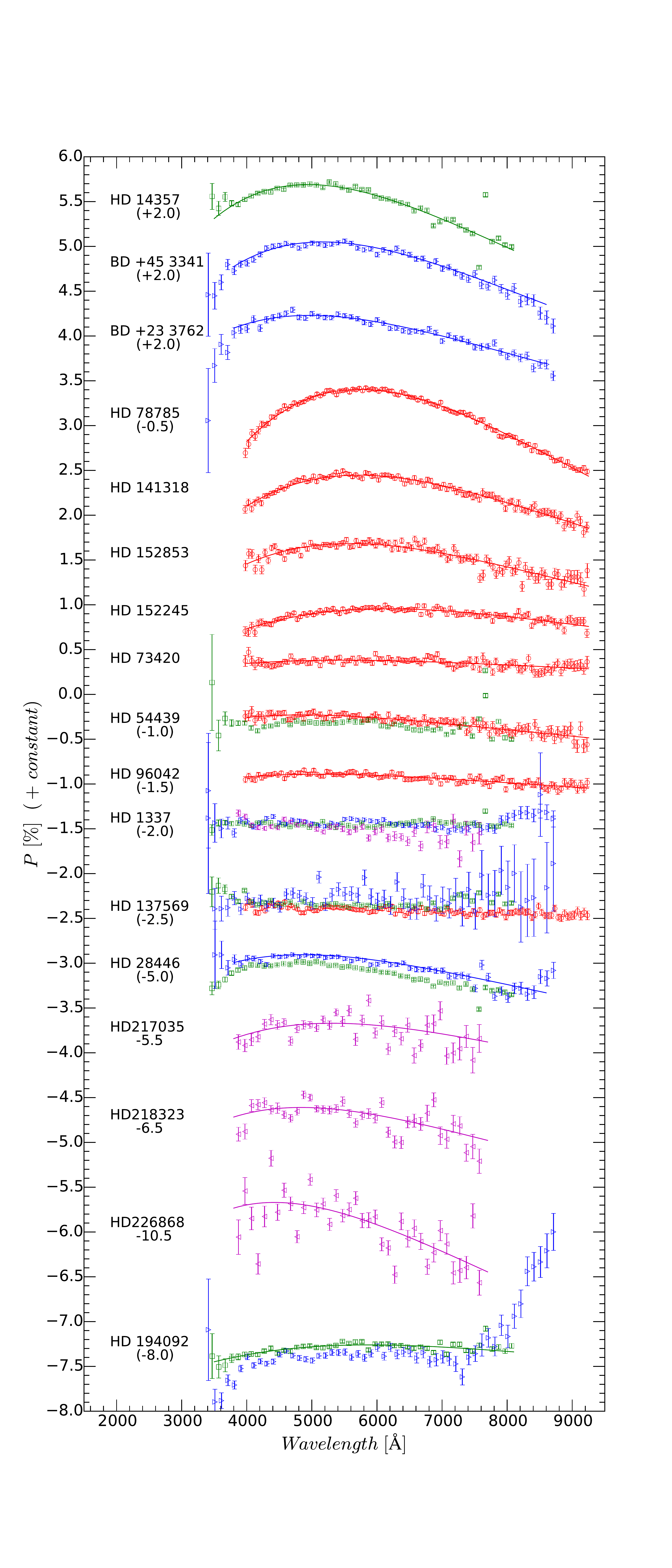}
\caption{Weighted averages of observed polarization curves for all science targets, derived with different instruments. The red circles show observations performed with FORS2, blue triangles with CAFOS, green squares with AFOSC, and left pointing purple triangles with HPOL. The full lines denote the Serkowski fits, as parametrized in Table~\ref{tab_Final_Serkowski_parameters}.}
\label{fig_all_SCI_stars}
\end{center}
\end{figure}

We extracted the spectra and calculated the polarization dependencies as described in Sect.~\ref{subsection_FORS2}. The correction for the instrumental polarization determined in \citet{2017MNRAS.464.4146C} was also applied to Stokes $Q$ and $U$. 
The targets have been observed with and without the GG435 filter. 
The GG435 filter blocks the blue light and thus prevents the second order spectrum. However, the effect in polarization is very small, and significant only for very blue spectral energy distributions and when measuring line polarization \citep{2010A&A...510A.108P}. For our reddened targets, the second order polarization is negligible. Therefore, for wavelengths $\lambda$>4250$\AA$ we calculated the weighted mean of all epochs, and for wavelengths $\lambda$<4250$\AA$ we calculated the weighted mean of all epochs taken without the GG435 filter. We then merged both ranges to one polarization spectrum and parameterized it by fitting a Serkowski curve (Eq.~\ref{eq_serkowski}) to the data. The individual results can be found in Table~\ref{tb_pol_individual}. The polarization dependencies are shown in Fig.~\ref{fig_all_SCI_stars}, and the Serkowski parameters are given in Table~\ref{tab_Final_Serkowski_parameters}.

\subsubsection{HD 137569}
\label{sub_HD_137569}

HD 137569 is an interesting case because it has relatively high reddening $E(B-V) \sim$ 0.40 mag \citep{2011A&A...527A..34M}, but its polarization degree is consistent with zero (Fig.~\ref{fig_all_SCI_stars}). Its mean Stokes Q and U are -0.07 $\pm$ 0.05 $\%$ and 0.01 $\pm$ 0.03 $\%$. HD\,137569 was also observed with CAFOS and AFOSC, and the results are consistent with the FORS2 observations. Furthermore, HD 137569 is a spectroscopic binary with a period of 529.8 days, and shows observational signatures normally seen in post-AGB stars \citep{2005A&A...443..297G}. This is interesting, because sightlines to post-AGB stars usually show  continuum polarization \citep{1991AJ....101.1735J}.

\subsection{CAFOS science data}
There are six stars with anomalous extinction sightlines observed with CAFOS. HD\,1337 and HD\,28446 were also observed with AFOSC, and HD\,137569 was additionally observed with both, FORS2 and AFOSC (see also Sect.~\ref{sub_HD_137569}). BD\,+45\,3341, BD\,+23\,3762 and HD\,194092 were observed with CAFOS only.

After the beams extraction using PyRAF's apextract.apall procedure, we bin the spectra in 100$\AA$ wide bins, and calculate the polarization as described in Sect.~\ref{subsection_CAFOS}. Finally, we fit the Serkowski curve in the range between 3800-8600$\AA$.
The individual results can be found in Table~\ref{tb_pol_individual_CAFOS}. The polarization dependencies are shown in Fig.~\ref{fig_all_SCI_stars}, and the Serkowski parameters are listed in Table~\ref{tab_Final_Serkowski_parameters}. 

Below we will discuss the interesting cases.

\subsubsection{HD 1337}
HD 1337 is a close spectroscopic binary star, with a period of 3.52 days \citep{2004A&A...424..727P}, and may be surrounded by a common-envelope, where significant dust amounts may be produced \citet{2013ApJ...768..193L}. HD 1337 has a constant polarization degree of $\sim$ 0.55 $\%$. The CAFOS observations are also consistent with the AFOSC observations, which confirms the polarization wavelength independence (see Fig.~\ref{fig_all_SCI_stars}). In this case the Serkowski fit is not meaningful.

\subsubsection{HD 194092}

For HD 194092, the polarization observed by CAFOS follows a Serkowski curve until 7250$\AA$ from where it starts to steeply increase from p$\sim$ 0.5 $\%$ to p$\sim$ 2 $\%$ at 8650 $\AA$. The steep increase is not present in the AFOSC observations, and is an artifact, which we could not explain. Thus, we fit the Serkowski curve in the range from 3800-7250$\AA$, and find $\lambda_{\rm max}$ = 5728 $\pm$ 235 $\AA$, $p_{\rm max}$ = 0.64 $\pm$ 0.01 $\%$ and $K$ = 1.46 $\pm$ 0.47.

\subsection{AFOSC science data}

Six stars with anomalous extinction sightlines have been observed with AFOSC. HD\,28446, HD\,1337 and HD\,194092 were also observed with CAFOS, HD\,54439 with FORS2, while HD\,137569 was additionally observed with CAFOS and FORS2 (see also Sect.~\ref{sub_HD_137569}). HD 14357 was observed with AFOSC only.

We extracted the beams from 3400-8150 $\AA$, using the same standard procedures in IRAF as for the extraction of FORS2 and CAFOS spectra, binned the data to 100$\AA$ wide bins, and calculated the polarization as described in Sect.~\ref{sec:subsection_AFOSC}. 

The polarization spectra from 3500-8150$\AA$ were fitted with a Serkowski curve, excluding the range from 7500-7700 $\AA$, which is contaminated by the telluric O$_{\rm 2}$ line.

The unpolarized standard stars are consistent with zero, which implies that there is no significant instrumental polarization (Sect.~\ref{sec:Standard stars with AFOSC}). However, 
based on the measurements of polarized standard stars HD 43384 and HD 21291 (see Sect.~\ref{sec:Standard stars with AFOSC}) which show a negative offset compared to the literature values, we conclude that the accuracy of the polarization measurements is within $\sim$\,0.4\,$\%$.
HD 28446 and HD 54439 show an offset of $\sim$\,-0.1\,$\%$ compared to the results achieved with CAFOS and FORS2 respectively, while the measurements of HD 1337 are consistent with the CAFOS measurements, and the measurements of HD 137569 are consistent with the FORS2 and CAFOS measurements (Fig.~\ref{fig_all_SCI_stars}). The individual results can be found in Table~\ref{tb_pol_individual_AFOSC}.

\subsubsection{HD 14357}

HD 14357 is the only AFOSC target which has no common observations with an other instrument. 
From the Serkowski fit, we determined $\lambda_{\rm max}$ = 4942 $\pm$ 31 $\AA$, $p_{\rm max}$ = 3.69 $\pm$ 0.01 $\%$ and $K$ = 0.91 $\pm$ 0.04. Based on HD\,43384 and other polarized stars, we believe that the $\lambda_{\rm max}$ and $K$ values are accurate, while there might be a negative offset, $\lesssim$ 0.4 $\%$, to the true value of $p_{\rm max}$.

\subsection{HPOL science data}

We found archival data for HD\,1337 (which was also observed with CAFOS and AFOSC), and three additional stars of the \citet{2011A&A...527A..34M} sample in the University of Wisconsin's Pine Bluff Observatory (PBO) HPOL spectropolarimeter (mounted at the 0.9 m f/13.5 cassegrain) data set. All targets were observed prior to the instrument update in 1995, when HPOL was providing spectropolarimetry over the range of 3200$\AA$ to 7750$\AA$, with a spectral resolution of 25$\AA$. A halfwave plate was rotated to 8 distinct angles to provide the spectropolarimetric modulation \citep{1996AJ....111..856W}. 

The HPOL data are available in the Mikulski Archive for Space Telescopes (MAST) at the Space Telescope Science Institute, and include the Stokes parameters $Q$ and $U$, and the error, as a function of wavelength.

We calculated the polarization and polarization angle using equations \ref{eqP} and \ref{eqtheta}, and fit a Serkowski curve to the data (equation \ref{eq_serkowski}). The results are given in Table~\ref{tab_Final_Serkowski_parameters}.

\subsection{Literature science data}

We find polarization measurements in the literature \citep{1974AJ.....79..581C, 1975ApJ...196..261S} of 8 stars of the \citet{2011A&A...527A..34M} sample. Three of the stars have been observed in this work with FORS2, AFOSC or CAFOS, and for two stars we found archival HPOL data. The stars were observed with broad band polarimeters, and characterized by fitting the Serkowski curve (Table~\ref{tab_literature}). However, because the authors assumed a fixed K=1.15, we could only partially use the measurements from \citet{1974AJ.....79..581C, 1975ApJ...196..261S} in our further analysis.

\begin{table}
{\tiny
\centering
\caption{$\lambda_{\rm max}$ and $P_{\rm max}$ from literature}
\label{tab_literature}
\begin{tabular}{lccl}
\hline\hline
Name & 	  $\lambda_{\rm max}$ & $P_{\rm max}$ &Reference\\
   	 &    ($\AA$)  & ($\%$) & \\
\hline
HD 2619   &  4900 $\pm$ 100 & 4.82 $\pm$ 0.32 & \citet{1974AJ.....79..581C}  \\
HD 37061  &  6300 $\pm$ 400 & 1.63 $\pm$ 0.19 & \citet{1974AJ.....79..581C} \\
HD 168021 &  5900 $\pm$ 100 & 2.13 $\pm$ 0.03 & \citet{1975ApJ...196..261S} \\
HD 28446\tablefootmark{b,c}  &  5500 $\pm$ 100 & 2.01 $\pm$ 0.10 & \citet{1974AJ.....79..581C}  \\
HD 226868\tablefootmark{d} &  5000 $\pm$ 100 & 5.04 $\pm$ 0.25 & \citet{1974AJ.....79..581C} \\
HD 218323\tablefootmark{d} &  5200 $\pm$ 200 & 1.95 $\pm$ 0.14 & \citet{1974AJ.....79..581C} \\
HD 78785\tablefootmark{a}  &  5800 $\pm$ 224 & 4.05 $\pm$ 0.04 & \citet{1975ApJ...196..261S} \\
HD 141318\tablefootmark{a} &  5700 $\pm$ 100 & 2.42 $\pm$ 0.08 & \citet{1975ApJ...196..261S} \\
\hline
\end{tabular}
\tablefoot{
\tablefoottext{a}{Also observed with FORS2.} 
\tablefoottext{b}{Also observed with AFOSC.} 
\tablefoottext{c}{Also observed with CAFOS.} 
\tablefoottext{d}{Also observed with HPOL.}
}
}
\end{table}

\begin{table}[h]
{\tiny
\centering
\caption{Final Serkowski parameters}
\label{tab_Final_Serkowski_parameters}
\begin{tabular}{lllll}
\hline\hline
&& \multicolumn{3}{c}{Serkowski parameters} \\\cline{3-5}
Name		       & Telescope  &  $\lambda_{\rm max}$ &  $p_{\rm max}$  & $K$  \\
	           &            &     ($\AA$)     &   ($\%$)    &    \\
\hline
HD 137569\tablefootmark{a} & VLT   & \dots & $\sim$ 0.1 & \dots                              \\
HD 54439       & VLT  & 4859 $\pm$ 129 & 0.77 $\pm$ 0.01 &  0.97 $\pm$ 0.10   \\
HD 73420       & VLT   & 5465 $\pm$ 175 & 0.38 $\pm$ 0.01 &	1.04 $\pm$ 0.21  \\
HD 78785       & VLT   & 5732 $\pm$ 8 & 3.90 $\pm$ 0.01 & 	1.25 $\pm$ 0.01  \\
HD 96042       & VLT  & 5109 $\pm$ 124 & 0.61 $\pm$ 0.01 &  0.84 $\pm$ 0.09   \\
HD 141318      & VLT   & 5719 $\pm$ 17 & 2.45 $\pm$ 0.01 & 	1.19 $\pm$ 0.03  \\
HD 152245      & VLT   & 6169 $\pm$ 33 & 0.96 $\pm$ 0.01 & 	1.41 $\pm$ 0.07  \\ 
HD 152853      & VLT   & 5584 $\pm$ 46 & 1.69 $\pm$ 0.01 & 	1.30 $\pm$ 0.07  \\ 
BD +23 3762    & CAHA  & 4965 $\pm$ 61 & 2.23 $\pm$ 0.01  & 0.92 $\pm$ 0.06   \\
BD +45 3341    & CAHA  & 5166 $\pm$ 31 & 3.05 $\pm$ 0.01 & 1.00 $\pm$ 0.05    \\
HD 1337\tablefootmark{a}  & CAHA & \dots & 0.55 $\pm$ 0.01 & \dots    \\
HD 28446       & CAHA  & 4865 $\pm$ 76 & 2.10 $\pm$ 0.01 & 	0.70 $\pm$ 0.07  \\
HD 194092      & Asiago & 5884 $\pm$ 107 & 0.74 $\pm$ 0.01 & 1.09 $\pm$ 0.18  \\
HD 14357       & Asiago & 4942 $\pm$ 31 & 3.69 $\pm$ 0.01 & 0.91 $\pm$ 0.04   \\
HD 226868    & HPOL & 4425 $\pm$ 262 & 4.83 $\pm$ 0.04 &  0.57 $\pm$  0.14    \\ 
HD 218323    & HPOL & 4837 $\pm$ 128 & 1.89 $\pm$ 0.01 &  1.00 $\pm$  0.20    \\ 
HD 217035    & HPOL &  5309 $\pm$ 126 & 1.83 $\pm$ 0.02 &  0.88 $\pm$  0.23   \\ 
\hline
\end{tabular}
\tablefoot{
\tablefoottext{a}{HD 137569 and HD 1337 have constant polarization curves which could not be fit with a Serkowski curve.}
}
}
\end{table}

\section{Data analysis}
\label{sec:data_analysis}

\begin{figure*}
\begin{center}
\includegraphics[trim=20mm 5mm 20mm 20mm, width=19cm, clip=true]{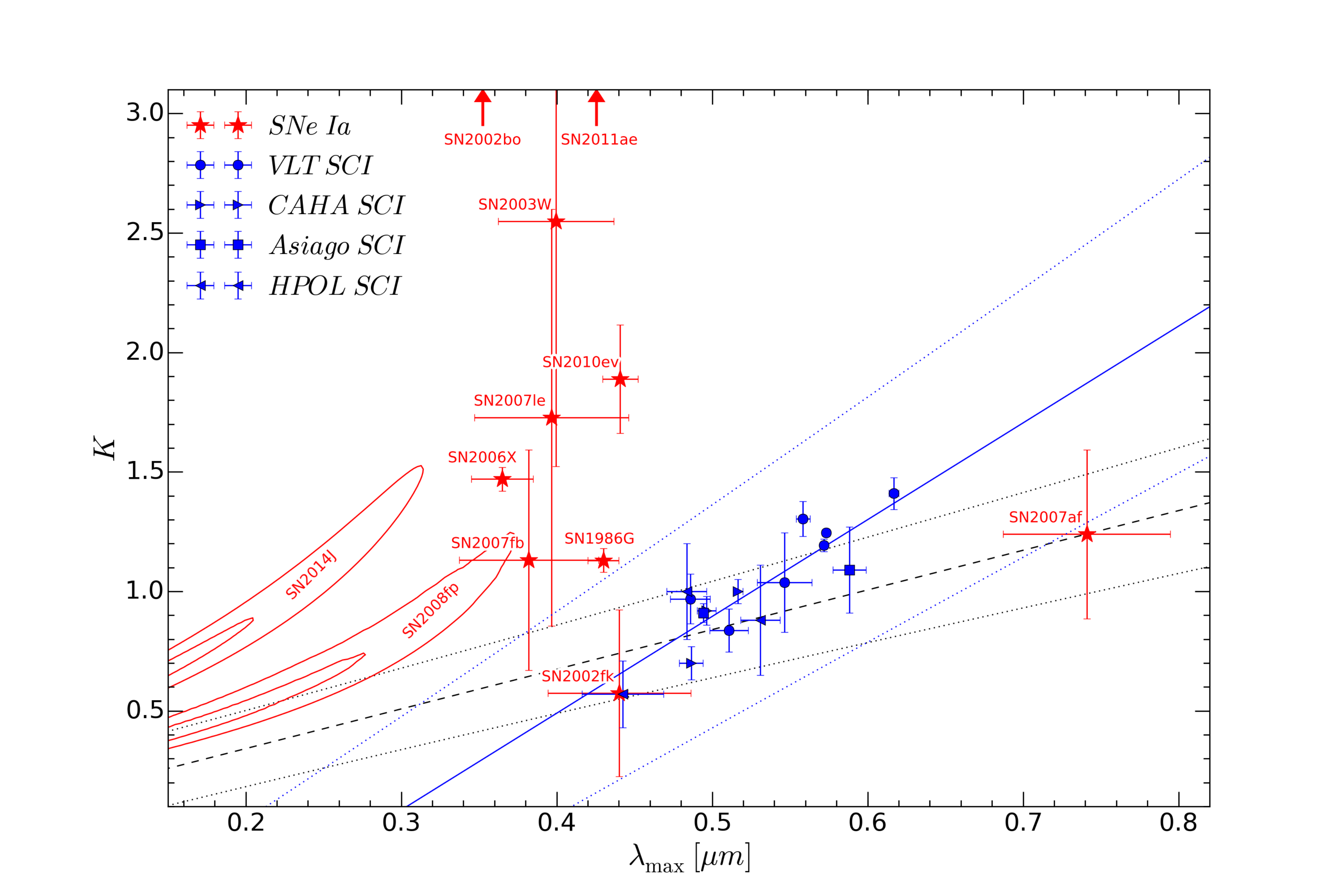}
\caption{Stars with anomalous extinction sightlines in the $\lambda_{\rm max}$-K plane. Blue color shows stars with anomalous sightlines observed with FORS2 (circles), CAHA (right pointing triangles), AFOSC (squared), and HPOL (left pointing triangles). The solid blue line represents the linear best fit to the sample, and it's 1$\sigma$ deviation (dotted). The dashed black line traces the \citet{1992ApJ...386..562W} relation and its 3$\sigma$ uncertainty (dotted). For comparison, a sample of SNe Ia from \citet{2015A&A...577A..53P} and \citealt{Zelaya2017arxiv} (see Appendix~\ref{appendix:fitserkowskitozelaya}) are marked with star symbols, and red contours, which indicate the 10 and 20 $\sigma$ confidence levels for SN 2008fp and SN 2014J.}
\label{fig_lmax-K}
\end{center}
\end{figure*}

Figure~\ref{fig_lmax-K} shows the sample of 15 anomalous sightlines (listed in Table~\ref{tab_Final_Serkowski_parameters}, excluding HD\,137569 and HD\,1337), compared to a sample of Galactic stars observed by \citet{1992ApJ...386..562W} and a sample SNe Ia from \citet{2015A&A...577A..53P} and \citealt{Zelaya2017arxiv} (see Appendix~\ref{appendix:fitserkowskitozelaya}).

\begin{figure}
\begin{center}
\includegraphics[trim=10mm 0mm 10mm 10mm, width=9cm, clip=true]{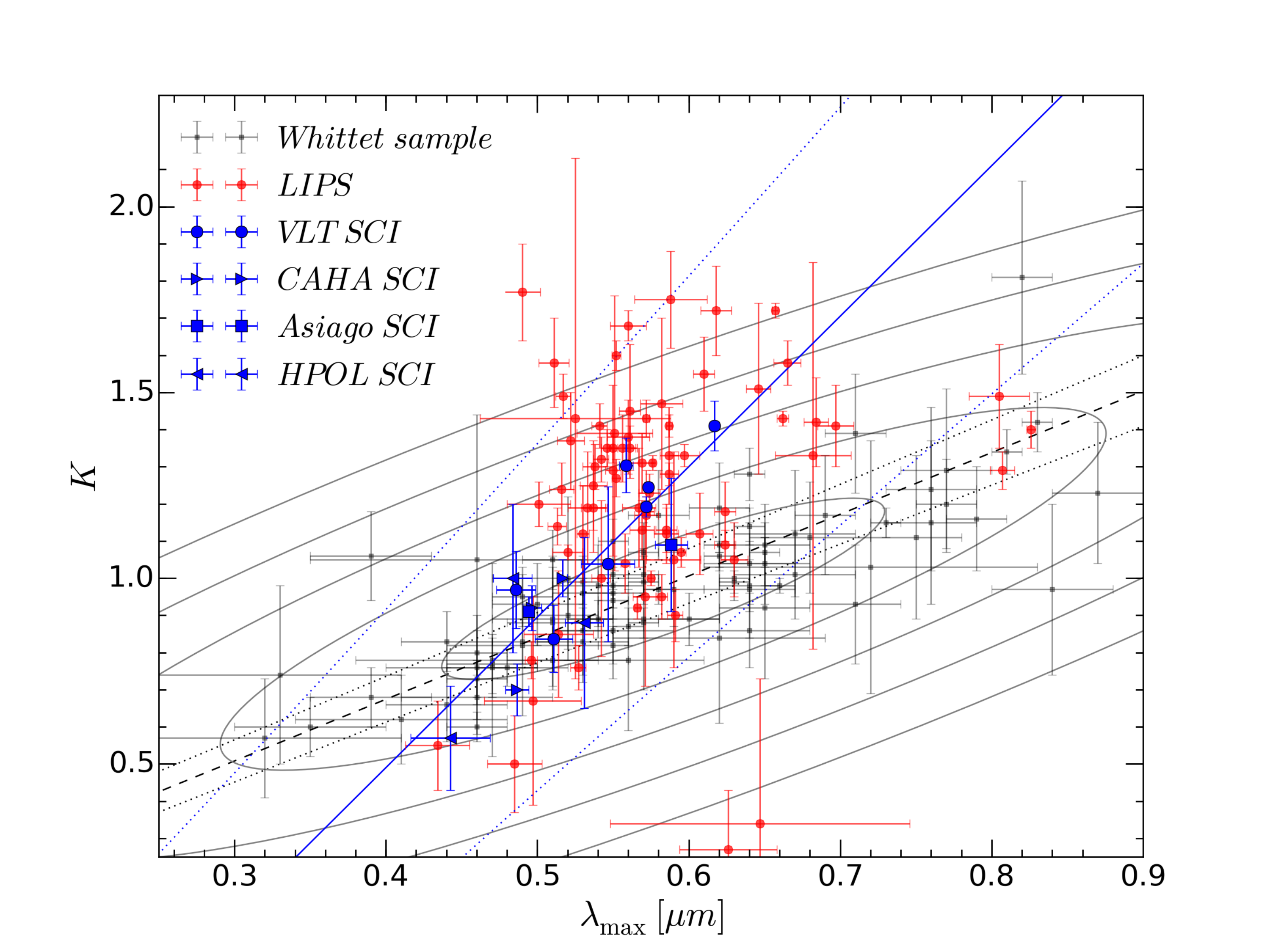}
\caption{Comparison of our sample of stars with anomalous extinction sightlines (blue symbols) to the \citet{1992ApJ...386..562W} sample (gray squares) and the LIPS sample (\citet{2017A&A...608A.146B}, red circles). The gray ellipses represent 1-5$\sigma$ confidence levels for the \citet{1992ApJ...386..562W} sample.  The dashed black line traces the \citet{1992ApJ...386..562W} relation and its 1$\sigma$ uncertainty (dotted). }
\label{fig_lmax-K-comparison}
\end{center}
\end{figure}

\citet{2011A&A...527A..34M} used models of \citet[][hereafter WD01]{Weingartner2001ApJ...548..296W} and the updates by \citet{Draine2007ApJ...657..810D} to compute grain-size distributions for spherical grains of amorphous Silicate and carbonaceous grains consisting of graphite grains and polycyclic aromatic hydrogenated (PAH) molecules. As described in \citet{2011A&A...527A..34M}, they best-fit the extinction curves with models as described above (Sect.~\ref{sec:stars}) to derive the properties of the dust in terms of dust-to-gas ratios, abundance ratios, and small-to-large grain size ratios of carbon and of silicon.
They excluded HD 1337 (a W UMa type variable) and HD 137569 (a post-AGB star) from their analysis before performing the modeling, because of extremely low CCM $R_V$ values of these sightlines, $R_V$ $\approx$ 0.6 and $R_V$ $\approx$ 1.1 respectively, well outside the range explored by CCM extinction curves.

We combine the $\lambda_{\rm max}$ values from this work with the results of WD01 best-fit models of \citet[][their Table 4]{2011A&A...527A..34M}, and compute the Pearson's correlation coefficient, $\rho$, and the p-value for testing non-correlation between $\lambda_{\rm max}$ and their derived dust-to-gas ratio ($\frac{\rho_d}{\rho_{\rm H}}$), carbon and silicon abundances compared to solar values ($\frac{\rm C}{\rm C_{\odot}}$, $\frac{\rm Si}{\rm Si_{\odot}}$), $R_V$ value, the ratio between reddening and total hydrogen column density ($\frac{E(B-V)}{N_{\rm H}}$), and small-to-large grain size ratios of carbon ($R_{\rm C}$) and silicon ($R_{\rm Si}$). They considered grains as small if their size is $\leqslant$ 0.01 ${\rm \mu m}$, and otherwise as large. Note that a detailed analysis and discussion of the WD01 best-fit model results for the whole sample of 64 anomalous sightlines is given in \citet{2011A&A...527A..34M}.

\begin{figure}[h!]
\begin{center}
\includegraphics[trim=7mm 0mm 13mm 10mm, width=9cm, clip=true]{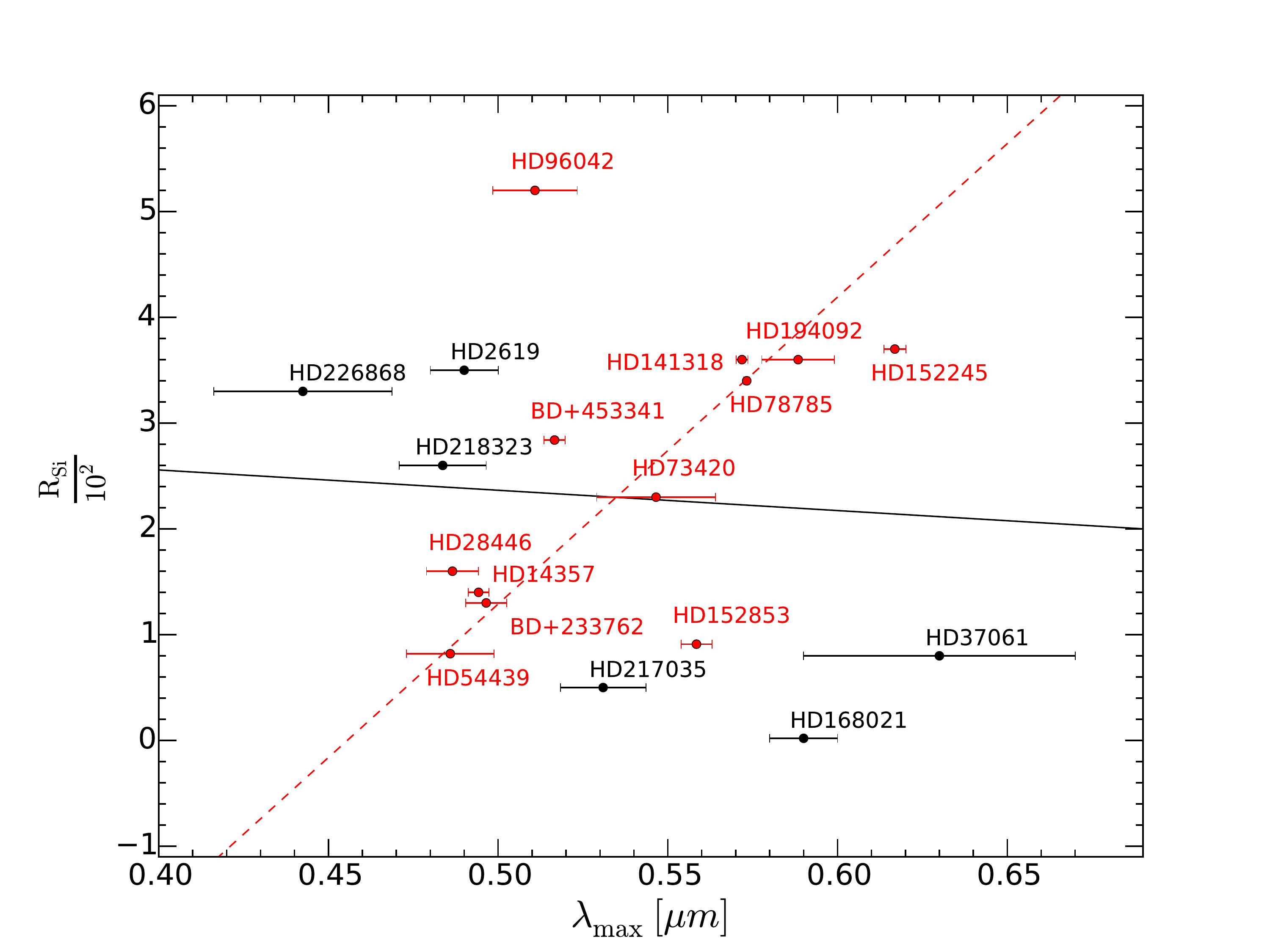}
\caption{$R_{\rm Si}$--$\lambda_{\rm max}$ relationship. The red dots are stars observed with FORS2, AFOSC and CAFOS, and the black dots are 7 additional measurements from HPOL or from the literature. The red dashed line is the linear least-square fit to the red dots, and the black solid line is the linear least-square fit to all data.}
\label{fig_lmax-RSi}
\end{center}
\end{figure}

The strongest correlation we found between $\lambda_{\rm max}$ and $R_{\rm Si}$, with a correlation factor of $\rho$ = 0.50, and the p-value for testing non-correlation of p=0.10, while there is no correlation between $\lambda_{\rm max}$ and other parameters ($\rho \lesssim 0.25$). 
Additionally we add three stars observed by HPOL, and three stars from the literature \citep{1974AJ.....79..581C, 1975ApJ...196..261S} to the sample, and undertake the same correlation tests. These additional stars have anomalous extinction sightlines \citep{2011A&A...527A..34M}, but larger $R_V$ values compared to our observed sample (see Table~\ref{tab_mazzeibarbaroparameters}).  
After including the three HPOL observations, and three stars from literature, the $R_{\rm Si}-\lambda_{\rm max}$ correlation factor drops to $\rho$ = -0.06. 
The $R_{\rm Si}-\lambda_{\rm max}$ correlation is shown in Fig.~\ref{fig_lmax-RSi}. However, these results should be taken with care, because when we fit the data, we assume no uncertainty in $R_{\rm Si}$. This may be problematic, because the realistic uncertainty in the model quantities should be considerable, probably larger than in $\lambda_{\rm max}$, which is a simple measurement. In this way, the fit is strongly driven by the few stars with very low $\lambda_{\rm max}$ errors. The relationship is further discussed in Sect.~\ref{discussion:Rsi-lmax}.

\begin{table*} 
{\tiny
\centering
\caption{Observational data and results of modeling}
\label{tab_mazzeibarbaroparameters}
\begin{tabular}{lllllllllllll}
\hline\hline
& \multicolumn{5}{c}{Mazzei $\&$ Barbaro (2011)} & & \multicolumn{3}{c}{Wegner (2002)}\\ \cline{2-6}  \cline{8-10}
Name &  Sp. & $E(B-V)$    &  CCM $R_V$   &  $R_V$  &  $\frac{R_{\rm Si}}{10^2}$ & & Sp. & E(B-V) &  $R_V$ & $\lambda_{\rm max}$ & $\lambda_{\rm max}$ Reference\\
    &  &  (mag) &     &		       &                & & &  (mag) &              &             ($\AA$)  &  \\
\hline
HD 54439   &B2III& 0.28&	2.13 $\pm$ 0.41& 1.98 &	 0.82 &  & B1V& 0.28 & 2.88 & 4859 $\pm$ 129  & This work (VLT) \\
HD 73420   &B2II/III& 0.37&2.47 $\pm$ 0.32& 2.24 & 2.3 &  & \dots &\dots & \dots & 5465 $\pm$ 175  & This work (VLT) \\
HD 78785   &B2II& 0.76&	2.55 $\pm$ 0.17& 2.29 &	 3.4  &  & B2II&0.67 & 3.08 & 5732 $\pm$ 8.2    & This work (VLT) \\
HD 96042   &O9.5V& 0.48&	1.97 $\pm$ 0.24& 1.87 &	 5.2  &  & B1V& 0.41 & 3.05 & 5109 $\pm$ 124  & This work (VLT) \\
HD 141318  &B2II& 0.30&	1.95 $\pm$ 0.18& 1.77 &	 3.6  &  & \dots &\dots & \dots & 5719 $\pm$ 17   &  This work (VLT) \\
HD 152245  &B0III& 0.42&	2.25 $\pm$ 0.29& 2.02 &	 3.7  &  &B0Ib &0.31 & 2.95 & 6169 $\pm$ 33   &  This work (VLT) \\
HD 152853  &B2II/III& 0.37&	2.50 $\pm$ 0.33& 2.19 &0.91 &  & \dots &\dots & \dots & 5584 $\pm$ 46   &  This work (VLT) \\
BD+23 3762 &B0.5III& 1.05&	2.47 $\pm$ 0.12& 2.15 &1.3  &  & \dots &\dots & \dots & 4965 $\pm$ 61   &  This work (CAHA) \\
BD+45 3341 &B1II& 0.74&	2.46 $\pm$ 0.17& 2.22 &	 2.84 &  &\dots &\dots & \dots & 5166 $\pm$ 31   & This work (CAHA) \\
HD 28446   &B0III& 0.46&	2.46 $\pm$ 0.26& 2.20 &	 1.6  &  & \dots &\dots & \dots & 4865 $\pm$ 76   & This work (CAHA) \\
HD 194092  &B0.5III& 0.41&	2.50 $\pm$ 0.30& 2.18 & 3.6  &  & \dots &\dots & \dots & 5884 $\pm$ 107  & This work (Asiago) \\
HD 14357   &B2II& 0.56&   2.31 $\pm$ 0.21& 2.12 &  1.4  &  & B1.5II& 0.49 & 2.88 & 4942 $\pm$ 31   & This work (Asiago) \\
HD 226868 &B0Ib&  1.08&  3.20 $\pm$ 0.14  &  2.78  & 3.3 & & B0Ib &1.03 & 3.32 & 4424.6 $\pm$ 262.4 & (This work, HPOL)\\
HD 218323 &B0III& 0.90&  2.55 $\pm$ 0.15  &  2.30  & 2.6 & & \dots &\dots & \dots & 4836.7 $\pm$ 128.2 & (This work, HPOL)\\
HD 217035 &B0V& 0.76&  2.77 $\pm$ 0.18  &  2.44  & 0.5 & & \dots &\dots & \dots & 5309.3 $\pm$ 125.9  & (This work, HPOL) \\
HD 2619   &B0.5III& 0.85&   2.55 $\pm$ 0.15  & 2.34& 3.5  & & \dots &\dots & \dots& 4900 $\pm$ 100 & \citet{1974AJ.....79..581C}  \\
HD 37061  &B1V& 0.52&   4.50 $\pm$ 0.38  & 3.82& 0.8  & & B0.5V& 0.47 & 4.14 & 6300 $\pm$ 400 & \citet{1974AJ.....79..581C} \\
HD 168021 &B0Ib& 0.55&   3.15 $\pm$ 0.27  & 2.74& 0.02 & & \dots &\dots & \dots& 5900 $\pm$ 100 & \citet{1975ApJ...196..261S}\\
\hline
\end{tabular}
\tablefoot{Spectral Type (Sp.), $E(B-V)$ and CCM $R_V$, $R_V$ and $R_{\rm Si}$ are taken from Table 1 and 4 of \citet{2011A&A...527A..34M}. The Sp. and $E(B-V)$ values in \citet{2011A&A...527A..34M} are taken from \citet{1985ApJS...59..397S}. "CCM $R_V$" is determined by fitting the IR observations with the CCM extinction curve, while "$R_V$" is determined from the best-fit of WD01 model to all observed data (see \citealt{2011A&A...527A..34M}). For comparison, cols. 7-9 report Sp., E(B-V) and $R_V$ values for seven common stars from \citet{2002BaltA..11....1W}.}
}
\end{table*}

\section{Dust properties inferred from simulations}
\label{sec:simulations}
 
The properties of dust grains towards the considered stars were obtained in \cite{2008MNRAS.390..706M} where the authors performed theoretical models fitting to the anomalous extinction curves. The authors found that to reproduce the anomalous extinction, silicate grains must be concentrated in small sizes of $a< 0.1\mum$ (see also \citealt{2011A&A...527A..34M}). Such small silicate grains cannot reproduce the normal $\lambda_{\max}$ that is measured. Therefore, in this section, we will infer the essential dust properties (i.e., size distribution and alignment) by fitting both, extinction curves with low $R_V$ values, and normal polarization curves. The obtained results will be used to interpret the $\lambda_{\rm max}$--$K$ relationship, the deviation of $K$ from the average value from a sample of "normal" Galactic stars, and to test the relationship between $R_{\rm Si}$ and $\lambda_{\rm max}$.

\subsection{Dust Model and Observational Constrains}

\subsubsection{Dust Model: Size distribution}
We adopt a mixed-dust model consisting of astronomical silicate and carbonaceous grains \citep[see][]{Weingartner2001ApJ...548..296W} (hereafter WD01). The same size distribution model was also used in \citet{2011A&A...527A..34M}.  We assume that grains have oblate spheroidal shapes, and let $a$ be the effective grain size defined as the radius of the equivalent sphere with the same volume as the grain.

Following WD01, the grain size distribution of dust component $j$ is described by an analytical function: 
\bea
\frac{dn_{j}}{n_{\rm H}da}&=& D_{j}(a) + \frac{C_{j}}{a}\left(\frac{a}{a_{t,j}}\right)^{\alpha_{j}}F(a;\beta_{j},a_{t,g})G(a;a_{t,j},a_{c,j}),\label{eq:dnda}
\ena
where $a$ is the grain size, $j=sil, carb$ for silicate and carbonaceous compositions, $D_{j}(a)$ is the size distribution for very small grains, $a_{t,j}$, $a_{c,j}$ are model parameters, and $C_{j}$ is a constant determined by the total gas-to-dust mass ratio. 

The coefficients $F$ and $G$ read:
\bea
F(a;\beta_{j},a_{t,j})&=& 1+\beta_{j} a/a_{t,j} {~\rm for~} \beta_j>0,\\
F(a;\beta_{j},a_{t,j})&=&(1-\beta_{j} a/a_{t,j})^{-1} {~\rm for~} \beta_j<0,
\ena
and
\bea
G(a;a_{t,j},a_{c,j})&=&1 {~\rm for~} a<a_{t,j}\label{eq:dnda1},\\
G(a;a_{t,j},a_{c,j})&=&\exp\left(-[(a-a_{t,j})/a_{c,j}]^{3}\right) {~\rm for~} a>a_{t,j}.
\ena

The term $D_{j}=0$ for $j= sil$. For very small carbonaceous grains (i.e., polycyclic aromatic hydrocarbons), $D_{j}(a)$ is described by a log-normal size distribution containing a parameter $b_{C}$ that denotes the fraction of C abundance present in very small sizes (see WD01 for more detail). Thus, the grain size distribution is completely described by a set of 11 parameters: $\alpha_{j},\beta_{j}, a_{t,j}, a_{c,j}, C_{j}$ where $j=sil,carb$ for silicate and carbonaceous compositions, and $b_{C}$. 

\subsubsection{Dust Model: Alignment function}

Let $f_{\ali}$ be the fraction of grains that are perfectly aligned with the symmetry axis $\hat{a}_{1}$ along the magnetic field $B$. The fraction of grains that are randomly oriented is thus $1-f_{\ali}$. To parameterize the dependence of $f_{\rm ali}$ on the grain size, we introduce the following function:
\bea
f_{\rm ali}(a;a_{\rm ali},f_{\rm min}, f_{\rm max}) = f_{\rm min} + \left[1-\exp\left(-\frac{a}{a_{\rm ali}}\right)^{3} \right] \left(f_{\rm max}-f_{\rm min}\right),\label{eq:fali}
\ena
where $a_{\rm ali}$ describes the minimum size of aligned grains, $f_{\rm max}$ describes the maximum degree of grain alignment, and $f_{\rm min}$ accounts for some residual small degree of alignment of very small grains. This alignment function reflects the modern understanding of grain alignment where large grains are efficiently aligned by radiative torques (see, e.g., \citealt{2016ApJ...831..159H}) and small grains are weakly aligned by paramagnetic relaxation \citep{2014ApJ...790....6H}.

\subsection{Model of Extinction and Polarization}
\subsubsection{Extinction}

The extinction of starlight due to scattering and absorption by interstellar grains in units of magnitude is given by
\bea
\frac{A({\lambda})}{N_{\rm H}}=1.086\sum_{j=sil,carb}\int_{a_{\rm min}}^{a_{\rm max}} C_{\rm ext}^{j}(a, \lambda)\left(\frac{dn_{j}}{da}\right)da,\label{eq:Aext}
\ena
where $C_{\rm ext}$ is the extinction cross-section, $a_{\rm min}$ and $a_{\rm max}$ are the lower and upper cutoffs of the grain size distribution, and $N_{\rm H}$ is the total gas column density along the sightline.

\subsubsection{Polarization}
\label{subsect:polarization}

Modeling the starlight polarization by aligned grains is rather complicated because it requires a detailed knowledge of the orientation of grains with the magnetic field and the magnetic field geometry along the line of sight. Specifically, a realistic modeling needs to take into account the nutation of the grain symmetry axis $\ahat_{1}$ around the angular momentum $J$, the precession of $J$ around $B$, and the distribution function of the cone angle between $J$ and $B$ (\citealt{1980A&A....88..194H}; see \citealt{2012JQSRT.113.2334V} for a review). However, an analytical distribution function for the cone angle is not known for the popular alignment mechanism by radiative torques (see \citealt{LAH15} and \citealt{Andersson:2015bq} for latest reviews). Therefore, in our paper, we adopt a picket-fence (PF) alignment model to compute the polarization, as used in previous works (\citealt{1995ApJ...444..293K}; \citealt{2006ApJ...652.1318D}; \citealt{2009ApJ...696....1D}; \citealt{2013ApJ...779..152H}; \citealt{2014ApJ...790....6H}). The essence of the PF model is as follows.

First, the oblate grain is assumed to be spinning around the symmetry axis $\ahat_{1}$ (i.e., having perfect internal alignment). The magnetic field $B$ is assumed to lie in the plane of the sky $\xhat\yhat$ with $B \| \xhat$, and the line of sight is directed along $\zhat$. Therefore, the polarization cross-section contributed by the perfectly aligned grains is $C_{x}-C_{y}=(C_{\|}-C_{\perp})f_{\ali}$ where $C_{\|}$ and $C_{\perp}$ are the cross-section for the incident electric field parallel and perpendicular to the symmetry axis, respectively (see \citealt{2013ApJ...779..152H}). Among $(1-f_{\ali})$ randomly oriented grains, the fraction of grains that are aligned with $\xhat,\yhat,\zhat$ are equal, of $(1-f_{\ali})/3$. The total polarization produced by grains with $\ahat_{1}\| B$ is then $C_{x}-C_{y}=(C_{\|}-C_{\perp})(f_{\ali} + (1-f_{\ali})/3)$. The polarization by grains aligned with $\ahat_{1}\|\yhat$ is $(1-f_{\ali})/3(C_{\|}-C_{\perp})/3$. Thus, the total polarization cross-section is $C_{x}-C_{y}= (C_{\|}-C_{\perp})[(1+2f_{\ali}) - (1-f_{\ali})]/3=C_{\rm pol}f_{\ali}$.

Because graphite grains are not aligned with the magnetic field (\citealt{2006ApJ...651..268C}; \citealt{2016ApJ...831..159H}), we assume that only silicate grains are aligned while carbonaceous grains are randomly oriented. Therefore, the degree of polarization of starlight due to differential extinction by aligned grains along the line of sight is computed by
\bea
\frac{p({\lambda})}{N_{\rm H}}= \int_{a_{\rm min}}^{a_{\rm max}} \frac{1}{2}C_{\rm pol}^{\rm sil}(a, \lambda)f_{\ali}(a)
\frac{dn_{\rm sil}}{da}da,\label{eq:Plam}
\ena
where $C_{\rm pol}^{\rm sil}$ is the polarization cross-section of silicate oblate grains, and $f_{\rm ali}$ is given by Equation (\ref{eq:fali}). Here we take $C_{\rm ext}$ and $C_{\rm pol}$ computed for different grain sizes and wavelengths from \cite{2013ApJ...779..152H}. 

Note that magnetic fields are perhaps varying for the different stars. However, in this paper, we do not attempt to infer a dust model for each specific sightline.
Instead, we only attempt to infer the general features of dust size distribution and alignment functions for this group of stars with anomalous $R_V$ and normal $\lambda_{max}$. Detailed modeling for each specific star is beyond the scope of this paper.

\subsection{Numerical Modeling and Results}\label{sec:method}

\subsubsection{Numerical Method}
Inverse modeling has frequently been used to infer the grain size distribution of dust grains in the ISM of the Milky Way (\citealt{1995ApJ...444..293K}), and in nearby galaxies (e.g, small Magellanic cloud \citep{2003ApJ...588..871C}. \cite{2009ApJ...696....1D} used Levenberg-Marquart (LM) method to infer both the grain size distribution and alignment function of interstellar grains in the Galaxy characterized by the typical values of $R_{V}=3.1$ and $\lambda_{\max}=0.55\mum$. A simulation-based inversion technique was developed in \cite{{2013ApJ...779..152H},{2014ApJ...790....6H}} to find best-fit grain size distribution and alignment function for interstellar grains in the SNe Ia hosted galaxies with anomalous extinction and polarization data. Although the Monte-Carlo simulations demonstrate some advantage (e.g., problem with local minima), its convergence is much slower than the LM method. Thus, in this paper, we adopt the LM method for our modeling.

The goodness of the fit of the model $F_{\rm mod}$ to observed data $F_{\rm obs}$ is governed by $\chi^{2}_{F}$ defined as follows:
\bea
\chi^{2}_{F}=\sum_{i} \frac{\left(F_{\rm mod}(\lambda_{i})-F_{\rm obs}(\lambda_{i})\right)^{2}}{F_{\rm err}(\lambda_{i})^{2}},\label{eq:chisq}
\ena
where $F_{\rm err}(\lambda)$ is the error in the measurement at wavelength $\lambda$. 

Assuming the same errors at all wavelengths, the total $\chi^{2}$ can be written as
\bea
\chi^{2} = \chi^{2}_{\rm ext} + \eta_{\rm pol}\chi_{\rm pol}^{2}+\chi_{\rm vol}^{2},\label{eq:chisq_tot}
\ena
where $\chi^{2}_{\rm ext}$ and $\chi^{2}_{\rm pol}$ are evaluated using Eq.~(\ref{eq:chisq}) for $F = A$ and $F=P$, respectively, $\chi_{\rm vol}^{2}$ desribes the volume constraint determined by the depletion of elements into dust, and $\eta_{\rm pol}$ is a coefficient introduced to adjust the fit to the polarization. The initial value of $\eta_{\rm pol}=1$ is chosen. When the fit to the polarization is poor, we can increase $\eta_{\rm pol}$. Here, we evaluate $\chi_{\rm vol}^2=\chi_{\rm vol,sil}^{2}+\chi_{\rm vol, carb}^{2}=\left(V_{\rm sil}/V_{\rm sil,0}-1\right)^{2}+\left(V_{\rm carb}/V_{\rm carb,0}-1\right)^{2}$ where $V_{\rm sil,0}=2.98\times 10^{-27}\rm cm^{3}$ per H nucleon and $V_{\rm carb,0}=2.07\times 10^{-27}\rm cm^{3}$ per H nucleon (see WD01).

We seek for the best-fit values of $\alpha_{j},\beta_{j}, a_{t,j}, a_{c,j}, c_{j}$ where $j=sil,carb$ and two parameters for grain alignment ($a_{\rm ali},f_{\rm min}$) by minimizing $\chi^{2}$ (Eq.~\ref{eq:chisq_tot}) using the Levenberg-Marquart method from the publicly available package lmfit-py\footnote{http://cars9.uchicago.edu/software/python/lmfit/index.html}. 
The errors from observed data are assumed to be $10\%$. 

We note that in WD01, the parameter $a_{c,\rm sil}$ is fixed to $0.1\mum$. However, \cite{2008MNRAS.390..706M} found that the best fit to the extinction for these anomalous stars requires $a_{c,\rm sil}$ to be reduced to $0.01\mum$, which corresponds to most Si being present in small grains of $a\le 0.01\mum$. In this paper, we treat $a_{c,\rm sil}$ as a model parameter. Furthermore, since we are dealing with $R_{V}<4$, grain growth is not expected, thus we constrain the size cutoff parameters $a_{c,\rm carb} \leq 0.5\mum$ and $a_{c,\rm sil} \leq 0.5\mum$.

\subsubsection{Model Setup}
The sightlines of the considered stars have anomalous extinction curves, with lower $R_{V}$ than the standard value of $R_{V}=3.1$ for the Milky Way. However, the polarization data appear to be normal, with a peak wavelength $\lambda_{\rm max}> 0.4\mum$. Thus, for our inverse modeling, we consider a fixed extinction curve described by a low value of $R_{V}=2.5$. For the polarization data, we consider six different values of $\lambda_{\rm max}=0.45, 0.51, 0.53, 0.55, 0.60, 0.65\mum$, which fully covers the range of $\lambda_{\max}$ inferred from observations shown in Table \ref{tab_mazzeibarbaroparameters}. For a given $R_{V}$, we generate (i.e. construct) the extinction data (hereafter, generated extinction curves) using the \cite{1989ApJ...345..245C} extinction law. For a given $\lambda_{\rm max}$, we generate the polarization data (hereafter, generated polarization curves) using the Serkowski curve with $K=k_{1}\lambda_{\rm max}+k_{2}$ (see \citealt{2015arXiv151001822H} for details). Here, we adopt a standard relationship with $k_{1}=1.66$ and $k_{2}=0.01$ from \cite{1992ApJ...386..562W}.

Because the extinction and polarization data in the far-UV ($\lambda<0.25\mum$) toward the considered stars are unavailable, we will not attempt to invert the data in the far-UV, which is mainly contributed by ultrasmall grains (including PAHs). Thus, we consider $\lambda=0.25-2.5\mum$ and compute the extinction and polarization model given by Equations (\ref{eq:Aext}) and (\ref{eq:Plam}), respectively. We use 32 bins of grain size in the range from $a_{\rm min}=3.5~$\AA~to $a_{\rm max}=1\mum$ and 32 wavelength bins. 

Furthermore, note that while we use the standard Serkowski curve to generate the polarization data, observational studies show differences in the amount of UV polarization relative to that in the visual Serkowski curve. \citet{1995ApJ...445..947C} found that UV polarimetry measurements of 7 out of 14 sightlines with $\lambda_{max} \geq$ 0.54 $\mu m$ agree well with an extrapolation of the Serkowski curve into the UV, while the other 7 sightlines with $\lambda_{max} \leq$ 0.53 $\mu m$ show polarization excess compared to the Serkowski extrapolation. They found a relationship between $\lambda_{max}^{-1}$ and the relative UV polarization p(6 $\mu m^{-1}$)/p$_{max}$ (see also \citealt{1999ApJ...510..905M}). \citet{1996AJ....112.2726A} found that at least half of their sample of 35 sightlines, for which they have reliable UV observations, do not agree well compared to the Serkowski extrapolation from visual and near-IR parameters. An increase/decrease in the UV polarization would lead to an increase/decrease in the degree of alignment of small grains inferred from simulations, whereas the alignment of large grains (a > 0.1 $\mu m$) that dominates the visible-IR polarization would be unchanged. The grain size distributions would be slightly changed (see \citealt{2014ApJ...790....6H}).

The important constraint for the polarization model (see Sect.~\ref{subsect:polarization}) and the alignment function $f_{\rm ali}(a)$ is that, for the maximum polarization efficiency $p_{\max}/A(\lambda_{\max})=3\%\,{\rm mag}^{-1}$ (see \citealt{2003ARA&A..41..241D} for a review), we expect that the conditions for grain alignment are optimal, which corresponds to the case in which the alignment of big grains can be perfect, and the magnetic field is regular and perpendicular to the line of sight. Thus, we set $f_{\rm ali}(a=a_{\max})=1$. 

\subsubsection{Results}\label{sec:result}
\begin{figure*}
\centering
\includegraphics[width=0.45\textwidth]{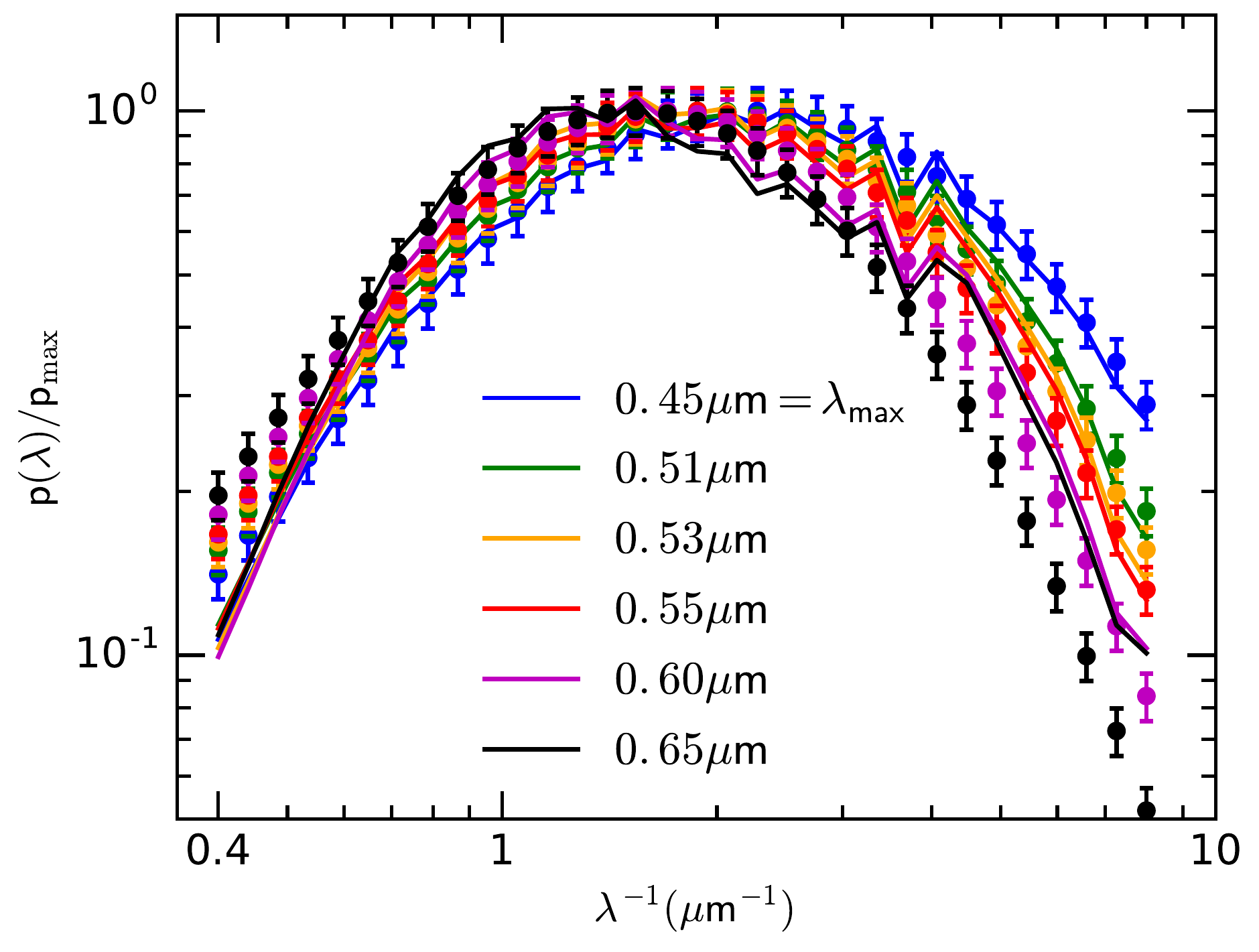}
\includegraphics[width=0.45\textwidth]{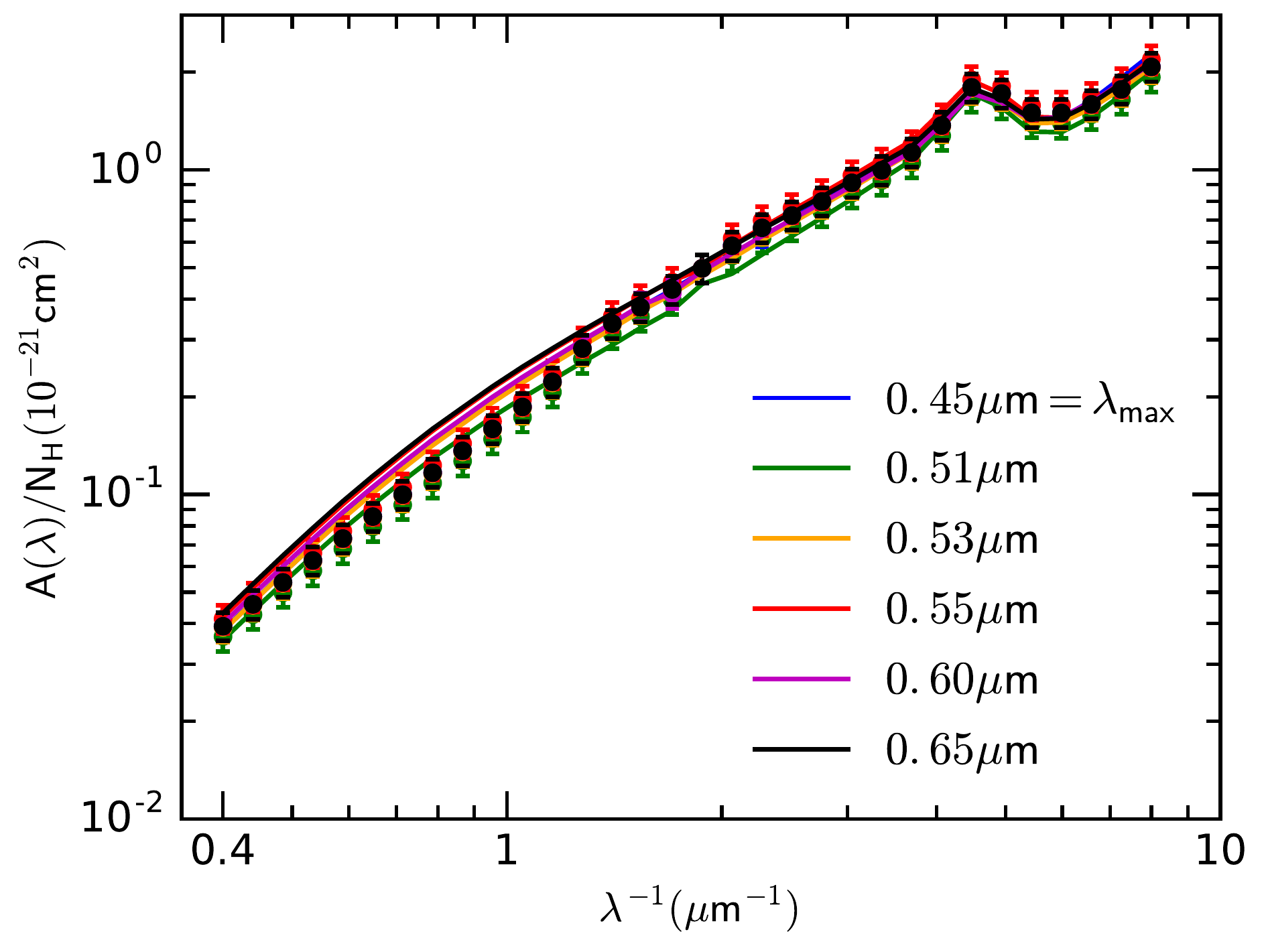}
\caption{Left panel: Best-fit models vs. generated (i.e. constructed) polarization curves for six different models described by $\lambda_{\max}$. Right panel: Best-fit models vs. generated extinction curves with a low $R_{V}=2.5$. Filled circle symbols show the generated data, and solid lines show our best-fit models.}
\label{fig:best_model}
\end{figure*}

Figure~\ref{fig:best_model} shows the best-fit polarization and extinction curves for the different $\lambda_{\rm max}$. The fit to the extinction curve is good, but the model overestimates the extinction for $\lambda \geq 1 \mum$ for $\lambda_{\max}=0.53 - 0.65\mum$.
For the polarization, the fit is excellent for $\lambda_{\max}<0.6\mum$, but the model (see Sect.~\ref{subsect:polarization}) overestimates the polarization at $\lambda<0.25\mum$ for $\lambda_{\max}=0.6\mum$ and $0.65\mum$.

\begin{figure*}
\centering
\includegraphics[width=0.45\textwidth]{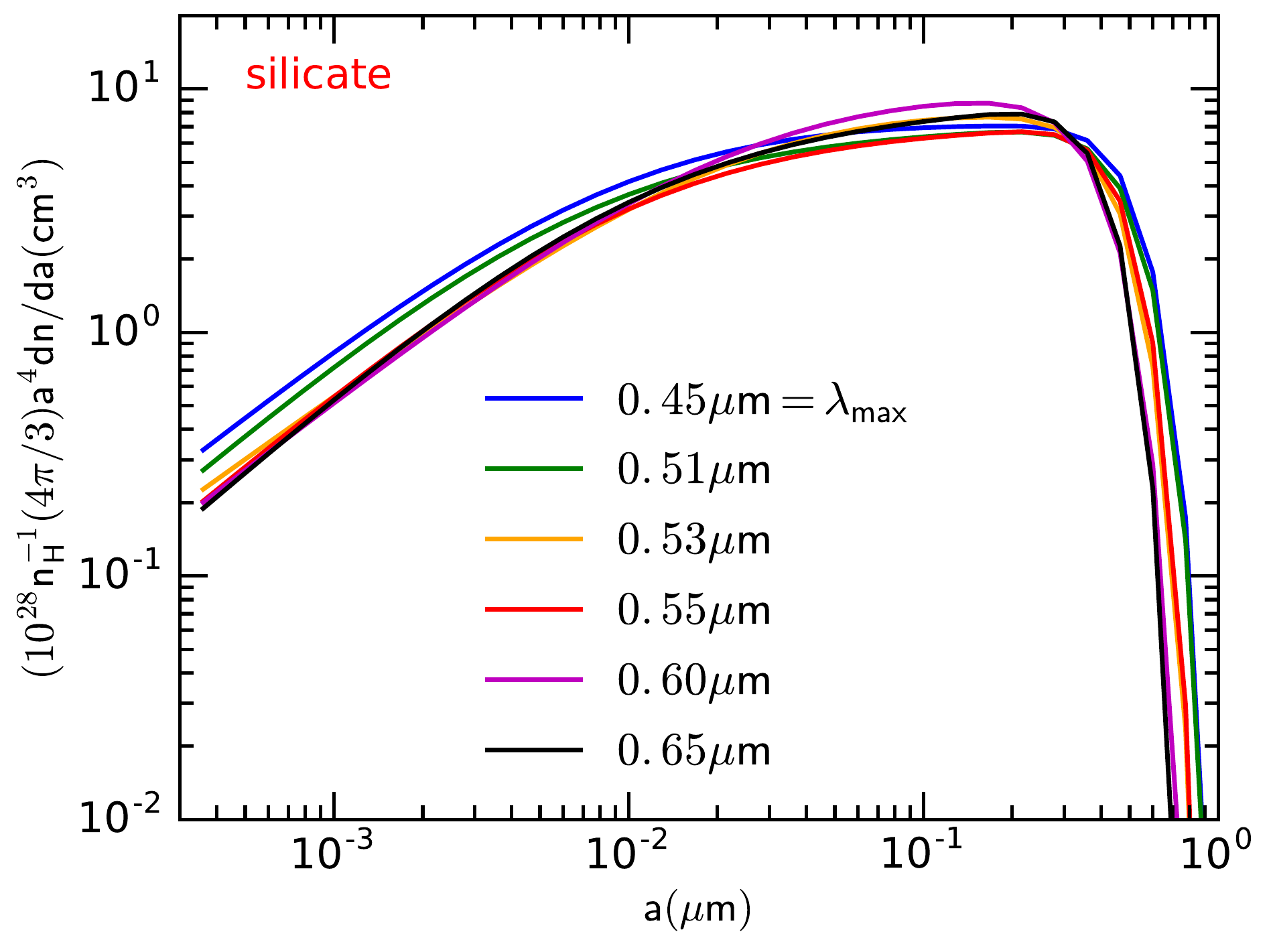}
\includegraphics[width=0.45\textwidth]{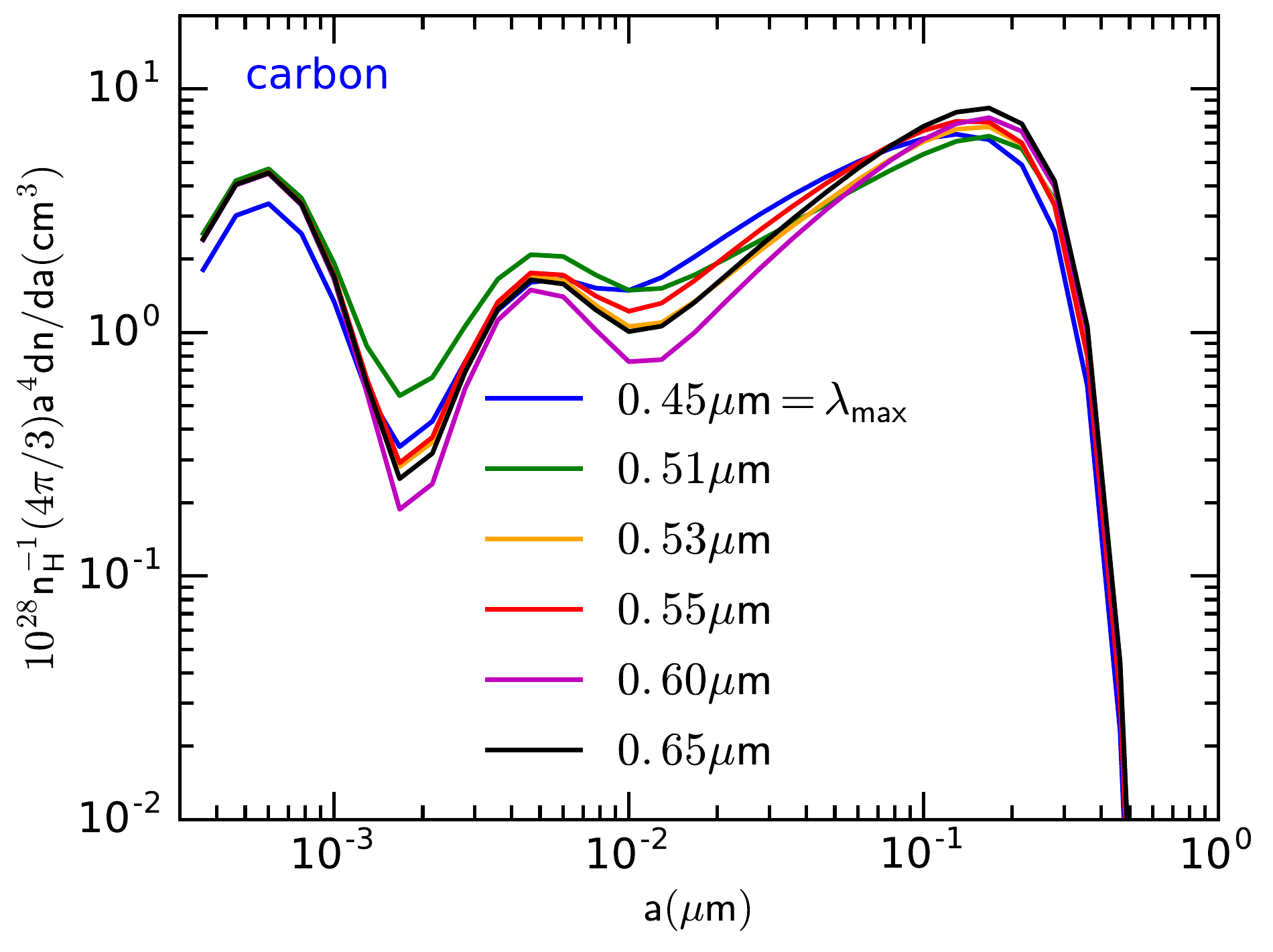}
\caption{Best-fit grain size distribution for silicate (left panel) and carbonaceous grains (right panel). Six different models described by $\lambda_{\max}$ are considered. Large silicate grains of size $a\ge 0.1\mum$ are present to reproduce normal $\lambda_{\max}$.}
\label{fig:best_dnda}
\end{figure*}

Figure \ref{fig:best_dnda} shows the best-fit size distributions for silicate and carbonaceous grains. The size distribution appears to change slightly with $\lambda_{\max}$, which is expected due to the fixed $R_{V}$. We see that, to reproduce the typical $\lambda_{\max}$, there must be a population of large silicate grains of $a\ge 0.1\mum$. This is different from the results obtained by \cite{2008MNRAS.390..706M} where the authors performed the fitting to the extinction curves only and found the lack of large silicate grains, however, the presence of large grains in the carbonacous grain size distribution.

\begin{figure}
\includegraphics[width=0.45\textwidth]{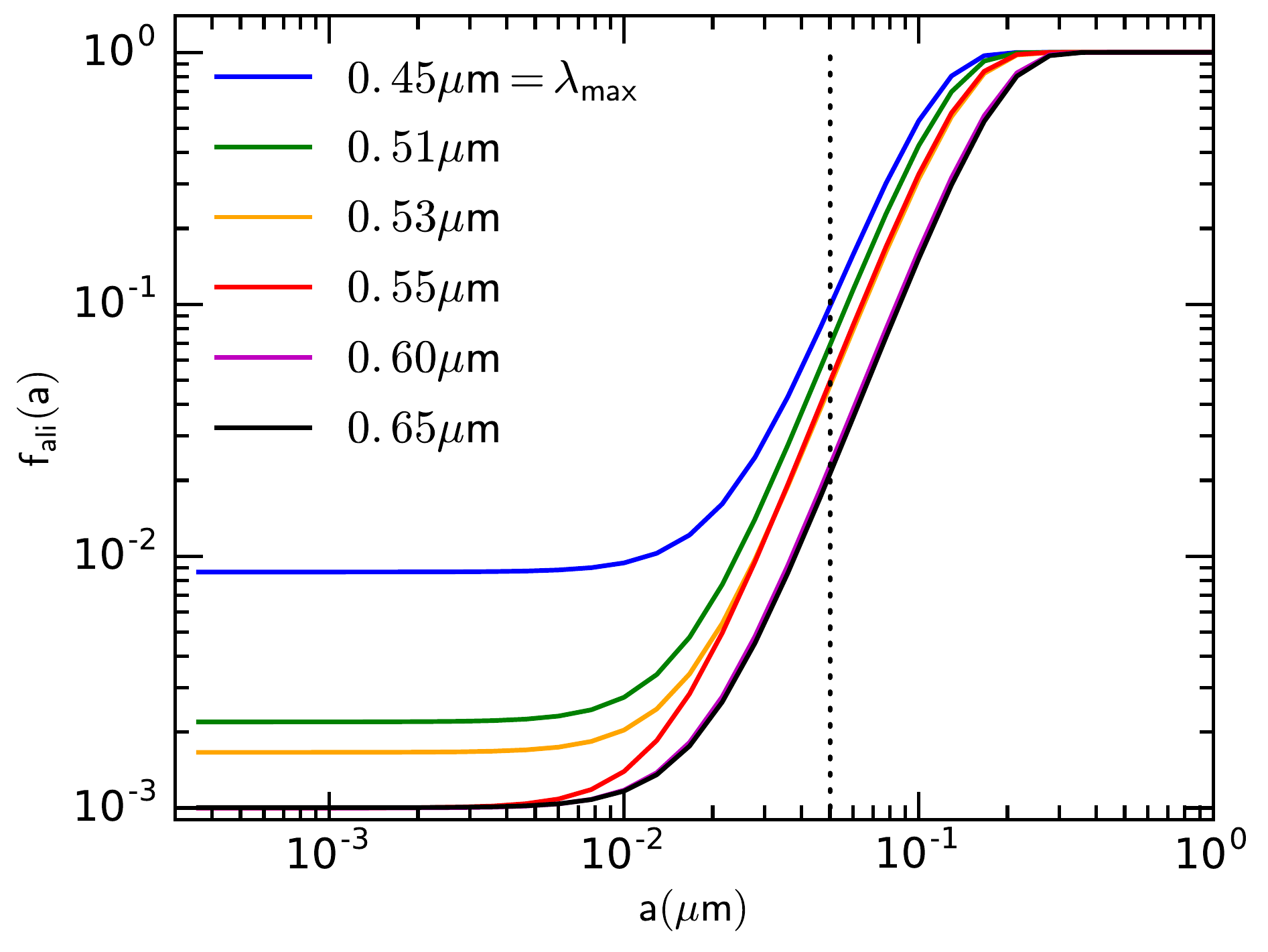}
\caption{Best-fit alignment function of silicates for the different models given by $\lambda_{\rm max}$. The dotted line marks the typical grain size a=0.05 ${\rm \mum}$. The alignment function tends to shift to smaller sizes as $\lambda_{\rm max}$ decreases.}
\label{fig:best_fali}
\end{figure}

Figure \ref{fig:best_fali} shows the best-fit alignment function for the different $\lambda_{\rm max}$. When $\lambda_{\rm max}$ decreases, the alignment function tends to shift to smaller sizes. Also, the alignment of small grains ($a<0.05\mum$) is increased with decreasing $\lambda_{\max}$. This trend is consistent with the results from \cite{2014ApJ...790....6H} where the modeling is done for the cases with normal extinction curves (i.e., $R_{V}\sim 3.1$) and excess UV polarization.
 
\section{Discussion}\label{sec:discussion}

\subsection{Comparison to Supernovae Ia and normal Galactic stars}

The main aim of this work is to investigate the polarization profiles of Galactic stars with low $R_V$ values, with the aim to try to find similar polarization behavior, as we observe in highly reddened SNe Ia with low $R_V$ values, with the polarization degree rising towards blue wavelengths (see e.g. Fig. 2 in \citealt{2015A&A...577A..53P}). However, none of the stars with anomalous extinction sightlines in our sample display such polarization curves, steeply rising towards the blue (Fig.~\ref{fig_all_SCI_stars}). 


Figure~\ref{fig_lmax-K} shows our sample of stars with anomalous extinction sightlines in the $\lambda_{\rm max}$--$K$ plane, compared to a sample of SNe Ia from \citet{2015A&A...577A..53P} and \citealt{Zelaya2017arxiv} (see Appendix~\ref{appendix:fitserkowskitozelaya}). Despite the low $R_V$ values, our sample has normal polarization curves with a mean $\lambda_{\rm max} \sim 0.53 {\rm \mum}$. The Serkowski parameters $K$ and $\lambda_{\rm max}$ are related ($\rho$=0.87, p=3$\times$10$^{-5}$), and can be described as a linear function of $\lambda_{\rm max}$: $K$= -1.13 $\pm$ 0.34 + (4.05 $\pm$ 0.64)$\lambda_{\rm max}$. This is steeper than compared to the empirical relationship found by \citet{1992ApJ...386..562W}: $K$=0.01 $\pm$ 0.05 + (1.66 $\pm$ 0.09)$\lambda_{\rm max}$ (see also \citealt{1980ApJ...235..905W, 1982AJ.....87..695W}). However, the $K$--$\lambda_{\rm max}$ relationship in \citet{1992ApJ...386..562W} was determined from a chosen sample of sightlines towards stars with a variety of interstellar environments, including dense clouds, diffuse clouds and low-density interstellar material.

Fig.~\ref{fig_lmax-K-comparison} shows a direct comparison of our sample with the \citet{1992ApJ...386..562W} sample, and the Large Interstellar Polarization Survey (LIPS) sample \citep{2017A&A...608A.146B}. Despite the difference in the slope, our sample is consistent within 3$\sigma$ with the \citet{1992ApJ...386..562W} sample, and also coincides well with the LIPS sample, which has many outliers from the $K$--$\lambda_{\rm max}$ relationship. 

For comparison, SNe Ia with low $R_V$ values have $\lambda_{\rm max} \lesssim 0.45 {\mum}$, and higher $K$ values, above the the \citet{1992ApJ...386..562W} $\lambda_{\rm max}$--$K$ relationship, due to the steep rise of the polarization curve towards the blue. 
There are two exceptions: SN 2002fk and SN 2007af, which are consistent (within the errors) with the Galactic stars sample and have $\lambda_{\rm max}$ of $\sim$ 0.44 ${\mum}$ and $\sim$ 0.74 ${\mum}$ respectively (Table~\ref{tab:SNefit}).


\citet{Cikota2017} noticed that some post-AGB stars (proto-planetary nebula, PPN) have polarization curves rising towards the blue, which are produced by CSM scattering \citep{2005ApJ...624..957O}. These polarization curves are remarkably similar to those observed towards highly reddened SNe Ia. They suggest that also these polarization curves observed towards highly reddened SNe Ia, might be produced by CSM dust scattering. Furthermore, those SNe Ia might explode within a PPN. The main caveat is that if the polarization is produced by scattering, the polarization angles, which carry the geometrical imprint of the dust distribution in the PPN, are expected to be randomly orientated, while the observed polarization angles in sight-lines of highly reddened SNe Ia show an alignment with the structure of their host galaxies, probably as a consequence of dust-grain alignment along the local magnetic field (\citealt{2015A&A...577A..53P}, see also \citealt{2015arXiv151001822H}).

\subsection{$R_{V} - \lambda_{\rm max}$ relationship}
\label{subsec:rv-lmax_relation}

\citet{1975ApJ...196..261S} found that $\lambda_{\rm max}$ is correlated with the ratios of color excess, e.g. $E(V-K)$/$E(B-V)$, and thus to the total-to-selective extinction ratio $R_V$. They found that $R_V$ = 5.5 $\lambda_{\rm max}$, where $\lambda_{\rm max}$ is in ${\rm \mu m}$. \citet{1978A&A....66...57W} deduced $R_V$ = (5.6 $\pm$ 0.3)$\lambda_{\rm max}$ using a sample of carefully selected normal stars and therewith confirm the result by \citet{1975ApJ...196..261S}. \citet{1988ApJ...327..911C} confirmed that the $\lambda_{\rm max}$--$R_V$ relationship is real, and derived $R_V$ = (-0.29 $\pm$ 0.74) + (6.67 $\pm$ 1.17)$\lambda_{\rm max}$, using a modified extinction law in which they forced the extinction to zero at infinite wavelengths. They also concluded that the variations in $\lambda_{\rm max}$ are produced by the dust grains' size distribution, rather than a variation in the alignment of the dust grains.

However, our sample of anomalous extinction sightlines does not show any significant correlation between $\lambda_{\rm max}$ and $R_V$. The correlation coefficient is $\rho \leq 0.26$. The $\lambda_{\rm max}$ values are higher than expected from the $\lambda_{\rm max}$--$R_V$ relationship given in e.g. \citet{1978A&A....66...57W}. 

The most likely explanation is that while not all dust types contributes to polarization, all dust types do contribute to extinction, and thus the $R_V$ value. The polarization curve mainly depends on the dust grain size distribution of silicates, because magnetic alignment is more efficient for silicates than, for instance, for carbonaceous dust grains \citep{1994ApJ...427L..47S}. 

It has already been shown in previous works that there is not necessarily a correlation between $R_{V}$ and $\lambda_{\rm max}$. \citet{1994MNRAS.268....1W} measured linear polarization towards the Chamaeleon I dark cloud, and found only a weak correlation between $R_{V}$ and $\lambda_{\rm max}$. \citet{2001ApJ...547..872W} presented observations of interstellar polarization for stars in the Taurus Dark Cloud, and found no clear trend of increasing $R_V$ with  $\lambda_{\rm max}$ (see their Fig. 9). Their sample shows normal optical properties, with $R_V\sim 3$, while the $\lambda_{\rm max}$ values are higher than expected from observations towards normal stars \citep[e.g.][]{1978A&A....66...57W}. They suggested that the poor $R_V$--$\lambda_{\rm max}$ correlation can be explained by dust grain size dependent variations in alignment capabilities of the dust grains. Also the LIPS data \citep{2017A&A...608A.146B} do not follow any $R_{V}$--$\lambda_{\rm max}$ relationship.

Another possibility is that the $R_V$ values presented in \citet{2011A&A...527A..34M} are lower than the true values. The CCM $R_V$ values (listed in Table~\ref{tab_mazzeibarbaroparameters}) were determined by best fitting the IR observations with the CCM law (Table 1 in \citealt{2011A&A...527A..34M}), and are consistent with estimates of $R_V$ values following the methods in \citet{1999PASP..111...63F}. The "$R_V$" values in Table~\ref{tab_mazzeibarbaroparameters} (taken from Table 4 in \citealt{2011A&A...527A..34M}) were determined by best-fitting the whole extinction curve with the WD01 model \citep[see][]{2011A&A...527A..34M}. It is important to note that \citet{2007ApJ...663..320F} showed that the relations between $R_V$ and UV extinction can arise from sample selection and methodology and, that there is generally no correlation between the UV and IR portions of the Galactic extinction curves. 

\citet{2002BaltA..11....1W} presented 436 extinction curves covering a wavelength range from UV to near-IR, including seven stars from our subsample: HD\,14357, HD\,37061, HD\,54439, HD\,78785, HD\,96042, HD\,152245 and HD\,226868. They determine the $R_V$ values by extrapolating the ratio E($\lambda$-V)/E(B-V) to 1/$\lambda$=0, where the extinction should be zero, and found slightly higher values. The $E(B-V)$ and $R_V$ values determined in \citet{2002BaltA..11....1W} of seven common stars are listed in Table~\ref{tab_mazzeibarbaroparameters}. The $R_V$ values determined by \citet{2002BaltA..11....1W} are 1.4 $\pm$ 0.2 times higher, compared to the $R_V$ values determined in \citet{2011A&A...527A..34M} by best fitting the WD01 model to observations, and 1.2 $\pm$ 0.2 times higher compared to the CCM $R_V$ values determined by best fitting the IR observations with the CCM law. A caveat of the extrapolation method is that IR emission from eventual CS shells around Be stars might suggest increased $R_V$ values \citep{2002BaltA..11....1W}. 

We also note that the $E(B-V)$ values used in \citet{2011A&A...527A..34M} (and originally taken from \citealt{1985ApJS...59..397S}) of the common stars are $\sim$1.1 $\pm$ 0.1 times higher than those in \citet{2002BaltA..11....1W}, which also contributes to lower $R_V$ values in \citet{2011A&A...527A..34M} compared to values in \citet{2002BaltA..11....1W}, and that the spectral types used in \citet{1985ApJS...59..397S} (used in \citealt{2011A&A...527A..34M}) are different than those in \citealt{2002BaltA..11....1W} (Table~\ref{tab_mazzeibarbaroparameters}). 
\citet{2002BaltA..11....1W} takes the Spectral classification from the SIMBAD database, and compared most recent estimate with most often. The author finds that for about 60\% of his sample, the most often and most recent spectral and luminosity classes are same, while there is a difference of 0.05 in the spectral class for about 18\% of stars, of 0.1 for 16\%, and of 0.2 spectral class for 6\%. Therefore, the extinction curves of same targets, computed by different authors, are slightly different.



\begin{figure}[h!]
\begin{center}
\includegraphics[trim=10mm 0mm 10mm 10mm, width=9cm, clip=true]{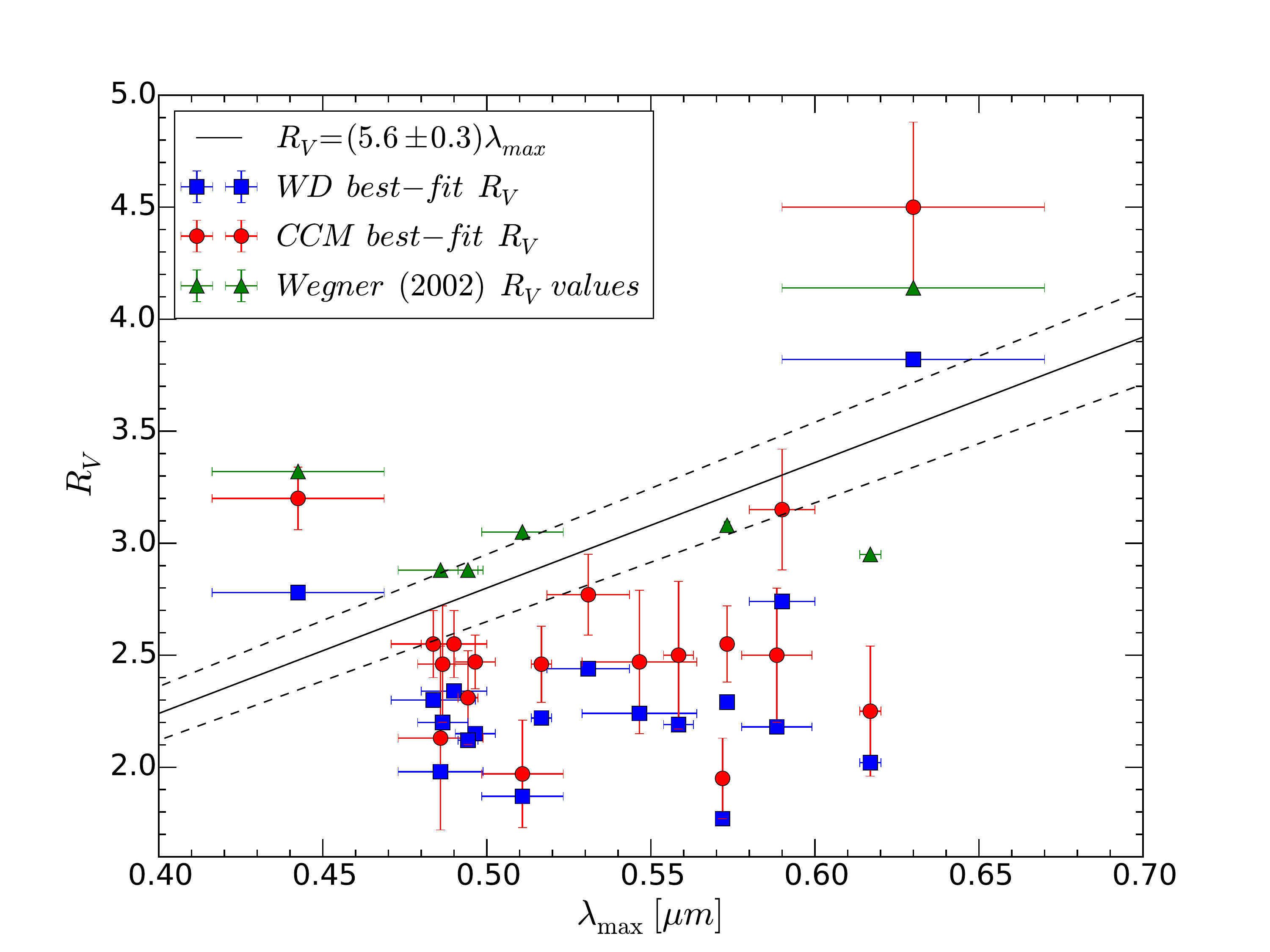}
\caption{$R_{V}$--$\lambda_{\rm max}$ plane. The red dots mark the "CCM $R_V$" values as a function of $\lambda_{\rm max}$ determined by best-fitting the IR extinction curves with the CCM extinction law, the blue dots mark the $R_V$ values determined by fitting the WD01 model to all observations, and the green triangles mark the $R_V$ values from \citet{2002BaltA..11....1W} determined by the extrapolation method (see Table~\ref{tab_literature}). The black line shows the $R_V$ relationship from \citet{1978A&A....66...57W} and its 1$\sigma$ uncertainty.}
\label{fig_lmax-RV}
\end{center}
\end{figure}

Figure~\ref{fig_lmax-RV} shows the $R_V$ values determined by best-fitting the IR extinction curves with the CCM law as a function of $\lambda_{\rm max}$, compared to the values determined from the best-fit of the observed extinction curves by the WD01 models (both taken from \citealt{2011A&A...527A..34M}), and the $R_V$ values for seven common stars from \citet{2002BaltA..11....1W}, determined by the extrapolation method (see Table~\ref{tab_literature}). Four of seven stars with $R_V$ values determined by \citet{2002BaltA..11....1W} lie within the \citet{1978A&A....66...57W} $R_V$ relationship, while most of the stars with $R_V$ values determined by best-fitting of the CCM law and the WD01 model \citep{2011A&A...527A..34M}, are below the \citet{1978A&A....66...57W} relationship.

\subsection{$p_{\max}$ -- $E(B-V)$ relationship}
\label{discussion:pmax-EBV_relation}

\begin{figure}[h!]
\begin{center}
\includegraphics[trim=10mm 0mm 10mm 10mm, width=9cm, clip=true]{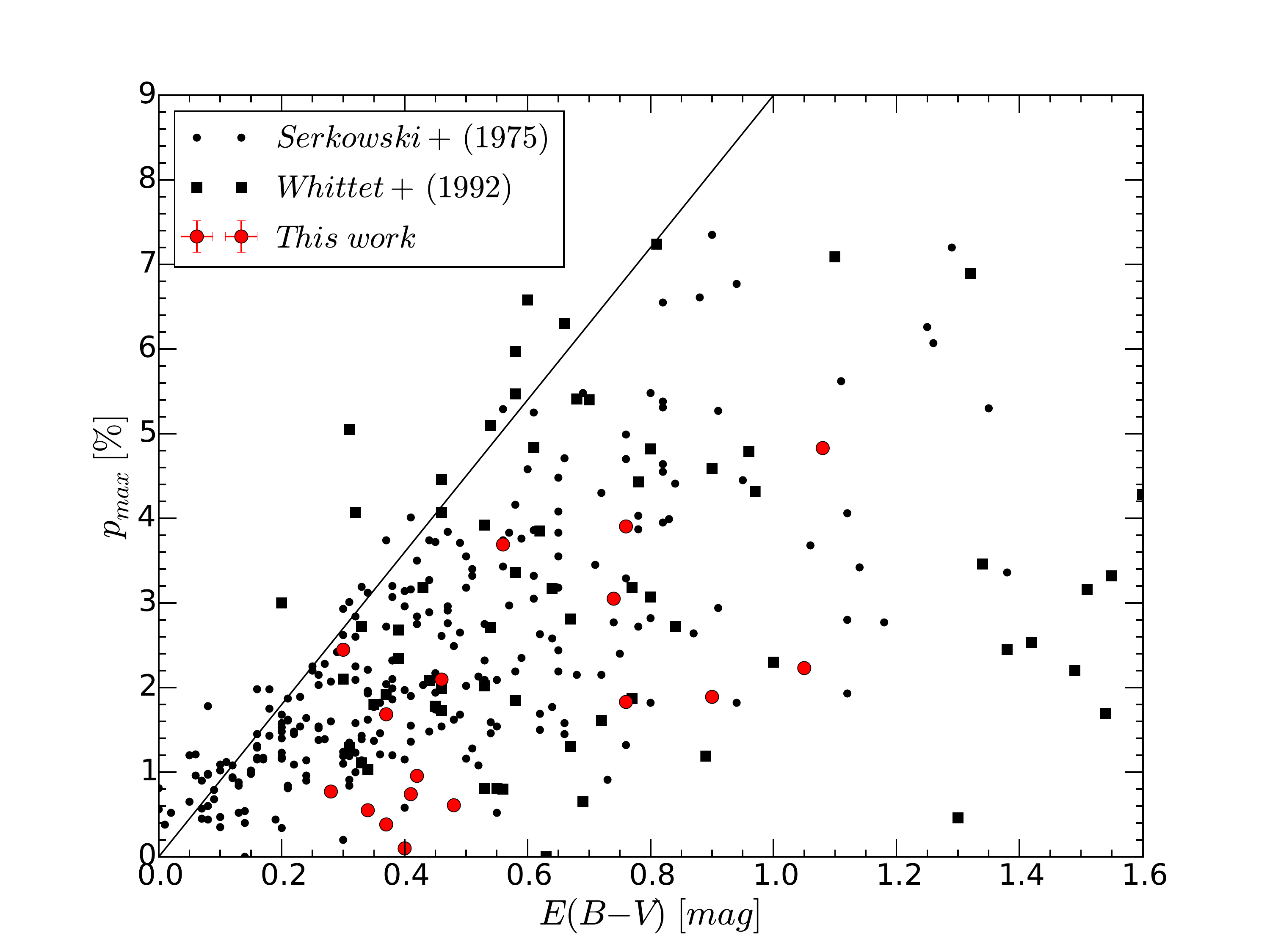}
\caption{Maximum interstellar polarization $p_{max}$ vs. color excess $E(B-V)$ of stars from \citealt{1992ApJ...386..562W} (black squares), \citealt{1975ApJ...196..261S} (black circles from), compared to our observed sample (red circles). The straight line denotes the upper limit $p_{max}(\%)$ = 9.0 $E(B-V)$ mag, defined by \citet{1975ApJ...196..261S}. }
\label{fig_pmax-EBV}
\end{center}
\end{figure}

There is no clear correlation between the maximum polarization and color excess. Figure~\ref{fig_pmax-EBV} shows the \citet{1975ApJ...196..261S} and \citet{1992ApJ...386..562W} sample in the $p_{\max}$ -- $E(B-V)$ plane compared to our sample of stars with anomalous sightlines. The scattered data in the plot shows there is no dependence of maximum polarization with reddening, however, there is an upper limit depending on reddening \citep{1975ApJ...196..261S}, which is rarely exceeded: $p_{\max}(\%)$ = 9.0 $E(B-V)$ mag. We calculated the mean of the ratio $\langle p_{max}$/$E(B-V) \rangle$ for the different samples: $\langle p_{max}$/$E(B-V) \rangle$=6.2 $\pm$ 3.8 $\%$ $mag^{-1}$ for the \citet{1975ApJ...196..261S} sample, $\langle p_{max}$/$E(B-V) \rangle$=4.6 $\pm$ 3.4 $\%$ $mag^{-1}$ for the \citet{1992ApJ...386..562W} sample, and $\langle p_{max}$/$E(B-V) \rangle$= 3.3 $\pm$ 2.1 $\%$ $mag^{-1}$ for our sample. The low $\langle p_{max}$/$E(B-V) \rangle$ ratio of our sample might also indicate that the silicate dust grains do not align as efficient compared to the \citet{1975ApJ...196..261S} and \citet{1992ApJ...386..562W} samples, however, because of a small number of stars in our sample, we can not draw such conclusions with high certainty. Another possible reason of low ratios of $p_{max}$/$E(B-V)$ is a small angle between the direction of the magnetic field and line of sight. The magnetic field wander is expected to reduce $p_{max}$/$E(B-V)$, as discussed in previous works, e.g. \citet{2014ApJ...790....6H}.

\subsection{What dust properties determine $\lambda_{\rm max}$?}
For a given grain shape and dust optical constant, Equation (\ref{eq:Plam}) reveals that the polarization spectrum is determined by $dn/da\times f_{\rm ali}$, which is considered the size distribution of aligned grains, while the extinction (i.e., $R_{V}$) in Eq.~(\ref{eq:Aext}) is only determined by $dn/da$. Thus, both a change in $dn/da$ and $f_{\rm ali}$ affect the polarization spectrum. 

Our simultaneous fitting to the extinction and polarization demonstrate that both, grain alignment and size distribution are required to change, in order to reproduce the variation of $\lambda_{\rm max}$ (see Figs.~\ref{fig:best_dnda} and \ref{fig:best_fali}). However, the change in grain alignment is more prominent. Indeed, Fig.~\ref{fig:best_fali} shows that the alignment of small grains required to reproduce $\lambda_{\rm max}=0.45\mum$ is an order of magnitude higher than that required for $\lambda_{\rm max}=0.55\mum$. We note that the modeling here is carried out for a constant $R_{V}$. In the lines of sight where grain growth can take place, resulting in the increase of $R_{V}$, we expect both grain evolution and alignment to contribute to the variation of $\lambda_{\rm max}$ and $K$.

To test whether grain evolution can reproduce the observed data, we rerun our simulations for the same six models by fixing the alignment function that reproduces the "standard" polarization curve with typical value $\lambda_{\max}=0.55\mum$. The size distributions $dn_{j}/da$ is varied. We find that the variation of $dn/da$ can reproduce the observed data to a satisfactory level only for the cases of $\lambda_{\rm max}=0.51-0.55\mum$, i.e., $\lambda_{\max}$ is not much different from the standard value. Meanwhile, the fit to the models is poor when $\lambda_{\rm max}$ differs much from the typical value of $0.55\mum$. It indicates that grain evolution alone cannot explain the wide range of $\lambda_{\max}$ as observed.

\subsection{Why is $K$ correlated to $\lambda_{\rm max}$?}
\label{discussion:K-lmax_relation}
The dependence of $K$ with $\lambda_{\rm max}$ appears to be an intrinsic property of the polarization. The Serkowski curve shows that a smaller $K$ corresponds to a broader polarization profile. From the inverse modeling for a constant $R_{V}$, we find that the grain alignment function becomes broader (narrower) for smaller (larger) values of $\lambda_{\rm max}$ as well as of $K$. This feature can be explained as follows. Each aligned grain of size $a$ produces an individual polarization profile $C_{\rm pol}$ with the peak at $\lambda \sim 2\pi a$ (see Fig. 1 in \citealt{2013ApJ...779..152H}). The polarization spectrum is the superimposition/integration over all grain sizes that are aligned. When the alignment function is broader, the superposition will produce a broader polarization profile, or smaller $K$.

\subsection{Deviation of $K$ from the standard value}

To explore the dust properties underlying the deviation of $K$ from the typical value, we perform the fit for a fixed $\lambda_{\rm max}$ and varying $k_{1}$ (in $K=k_{1}\lambda_{\rm max}+ 0.01$). 

\begin{figure*}
\centering
\includegraphics[width=0.3\textwidth]{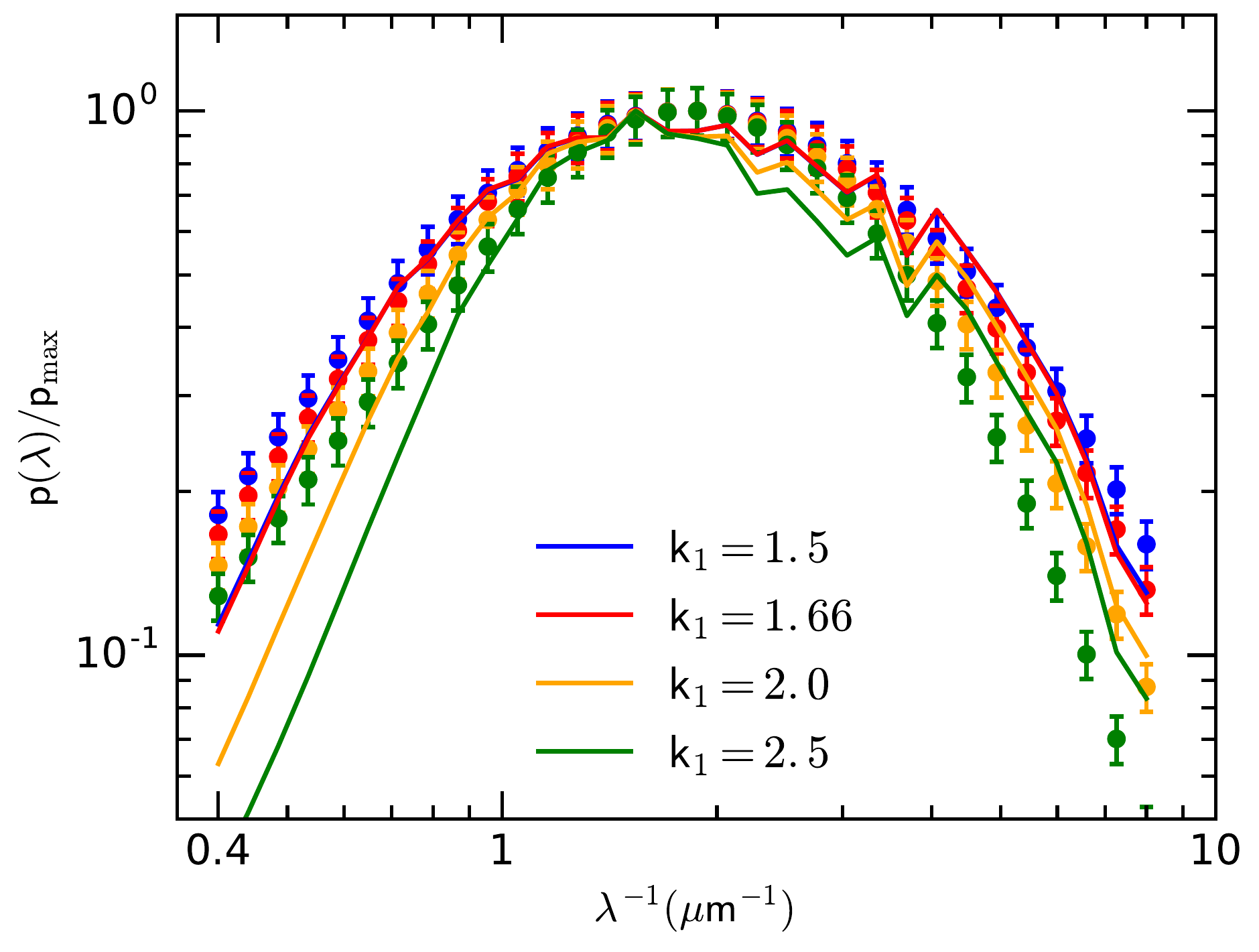}
\includegraphics[width=0.3\textwidth]{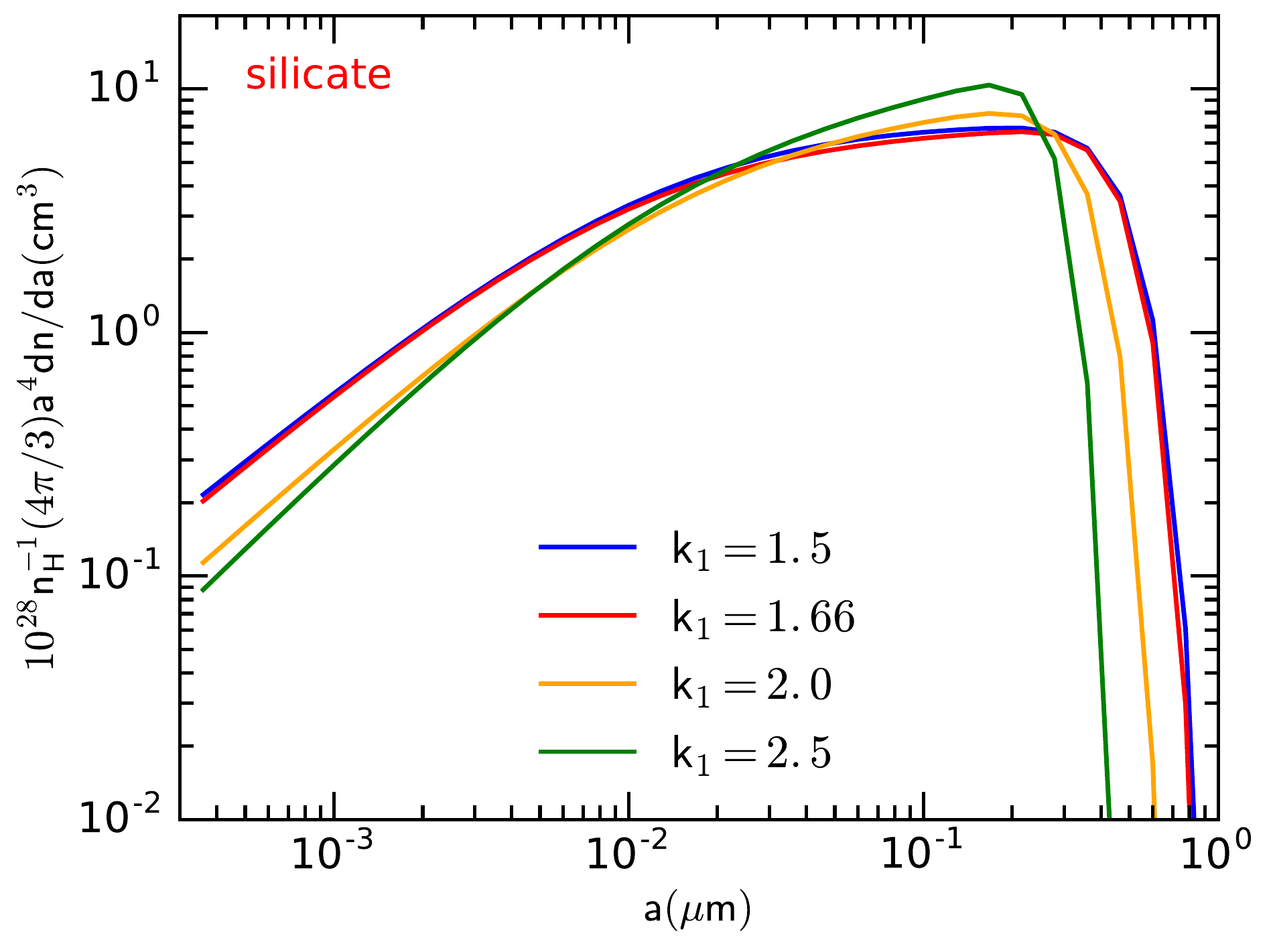}
\includegraphics[width=0.3\textwidth]{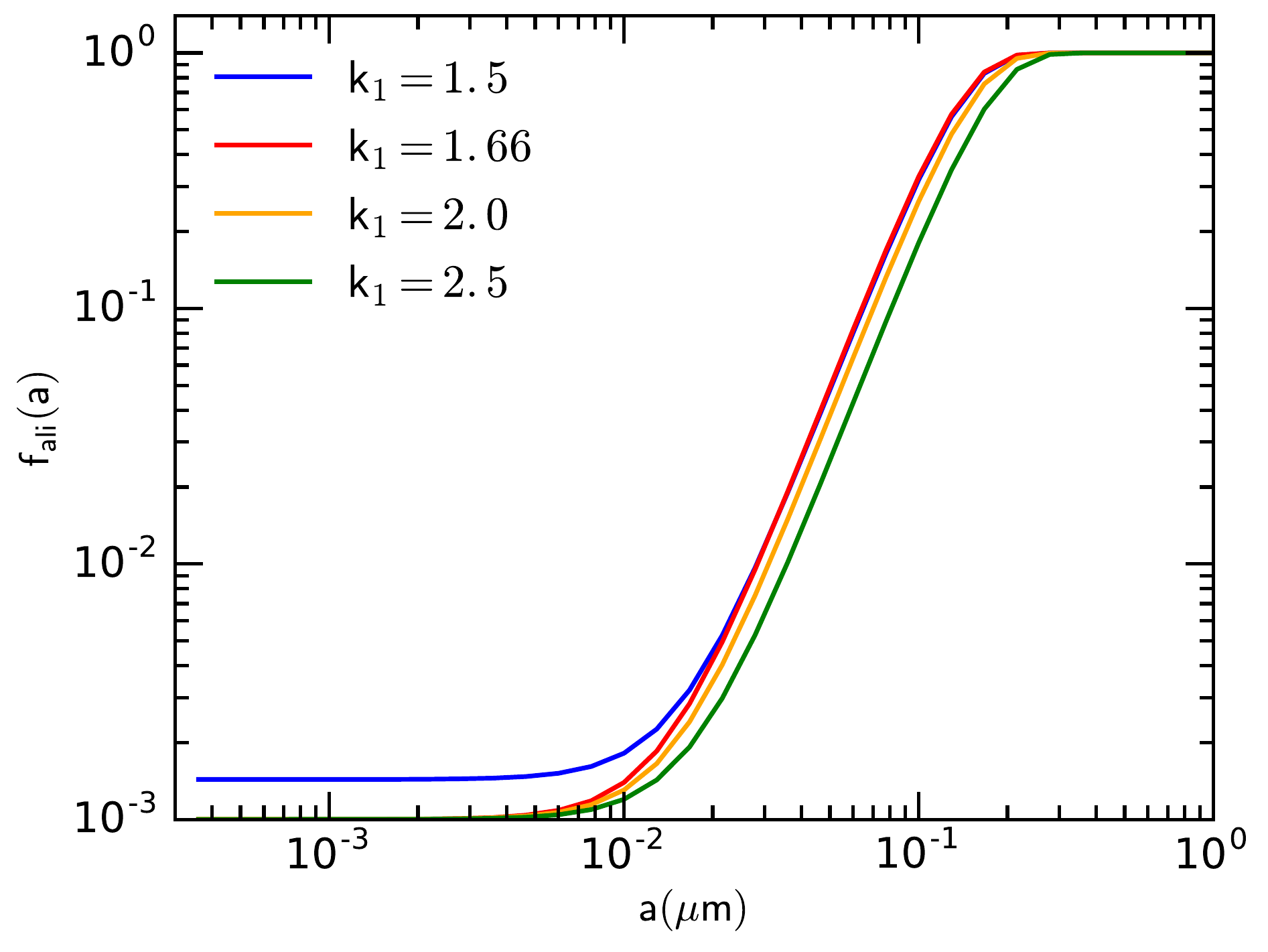}
\caption{Polarization curves (left panel), best-fit size distribution (middle panel), and best-fit alignment function (right panel) for the different values of $k_{1}$ where $K=k_{1}\lambda_{\rm max}+ 0.01$. The value of $\lambda_{\rm max}=0.55\mum$ is fixed.}
\label{fig:bestfit-K_w55}
\end{figure*}

Figure~\ref{fig:bestfit-K_w55} shows the best-fit models (left panel), alignment function (right panel), and size distribution (middle panel) for a fixed $\lambda_{\rm max}=0.55\mum$ and varying $K$. We find that the increase of $K$ is produced by the decrease of large Si grains ($a>0.2\mum$). At the same time, the alignment is shifted to the larger size when $K$ increases, which is seen in Fig.~\ref{fig:best_fali} when $\lambda_{\rm max}$ (and $K$) is increased.

\subsection{Relationship between $R_{\rm Si}$ and $\lambda_{\rm max}$}
\label{discussion:Rsi-lmax}

\begin{figure}
\includegraphics[width=0.45\textwidth]{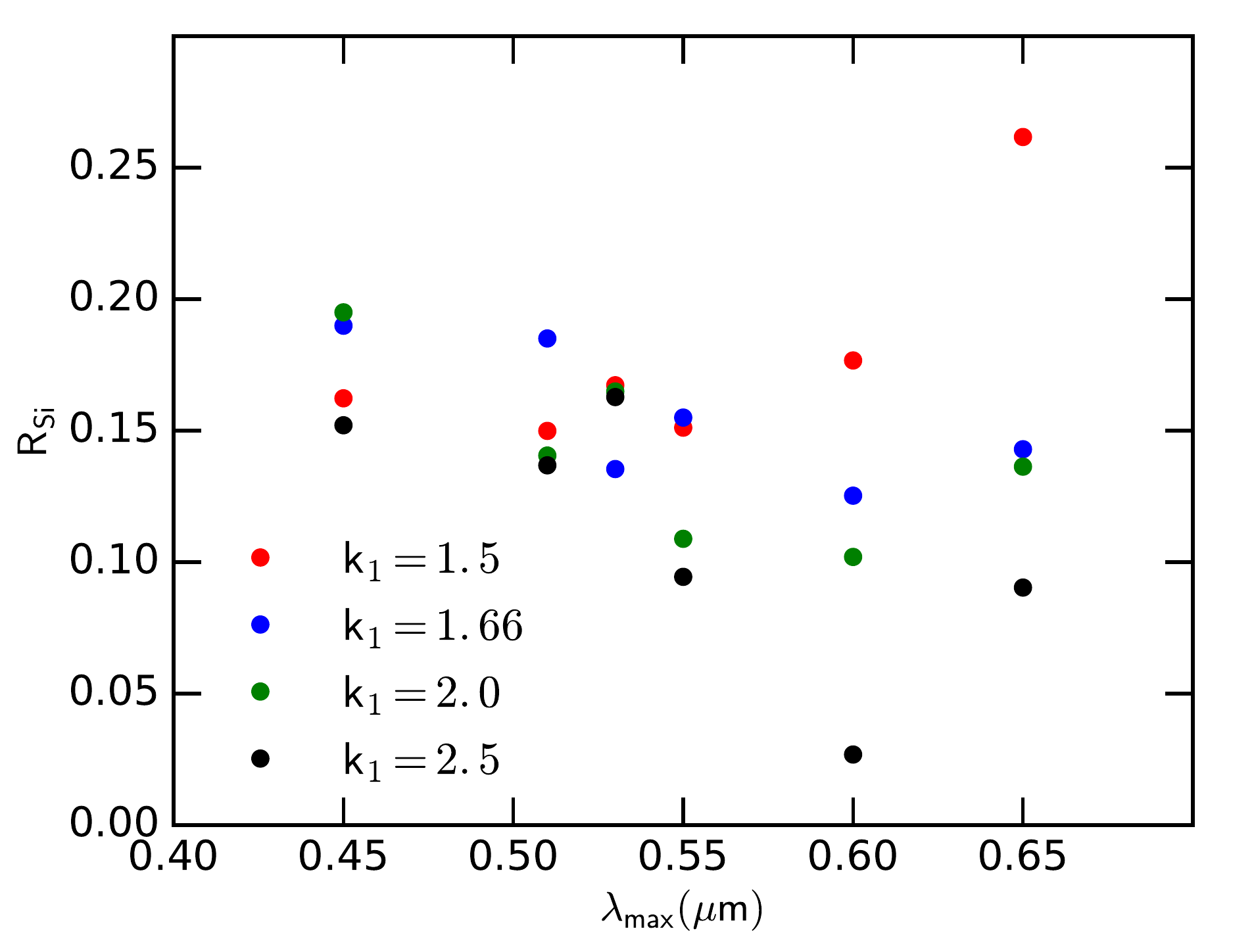}
\caption{The relative ratio of Si abundance in very small and large grain sizes vs. $\lambda_{\rm max}$.}
\label{fig:RSi_wavemax}
\end{figure}

Figure~\ref{fig:RSi_wavemax} shows the dependence of $R_{\rm Si}$ on $\lambda_{\rm max}$. There is a slight decrease of $R_{\rm Si}$ with increasing $\lambda_{\rm max}$. This is straightforward because small grains are required to reproduce larger polarization in the UV when $\lambda_{\rm max}$ decreases. 

The values $R_{\rm Si}$ from Fig.~\ref{fig_lmax-RSi} were inferred from fitting the extinction data only \citep{2011A&A...527A..34M}.  Thus, there is no direct relation between such inferred $R_{\rm Si}$ and $\lambda_{\rm max}$ that describes the polarization curve. 
It is known that a dust model that fits well the extinction may not reproduce the polarization data (e.g., dn/da $\sim$ a$^{-3.5}$ law by \citealt{Mathis:1977p3072}; see e.g. \citealt{1995ApJ...444..293K,2009ApJ...696....1D}).

The scale difference between $R_{\rm Si}$ shown in Fig.~\ref{fig_lmax-RSi} and Fig.~\ref{fig:RSi_wavemax} arises from the different ways of modeling. In \citet{2011A&A...527A..34M}, to reproduce the extinction curve with low $R_V$ values, dust grains are found to be rather small ($\lesssim$ 0.05 ${\rm \mu m}$), leading to a high value of $R_{\rm Si}$. Such a dust model cannot reproduce the polarization data with standard $\lambda_{\max}$ that require the presence of aligned large grains (i.e., larger 0.1 ${\rm \mu m}$). Our best-fit models that simultaneously fits extinction and polarization data provide the different size distributions that have large grains, leading to smaller $R_{\rm Si}$.

\section{Summary and conclusions}
\label{sec:summary}

We investigated linear polarization of seventeen sightlines to Galactic stars with anomalous extinction laws and low total-to-selective visual extinction ratio, $R_V$, selected from the \citet{2011A&A...527A..34M} sample, and adopt a simple dust model that can reproduce the observed sightlines with low $R_V$ values and normal polarization curves. Thereby, we adopt the \citet{Weingartner2001ApJ...548..296W} dust model and a picket-fence alignment model to compute extinction and polarization curves (see Sect.~\ref{subsect:polarization}). 

Our results can be summarized as follows:

\begin{enumerate}
\item The Galactic stars with anomalous extinction sightlines, with low $R_V$ values, show "normal" polarization curves with a mean $\lambda_{\rm max} \sim 0.53 {\rm \mu m}$. This can be explained by considering that not all dust which contributes to extinction, contributes to polarization. The polarization mainly depends on the dust grain size distribution of silicates, because grain alignment is more efficient for silicates than, for instance, for carbonaceous dust grains \citep{1994ApJ...427L..47S}, whereas $R_V$ is strongly dependent on carbonaceous grains too. 
\item There is no significant $R_V$--$\lambda_{\rm max}$ relation in our sample (Fig.~\ref{fig_lmax-RV}). The $\lambda_{\rm max}$ values in our sample are higher than compared to normal stars which follow the empirical $R_V$--$\lambda_{\rm max}$ relationship by e.g. \citet{1978A&A....66...57W}.
\item Despite the low $R_V$ value, there is no similarity between the polarization curves in the investigated sample and the polarization curves observed in reddened SNe Ia with low $R_V$ values. The polarization curves are consistent with a sample of Galactic stars observed by \citet{1992ApJ...386..562W} within 3$\sigma$ (Fig.~\ref{fig_lmax-K}).
\item The Serkowski parameters $K$ and $\lambda_{\rm max}$ are correlated. However, we find a steeper slope ($K$= -1.13 $\pm$ 0.34 + (4.05 $\pm$ 0.64)$\lambda_{\rm max}$) in our sample, compared to the empirical relationship found by e.g. \citet{1992ApJ...386..562W}.

\item Simulations show that, to reproduce a polarization curve with the normal $\lambda_{\max}$ and low $R_V$, there must be a population of large interstellar silicate grains of size $a\ge 0.1\mum$. This is different compared to results by \citet{2008MNRAS.390..706M, 2011A&A...527A..34M}, who best-fit the extinction curves only, and found a lack of such large Si grains, however the presence of large carbonaceous grains.
Moreover, both, variations in grain alignment and size distribution are required to reproduce the variation in $\lambda_{\rm max}$, for a fixed, low, $R_{V}$ value. However, a change in grain alignment has a greater impact.

\item By comparing the $R_V$ values of the sample here considered with those in  \citet{2002BaltA..11....1W}, for a subset of our stars, we find some differences probably due to a different spectral classification and/or luminosity class adopted to derive their extinction curves (see Sect.~\ref{subsec:rv-lmax_relation}). The $\lambda_{\rm max}$ value that we measure, and the deviation from the empirical $R_V$--$\lambda_{\rm max}$ relationship, may also suggest a spectral misclassification of some stars by \citet{1985ApJS...59..397S}.
\item The $K$--$\lambda_{\rm max}$ relation appears to be an intrinsic property of the polarization. Simulations show that, for a fixed $R_V$, the grain alignment function becomes narrower (broader) for a smaller (larger) value of $\lambda_{\rm max}$ and $K$ (see Sect.~\ref{discussion:K-lmax_relation}).
\item An increase of the Serkowski parameter $K$, and deviation from the standard value in the $K$--$\lambda_{\rm max}$ plane, can be reproduced by decreasing contribution of large Si grains (Fig.~\ref{fig:bestfit-K_w55}). 
\end{enumerate}

\begin{acknowledgements}
We thank an anonymous reviewer for useful comments that significantly improved the paper. 
This work is partially based on observations collected at the German-Spanish Astronomical Center, Calar Alto, jointly operated by the Max-Planck-Institut f\"{u}r Astronomie Heidelberg and the Instituto de Astrofisica de Andalucia (CSIC). T.H. acknowledges the support from the Basic Science Research Program through the National Research Foundation of Korea (NRF), funded by the Ministry of Education (2017R1D1A1B03035359).
This work is based on observations made with ESO Telescopes at the Paranal Observatory under Program ID 094.C-0686, and partially based on observations collected with the Copernico telescope (Asiago, Italy) of the INAF - Osservatorio Astronomico di Padova. We also thank P. Ochner for taking some observations with AFOSC in service time.
Some of the data presented in this paper were obtained from the Mikulski Archive for Space Telescopes (MAST). STScI is operated by the Association of Universities for Research in Astronomy, Inc., under NASA contract NAS5-26555. Support for MAST for non-HST data is provided by the NASA Office of Space Science via grant NNX09AF08G and by other grants and contracts.
ST acknowledges support from TRR33 "The Dark Universe" of the German Research Foundation.

\end{acknowledgements}

\bibliographystyle{aa} 
\bibliography{specpol.bib} 

\begin{appendix}

\section{Standard stars}
\label{sec:Standard stars}

\subsection{Standard stars with CAFOS}
\label{sec:Standard stars with CAFOS}

Two unpolarized standard stars were observed with CAFOS: HD 144579 (2 epochs), and HD 90508 (1 epoch); and two polarized standard stars, each at two epochs: HD 154445 and HD 43384.

\begin{figure}[h!]
\begin{center}
\includegraphics[trim=10mm 0mm 0mm 10mm, width=9cm, clip=true]{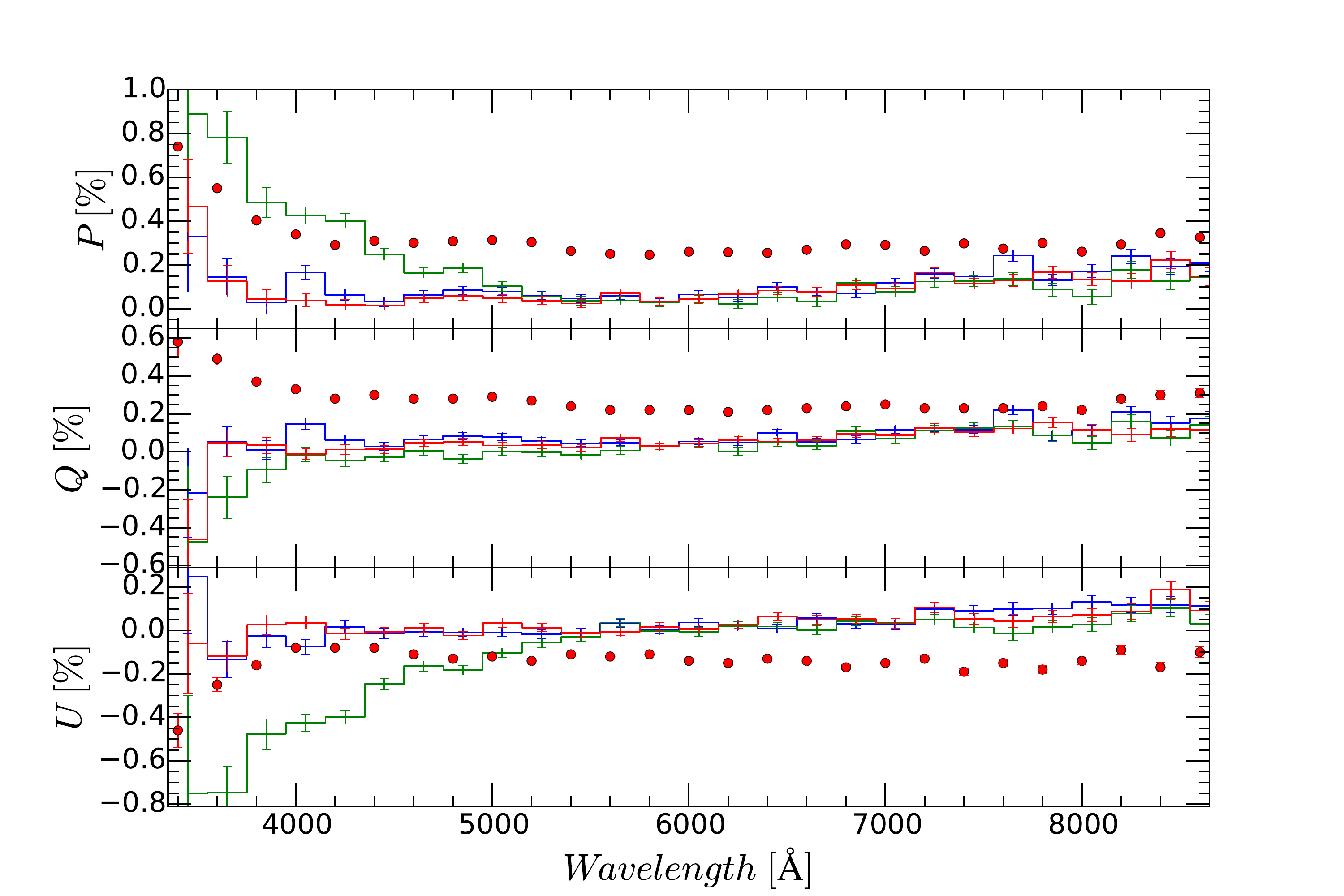}
\caption{Unpolarized standard stars observed with CAFOS. HD 144579 was observed at two epochs, on 2015-04-30 at 00:51 UT (blue line) and 04:22 UT (green line). The red line indicates HD 90508 observed on 2015-04-29 at 21:28 UT. For comparison, the red dots indicate the instrumental polarization determined in \citet{2011A&A...529A..57P}.}
\label{fig_std_stars_unpol}
\end{center}
\end{figure}

We use the unpolarized standard stars to investigate possible instrumental effects. The observations have been binned in 200$\AA$ bins, and the Stokes parameters, the polarization degree and polarization angle were calculated as described in Sect.~\ref{subsection_CAFOS}.
Figure~\ref{fig_std_stars_unpol} shows the derived Stokes parameters and polarization of the two unpolarized standard stars at three epochs, compared to the instrumental polarization determined in \citet{2011A&A...529A..57P}. Our measurements show consistent values between the three epochs, with average $Q$ and $U$ vales of 0.07\,$\%$ and 0.004\,$\%$ respectively, in a wavelength range 3850-8650$\AA$, leading to an average polarization of $P \approx$ 0.10\,$\%$. The standard deviations are $\sigma_Q \approx$ 0.06 $\%$, $\sigma_U \approx$ 0.11 $\%$ and $\sigma_P \approx$ 0.07 $\%$. The average uncertainty per 200$\AA$ bin is $\sim$0.03 $\%$. 
Our values are more consistent with zero than the values determined in \citet{2011A&A...529A..57P}. They analyzed observations of the unpolarized star HD 14069 observed at 16 half-wave plate angle, and claimed average values of the instrumental Stokes parameters in a wavelength range above 4000$\AA$ of $\langle Q_{ins} \rangle$ = 0.25 $\pm$ 0.03$\%$ and $\langle U_{ins} \rangle$ = -0.13 $\pm$ 0.03$\%$, leading to an average polarization $P_{ins}$ = 0.28 $\pm$ 0.03$\%$.

\begin{figure}[h!]
\begin{center}
\includegraphics[trim=10mm 0mm 0mm 25mm, width=9cm, clip=true]{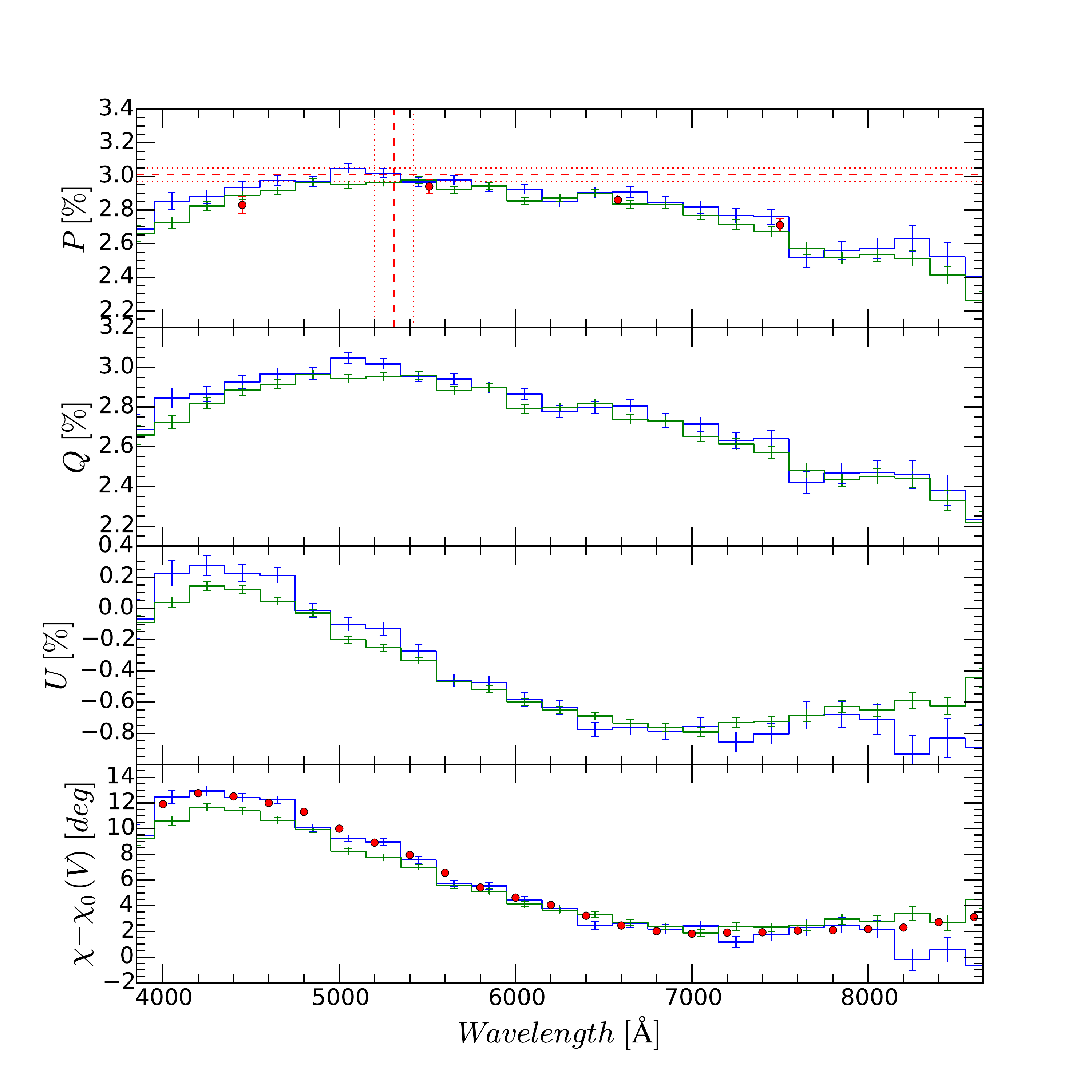}
\caption{Polarized standard star HD 43384 observed with CAFOS at two epochs, on 2015-04-29 at 20:14 UT (green line) and 21:14 UT (blue line). Both epochs are consistent to each other with a RMS of $\sim$ 0.04 $\%$. The red dashed lines in the polarization panel indicate $p_{\rm max}$ and $\lambda_{\rm max}$ and their errors (dotted line) determined by \citet{1982ApJ...262..732H}, and the red dots are their individual measurements. The red dots in the $\chi$ - $\chi_0(V)$ panel indicate the phase retardance variance determined in \citet{2011A&A...529A..57P}.}
\label{fig_std_stars_pol}
\end{center}
\end{figure}

There were two polarized standard stars observed with CAFOS, each at two epochs: HD 154445 and HD 43384. The calculated polarization spectra of HD 43384 are consistent to each other with an RMS of $\sim$ 0.04 $\%$.
Our Serkowski parameters (see Table~\ref{tb_pol_individual_CAFOS}) are fully consistent with $p_{\rm max}$ = 3.01 $\pm$ 0.04 $\%$ and $\lambda_{\rm max}$ = 0.531 $\pm$ 0.011 ${\rm \mu m}$, determined in \citet{1982ApJ...262..732H}.
The wavelength dependent phase retardance variation of the half-wave plate deployed in CAFOS was quantified in \citet{2011A&A...529A..57P}. HD 43384 has a variable (+0.6 $^{\circ}$/100 yr), and slightly wavelength dependent (+2.5$\pm$1.3 $^{\circ}$/${\rm \mu m}$) polarization position angle \citep{1982ApJ...262..732H}. Therefore it is not the best standard star for HWP chromatism investigation. However, for comparison reasons, we use $\chi_0(V)$= 169.8 $\pm$ 0.7 degrees to compute the phase retardance variance and find that it is consistent with \citet{2011A&A...529A..57P} (see Fig.~\ref{fig_std_stars_pol}). The average deviation of our phase retardance variation compared to \citet{2011A&A...529A..57P} is $\sim$ +0.3 degrees, which is within the errors of $\chi_0$.

Also, both observed polarization spectra of HD 154445 are consistent to each other with an RMS of 0.07 $\%$. The average $\lambda_{\rm max}$=5579 $\pm$ 11 $\AA$ and $p_{\rm max}$= 3.64 $\pm$ 0.01 $\%$ matches with the literature values $p_{\rm max}$ = 3.66$\pm$ 0.01 $\%$ and $\lambda_{\rm max}$=5690 $\pm$ 10 $\AA$ \citep{1996AJ....111..856W}. Our average $\theta$ after the HWP chromatism correction is 89.2 $\pm$ 0.4 degree, which is similar to the literature values of $\theta_{V}$ = 88.8 $\pm$ 0.1 \citep{1992AJ....104.1563S}, $\theta_{V}$ = 90.1 $\pm$ 0.1 and $\theta_{\rm max}$=88.3 $\pm$ 0.1 degrees \citep{1982ApJ...262..732H}.

\subsection{Standard stars with AFOSC}
\label{sec:Standard stars with AFOSC}

Because of lack of space in AFOSC, for polarimetry purposes, the Wollaston prism gets inserted into the filter wheel in place of a filter. Therefore, during each observing run it is necessary to calibrate the instrument's zero point rotation angle using polarized standard stars.

\begin{figure}[h!]
\begin{center}
\includegraphics[trim=0mm 0mm 0mm 10mm, width=9cm, clip=true]{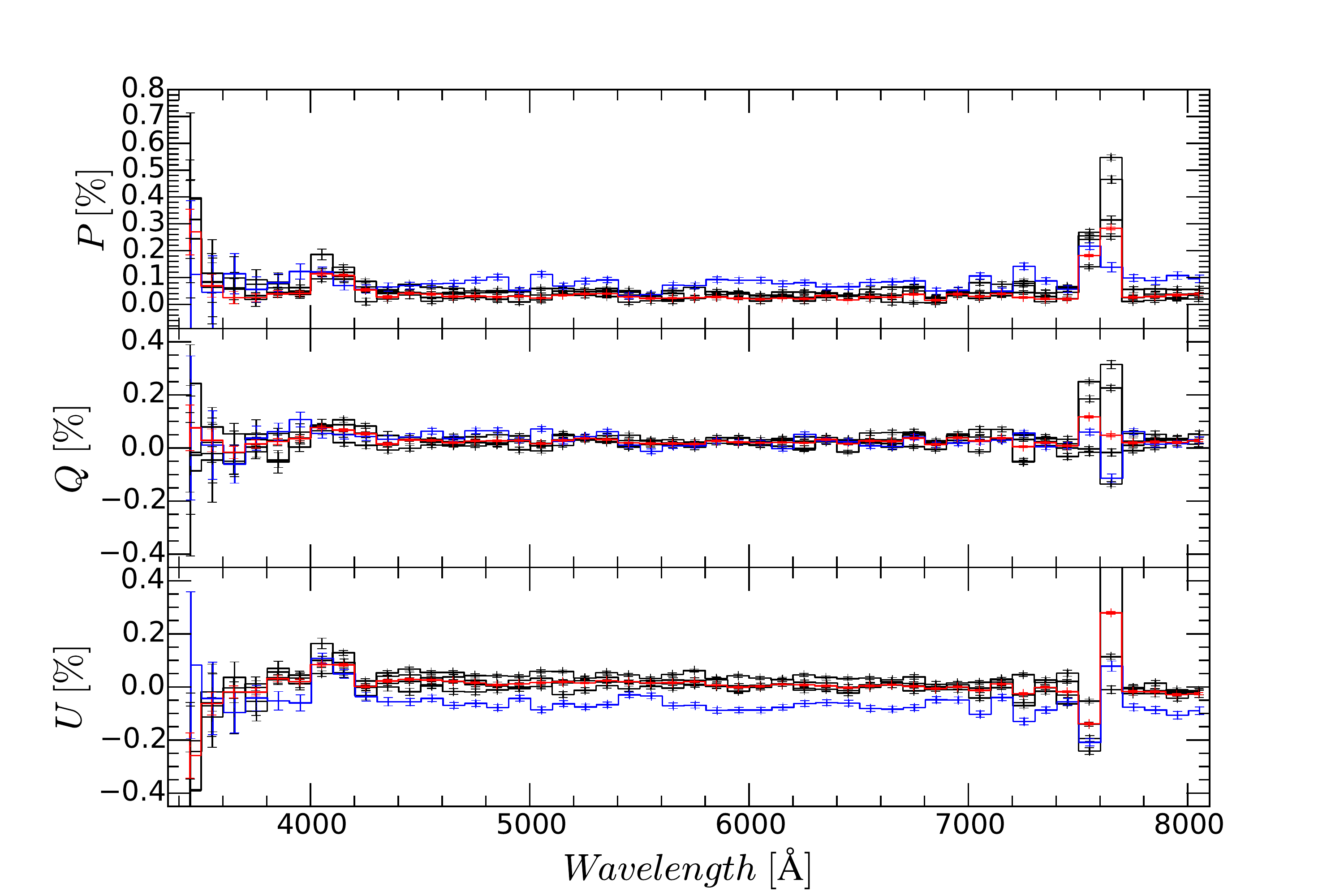}
\caption{Unpolarized standard stars observed with AFOSC. The black lines indicates observations observed at 4 rotation angles, and the blue lines indicates HD 90508 observed at two rotation angles only. The red line is the average of all measurements.}
\label{fig_std_stars_unpol_asiago}
\end{center}
\end{figure}

We use observations of 3 unpolarized standard stars observed from 2015-02-09 to 2015-03-11 to investigate possible instrumental polarization of AFOSC: HD 90508 (2 epochs), HD 39587 (2 epochs) and HD 144579 (1 epoch). They have all been observed at 4 rotation angles of the adapter (-45, 0, 45 and 90 degrees), except for one epoch of HD 90508 which has been observed at only 2 rotation angles (0 and 90 degree).
Figure~\ref{fig_std_stars_unpol_asiago} shows the derived Stokes parameters $Q$ and $U$, and the polarization for all unpolarized standard stars. The black lines indicate the epochs observed at 4 rotation angles, and the blue line indicates HD 90508 observed at 2 rotation angles.
The average stokes parameters at a wavelength range above 3600 $\AA$, excluding the range from 7500-7700$\AA$, which is contaminated by the strong telluric 7605.0 $\AA$ ${\rm O_2}$ line, are 0.03 $\%$ and -0.002 $\%$ for $Q$ and $U$ respectively, leading to a polarization of 0.05 $\%$, with a standard deviation of 0.05 $\%$.

When ignoring the observations performed with the adapter rotation angles of -45 and 45 degrees, and calculating the polarization using only the adapter rotation angles of 0 and 90 degrees for all 5 epochs of the 3 standard stars, the average stokes parameters, are 0.05 $\%$ and -0.04 $\%$ for Q and U respectively, leading to a polarization of 0.08 $\%$, with a standard deviation of 0.10 $\%$.

Also HD 185395, observed on 2016-08-02, is consistent with zero, with an average polarization above 3600 $\AA$, excluding the range from 7500-7700$\AA$, of 0.08 $\%$ and an RMS of 0.06 $\%$.

There were three polarized standard stars observed with AFOSC: HD 43384 (3 epochs at two different runs), HD 21291 (1 epoch), and HD 198478 (1 epoch), all at 4 rotation angles of the adapter. However, because most of the science data was taken only with two rotation angles, to be consistent, we use only two rotation angles.

\begin{figure}[h!]
\begin{center}
\includegraphics[trim=0mm 10mm 0mm 30mm, width=9cm, clip=true]{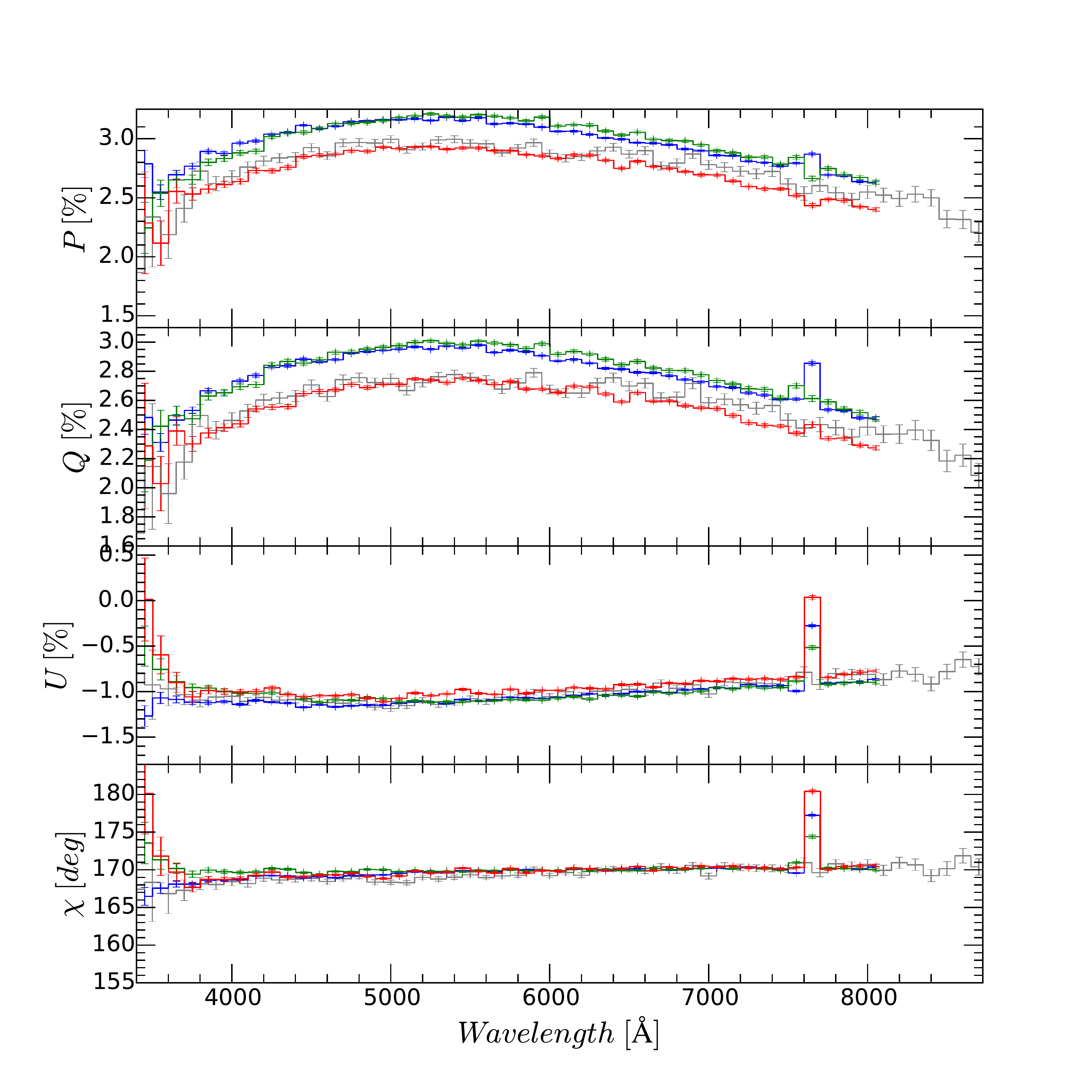}
\caption{Polarized standard star HD 43384 observed at three different nights with AFOSC. The  peak polarization values, $p_{\rm max}$, are not fully consistent and range from $\sim$2.92$\%$ to $\sim$3.19$\%$. For reference, the gray line indicates HD 43384 observed with CAFOS.}
\label{fig_std_HD43384_asiago}
\end{center}
\end{figure}

Figure~\ref{fig_std_HD43384_asiago} shows HD 43384 at 3 different epochs. Although the shapes of the polarization spectra are similar, i.e. $\lambda_{\rm max}$ and $K$ of the Serkowski fit are similar to each other, the peak polarization values, $p_{\rm max}$, are not consistent and range from $\sim$2.92$\%$ to $\sim$3.19$\%$ (see Table~\ref{tb_pol_individual_AFOSC}), while the literature value is $p_{\rm max}$ = 3.01 $\pm$ 0.04 $\%$ \citep{1982ApJ...262..732H}.

For HD 21291 our determined peak polarization value is lower than the literature value. By fitting the Serkowski curve, we find $\lambda_{\rm max}$ = 5166 $\pm$ 27, $p_{\rm max}$ = 2.95 $\pm$ 0.01 $\%$, while the literature values are $p_{\rm max}$ $\sim$ 3.53 $\pm$ 0.02 $\%$ and $\lambda_{\rm max}$ = 5210 $\pm$ 30 $\AA$ \citep{1982ApJ...262..732H}.
 
HD 198478 was observed on 2016-08-02. Our determined peak polarization value is almost consistent with the literature value. By fitting the Serkowski curve, we find $\lambda_{\rm max}$ = 5132 $\pm$ 42, $p_{\rm max}$ = 2.76 $\pm$ 0.01 $\%$, while the literature values are $p_{\rm max}$ $\sim$ 2.72 $\pm$ 0.02 $\%$ and $\lambda_{\rm max}$ = 5220 $\pm$ 80 $\AA$ \citep{1982ApJ...262..732H}.

We use polarized standard stars to determine the correction of the instrument's rotation angle zero point for each of the three runs separately. 

During the first run (2015-02-09) two stars have been observed: HD 43384 (2 epochs) and HD 21291 (1 epoch). Using the literature values of $\chi_V$ = 169.8 $\pm$ 0.7 degrees for HD 43384, and $\chi_V$ = 116.6 $\pm$ 0.2 degrees for HD 21291, we calculated an weighted average of the offset $\Delta \theta_V$ = 136.4 $\pm$ 0.3 degrees. 

During the second run (2015-03-09), HD 43384 was observed at one epoch, from which we calculated the instrument's rotation angle zero point offset $\Delta \theta_V$ = 134.0 $\pm$ 0.8 degrees.

Finally, from HD 198478 observed during the third run (2016-08-02), we calculated $\Delta \theta_V$ = 138.3 $\pm$ 0.4 degrees.

\section{Serkowski fit to SNe Ia}
\label{appendix:fitserkowskitozelaya}

We determine the Serkowski parameters of the sodium sample in \citet[][]{Zelaya2017arxiv} by fitting Eq.~(\ref{eq_serkowski}) to the data shown in their Fig. 1. The polarized lines have been excluded from the wavelength range. The results are given in Table~\ref{tab:SNefit}.

\begin{table}[h!]
\caption{Serkowski paramaters of the SNe Ia Sodium-sample (Zelaya et al. 2017)}
\vspace{-0.3cm}
\label{tab:SNefit}
\begin{tabular}{llll}
\hline\hline
SN Name & $\lambda_{\rm max}$ $(\AA)$ & $p_{\rm max}$ $(\%)$ & $K$ \\
\hline
SN 2007le  & 3967 $\pm$ 494  & 1.93 $\pm$ 0.09  & 1.73 $\pm$ 0.87  \\
SN 2010ev  & 4408 $\pm$ 114  & 1.88 $\pm$ 0.02  & 1.89 $\pm$ 0.23  \\
SN 2007fb  & 3821 $\pm$ 447  & 0.76 $\pm$ 0.03  & 1.13 $\pm$ 0.46  \\
SN 2003W   & 3996 $\pm$ 371  & 0.78 $\pm$ 0.04  & 2.56 $\pm$ 1.03  \\
SN 2007af  & 7409 $\pm$ 537  & 0.64 $\pm$ 0.01  & 1.24 $\pm$ 0.35  \\
SN 2002fk  & 4403 $\pm$ 460  & 0.40 $\pm$ 0.01  & 0.57 $\pm$ 0.35  \\
SN 2002bo  & 3525 $\pm$ 137  & 1.11 $\pm$ 0.05  & 3.95 $\pm$ 0.49  \\
SN 2011ae  & 4256 $\pm$ 295  & 0.33 $\pm$ 0.01  & 3.56 $\pm$ 1.87  \\
SN 2005hk\tablefootmark{a}  & 6731 $\pm$ 2116 & 0.17 $\pm$ 0.02  & -1.36 $\pm$ 1.51  \\
  \hline
\end{tabular} 
\tablefoot{\tablefoottext{a}{ because of a low polarization degree and low signal to noise, a good fit with the Serkowski curve was not possible.} }
\end{table}

\section{Individual observations}

\onecolumn

\begin{landscape}
{\tiny
\longtab{
\begin{longtable}{lllllllllll}
\caption{\label{tb_pol_individual} Individual epochs with FORS2}\\
\hline\hline
&&&&&&&& \multicolumn{3}{c}{Serkowski curve} \\\cline{9-11}
Name & Filter & Epoch & Passband &  $P \hspace{0.3mm} (\%)$ &  $P_Q \hspace{0.3mm} (\%)$ &  $P_U \hspace{0.3mm} (\%)$ & $\theta \hspace{0.3mm} (^{\circ})$ & $\lambda_{\rm max} \hspace{0.3mm} (\AA)$ &  $P_{\rm max} \hspace{0.3mm} (\%)$ & $K$ \\
\hline
\endfirsthead
\caption{continued.}\\
\hline
&&&&&&&& \multicolumn{3}{c}{Serkowski curve} \\\cline{9-11}
Name & Filter & Epoch & Passband &  $P \hspace{0.3mm} (\%)$ &  $P_Q \hspace{0.3mm} (\%)$ &  $P_U \hspace{0.3mm} (\%)$ & $\theta \hspace{0.3mm} (^{\circ})$ & $\lambda_{\rm max} \hspace{0.3mm} (\AA)$ &  $P_{\rm max} \hspace{0.3mm} (\%)$ & $K$ \\
\hline
\endhead
\hline
\endfoot
HD 141318& free & 2014-10-10 00:23:57& &   &  &   &  &  5827 $\pm$ 40 & 2.43 $\pm$ 0.01 & 1.11 $\pm$ 0.06 \\ 
         &                &  &     B &2.28 $\pm$ 0.02 & -0.41 $\pm$ 0.01 & 2.24 $\pm$ 0.02 & 50.2 $\pm$ 0.2 & & &\\ 
         &                &  &     V &2.43 $\pm$ 0.01 & -0.52 $\pm$ 0.01 & 2.38 $\pm$ 0.01 & 51.2 $\pm$ 0.2& & &\\ 
         &                &  &     R &2.40 $\pm$ 0.01 & -0.55 $\pm$ 0.01 & 2.34 $\pm$ 0.01 & 51.7 $\pm$ 0.1& & &\\ 
         &                &  &     I &2.18 $\pm$ 0.01 & -0.50 $\pm$ 0.01 & 2.12 $\pm$ 0.01 & 51.7 $\pm$ 0.1& & &\\ 
HD 141318& free & 2014-10-10 00:27:27& &   &  &   &  &  5770 $\pm$ 23 & 2.46 $\pm$ 0.01 & 1.29 $\pm$ 0.04 \\ 
         &                &  &     B &2.30 $\pm$ 0.01 & -0.40 $\pm$ 0.01 & 2.26 $\pm$ 0.01 & 50.0 $\pm$ 0.1 & & &\\ 
         &                &  &     V &2.46 $\pm$ 0.01 & -0.52 $\pm$ 0.01 & 2.40 $\pm$ 0.01 & 51.1 $\pm$ 0.1& & &\\ 
         &                &  &     R &2.41 $\pm$ 0.01 & -0.55 $\pm$ 0.01 & 2.35 $\pm$ 0.01 & 51.6 $\pm$ 0.1& & &\\ 
         &                &  &     I &2.15 $\pm$ 0.01 & -0.48 $\pm$ 0.01 & 2.10 $\pm$ 0.01 & 51.4 $\pm$ 0.1& & &\\ 
HD 141318& GG435 & 2014-10-10 00:37:56& &   &  &   &  &  5544 $\pm$ 57 & 2.43 $\pm$ 0.01 & 1.24 $\pm$ 0.08 \\ 
         &                &  &     B &2.35 $\pm$ 0.02 & -0.49 $\pm$ 0.02 & 2.30 $\pm$ 0.02 & 51.0 $\pm$ 0.2 & & &\\ 
         &                &  &     V &2.44 $\pm$ 0.01 & -0.50 $\pm$ 0.01 & 2.39 $\pm$ 0.01 & 50.9 $\pm$ 0.1& & &\\ 
         &                &  &     R &2.36 $\pm$ 0.01 & -0.50 $\pm$ 0.01 & 2.31 $\pm$ 0.01 & 51.1 $\pm$ 0.1& & &\\ 
         &                &  &     I &2.06 $\pm$ 0.01 & -0.39 $\pm$ 0.01 & 2.02 $\pm$ 0.01 & 50.6 $\pm$ 0.1& & &\\ 
HD 141318& GG435 & 2014-10-10 00:46:02& &   &  &   &  &  5851 $\pm$ 79 & 2.38 $\pm$ 0.01 & 0.81 $\pm$ 0.08 \\ 
         &                &  &     B &2.33 $\pm$ 0.02 & -0.46 $\pm$ 0.02 & 2.28 $\pm$ 0.02 & 50.7 $\pm$ 0.2 & & &\\ 
         &                &  &     V &2.40 $\pm$ 0.01 & -0.44 $\pm$ 0.01 & 2.36 $\pm$ 0.01 & 50.3 $\pm$ 0.1& & &\\ 
         &                &  &     R &2.37 $\pm$ 0.01 & -0.46 $\pm$ 0.01 & 2.32 $\pm$ 0.01 & 50.6 $\pm$ 0.1& & &\\ 
         &                &  &     I &2.20 $\pm$ 0.01 & -0.46 $\pm$ 0.01 & 2.15 $\pm$ 0.01 & 51.0 $\pm$ 0.1& & &\\ 
HD 152245& free & 2014-10-13 00:22:31& &   &  &   &  &  6360 $\pm$ 63 & 0.93 $\pm$ 0.01 & 1.32 $\pm$ 0.12 \\ 
         &                &  &     B &0.80 $\pm$ 0.01 & -0.10 $\pm$ 0.01 & 0.79 $\pm$ 0.01 & 48.4 $\pm$ 0.4 & & &\\ 
         &                &  &     V &0.91 $\pm$ 0.01 & -0.10 $\pm$ 0.01 & 0.90 $\pm$ 0.01 & 48.3 $\pm$ 0.3& & &\\ 
         &                &  &     R &0.92 $\pm$ 0.01 & -0.10 $\pm$ 0.01 & 0.92 $\pm$ 0.01 & 48.2 $\pm$ 0.2& & &\\ 
         &                &  &     I &0.84 $\pm$ 0.01 & -0.06 $\pm$ 0.01 & 0.84 $\pm$ 0.01 & 47.2 $\pm$ 0.2& & &\\ 
HD 152245& GG435 & 2015-02-06 08:26:24& &   &  &   &  &  6174 $\pm$ 64 & 1.01 $\pm$ 0.01 & 1.25 $\pm$ 0.12 \\ 
         &                &  &     B &0.92 $\pm$ 0.01 & 0.18 $\pm$ 0.01 & 0.90 $\pm$ 0.01 & 39.3 $\pm$ 0.4 & & &\\ 
         &                &  &     V &0.98 $\pm$ 0.01 & 0.24 $\pm$ 0.01 & 0.95 $\pm$ 0.01 & 38.0 $\pm$ 0.2& & &\\ 
         &                &  &     R &0.98 $\pm$ 0.01 & 0.25 $\pm$ 0.01 & 0.94 $\pm$ 0.01 & 37.6 $\pm$ 0.1& & &\\ 
         &                &  &     I &0.88 $\pm$ 0.01 & 0.26 $\pm$ 0.01 & 0.84 $\pm$ 0.01 & 36.4 $\pm$ 0.2& & &\\ 
HD 152245& GG435 & 2015-02-06 08:31:36.880& &   &  &   &  &  6465 $\pm$ 66 & 1.0 $\pm$ 0.01 & 1.13 $\pm$ 0.11 \\ 
         &                &  &     B &0.88 $\pm$ 0.01 & 0.22 $\pm$ 0.01 & 0.85 $\pm$ 0.01 & 37.9 $\pm$ 0.5 & & &\\ 
         &                &  &     V &0.95 $\pm$ 0.01 & 0.28 $\pm$ 0.01 & 0.91 $\pm$ 0.01 & 36.4 $\pm$ 0.2& & &\\ 
         &                &  &     R &0.97 $\pm$ 0.01 & 0.30 $\pm$ 0.01 & 0.92 $\pm$ 0.01 & 36.1 $\pm$ 0.1& & &\\ 
         &                &  &     I &0.91 $\pm$ 0.01 & 0.30 $\pm$ 0.01 & 0.85 $\pm$ 0.01 & 35.3 $\pm$ 0.2& & &\\ 
HD 54439& free & 2014-10-27T05:39:01& &   &  &   &  &  4802 $\pm$ 196 & 0.8 $\pm$ 0.01 & 0.82 $\pm$ 0.13 \\ 
         &                &  &     B &0.76 $\pm$ 0.01 & 0.04 $\pm$ 0.01 & -0.76 $\pm$ 0.01 & 136.6 $\pm$ 0.5 & & &\\ 
         &                &  &     V &0.76 $\pm$ 0.01 & 0.09 $\pm$ 0.01 & -0.76 $\pm$ 0.01 & 138.2 $\pm$ 0.4& & &\\ 
         &                &  &     R &0.72 $\pm$ 0.01 & 0.11 $\pm$ 0.01 & -0.71 $\pm$ 0.01 & 139.3 $\pm$ 0.2& & &\\ 
         &                &  &     I &0.64 $\pm$ 0.01 & 0.07 $\pm$ 0.01 & -0.64 $\pm$ 0.01 & 138.2 $\pm$ 0.4& & &\\ 
HD 54439& GG435 & 2014-10-27 05:50:15& &   &  &   &  &  4756 $\pm$ 268 & 0.79 $\pm$ 0.01 & 1.11 $\pm$ 0.22 \\ 
         &                &  &     B &0.74 $\pm$ 0.02 & 0.10 $\pm$ 0.02 & -0.73 $\pm$ 0.02 & 139.0 $\pm$ 0.7 & & &\\ 
         &                &  &     V &0.75 $\pm$ 0.01 & 0.11 $\pm$ 0.01 & -0.74 $\pm$ 0.01 & 139.3 $\pm$ 0.4& & &\\ 
         &                &  &     R &0.69 $\pm$ 0.01 & 0.15 $\pm$ 0.01 & -0.68 $\pm$ 0.01 & 141.2 $\pm$ 0.3& & &\\ 
         &                &  &     I &0.58 $\pm$ 0.01 & 0.15 $\pm$ 0.01 & -0.56 $\pm$ 0.01 & 142.4 $\pm$ 0.4& & &\\ 
HD 73420& free & 2014-10-27 06:04:31& &   &  &   &  &  6756 $\pm$ 323 & 0.55 $\pm$ 0.01 & 0.65 $\pm$ 0.23 \\ 
         &                &  &     B &0.35 $\pm$ 0.01 & 0.35 $\pm$ 0.01 & -0.04 $\pm$ 0.02 & 176.4 $\pm$ 1.0 & & &\\ 
         &                &  &     V &0.37 $\pm$ 0.01 & 0.36 $\pm$ 0.01 & -0.10 $\pm$ 0.01 & 172.5 $\pm$ 0.7& & &\\ 
         &                &  &     R &0.36 $\pm$ 0.01 & 0.35 $\pm$ 0.01 & -0.08 $\pm$ 0.01 & 173.4 $\pm$ 0.4& & &\\ 
         &                &  &     I &0.33 $\pm$ 0.01 & 0.32 $\pm$ 0.01 & -0.08 $\pm$ 0.01 & 172.5 $\pm$ 0.6& & &\\ 
HD 73420& GG435 & 2014-10-27 06:41:43& &   &  &   &  &  6482 $\pm$ 150 & 0.42 $\pm$ 0.01 & 1.22 $\pm$ 0.28 \\ 
         &                &  &     B &0.35 $\pm$ 0.02 & 0.34 $\pm$ 0.02 & -0.08 $\pm$ 0.02 & 173.3 $\pm$ 1.3 & & &\\ 
         &                &  &     V &0.38 $\pm$ 0.01 & 0.35 $\pm$ 0.01 & -0.15 $\pm$ 0.01 & 168.4 $\pm$ 0.6& & &\\ 
         &                &  &     R &0.38 $\pm$ 0.01 & 0.36 $\pm$ 0.01 & -0.14 $\pm$ 0.01 & 169.2 $\pm$ 0.4& & &\\ 
         &                &  &     I &0.30 $\pm$ 0.01 & 0.28 $\pm$ 0.01 & -0.11 $\pm$ 0.01 & 168.8 $\pm$ 0.6& & &\\ 
HD 78785& free & 2014-11-14 07:04:31& &   &  &   &  &  5771 $\pm$ 10 & 3.96 $\pm$ 0.01 & 1.22 $\pm$ 0.02 \\ 
         &                &  &     B &3.64 $\pm$ 0.01 & 3.39 $\pm$ 0.01 & 1.30 $\pm$ 0.01 & 10.5 $\pm$ 0.1 & & &\\ 
         &                &  &     V &3.88 $\pm$ 0.01 & 3.63 $\pm$ 0.01 & 1.37 $\pm$ 0.01 & 10.3 $\pm$ 0.1& & &\\ 
         &                &  &     R &3.80 $\pm$ 0.01 & 3.53 $\pm$ 0.01 & 1.39 $\pm$ 0.01 & 10.8 $\pm$ 0.1& & &\\ 
         &                &  &     I &3.40 $\pm$ 0.01 & 3.16 $\pm$ 0.01 & 1.26 $\pm$ 0.01 & 10.9 $\pm$ 0.1& & &\\ 
HD 78785& GG435 & 2014-11-14 06:54:19& &   &  &   &  &  5766 $\pm$ 16 & 3.97 $\pm$ 0.01 & 1.18 $\pm$ 0.02 \\ 
         &                &  &     B &3.73 $\pm$ 0.02 & 3.48 $\pm$ 0.02 & 1.35 $\pm$ 0.02 & 10.6 $\pm$ 0.1 & & &\\ 
         &                &  &     V &3.89 $\pm$ 0.01 & 3.65 $\pm$ 0.01 & 1.34 $\pm$ 0.01 & 10.0 $\pm$ 0.1& & &\\ 
         &                &  &     R &3.81 $\pm$ 0.01 & 3.55 $\pm$ 0.01 & 1.38 $\pm$ 0.01 & 10.6 $\pm$ 0.1& & &\\ 
         &                &  &     I &3.41 $\pm$ 0.01 & 3.18 $\pm$ 0.01 & 1.24 $\pm$ 0.01 & 10.6 $\pm$ 0.1& & &\\ 
HD 78785& GG435 & 2014-11-14 07:13:33& &   &  &   &  &  5754 $\pm$ 17 & 3.94 $\pm$ 0.01 & 1.18 $\pm$ 0.02 \\ 
         &                &  &     B &3.71 $\pm$ 0.02 & 3.45 $\pm$ 0.02 & 1.37 $\pm$ 0.02 & 10.8 $\pm$ 0.1 & & &\\ 
         &                &  &     V &3.86 $\pm$ 0.01 & 3.62 $\pm$ 0.01 & 1.34 $\pm$ 0.01 & 10.2 $\pm$ 0.1& & &\\ 
         &                &  &     R &3.78 $\pm$ 0.01 & 3.52 $\pm$ 0.01 & 1.38 $\pm$ 0.01 & 10.7 $\pm$ 0.1& & &\\ 
         &                &  &     I &3.38 $\pm$ 0.01 & 3.15 $\pm$ 0.01 & 1.24 $\pm$ 0.01 & 10.8 $\pm$ 0.1& & &\\ 
HD 96042& free & 2014-12-21 08:14:55& &   &  &   &  &  4816 $\pm$ 255 & 0.58 $\pm$ 0.01 & 0.93 $\pm$ 0.19 \\ 
         &                &  &     B &0.59 $\pm$ 0.01 & -0.38 $\pm$ 0.01 & -0.45 $\pm$ 0.01 & 115.1 $\pm$ 0.5 & & &\\ 
         &                &  &     V &0.60 $\pm$ 0.01 & -0.40 $\pm$ 0.01 & -0.45 $\pm$ 0.01 & 114.0 $\pm$ 0.4& & &\\ 
         &                &  &     R &0.57 $\pm$ 0.01 & -0.38 $\pm$ 0.01 & -0.42 $\pm$ 0.01 & 114.0 $\pm$ 0.3& & &\\ 
         &                &  &     I &0.51 $\pm$ 0.01 & -0.34 $\pm$ 0.01 & -0.38 $\pm$ 0.01 & 114.2 $\pm$ 0.4& & &\\ 
HD 96042& free & 2014-12-21 08:22:47& &   &  &   &  &  4850 $\pm$ 245 & 0.60 $\pm$ 0.01 & 0.90 $\pm$ 0.18 \\ 
         &                &  &     B &0.60 $\pm$ 0.01 & -0.39 $\pm$ 0.01 & -0.46 $\pm$ 0.01 & 114.7 $\pm$ 0.5 & & &\\ 
         &                &  &     V &0.62 $\pm$ 0.01 & -0.41 $\pm$ 0.01 & -0.46 $\pm$ 0.01 & 114.1 $\pm$ 0.4& & &\\ 
         &                &  &     R &0.59 $\pm$ 0.01 & -0.39 $\pm$ 0.01 & -0.44 $\pm$ 0.01 & 114.2 $\pm$ 0.3& & &\\ 
         &                &  &     I &0.54 $\pm$ 0.01 & -0.36 $\pm$ 0.01 & -0.40 $\pm$ 0.01 & 114.1 $\pm$ 0.4& & &\\ 
HD 96042& GG435 & 2015-01-02 07:25:12& &   &  &   &  &  3494 $\pm$ 955 & 0.64 $\pm$ 0.05 & 0.51 $\pm$ 0.24 \\ 
         &                &  &     B &0.62 $\pm$ 0.02 & -0.34 $\pm$ 0.02 & -0.52 $\pm$ 0.02 & 118.5 $\pm$ 0.8 & & &\\ 
         &                &  &     V &0.60 $\pm$ 0.01 & -0.36 $\pm$ 0.01 & -0.49 $\pm$ 0.01 & 116.9 $\pm$ 0.5& & &\\ 
         &                &  &     R &0.56 $\pm$ 0.01 & -0.33 $\pm$ 0.01 & -0.46 $\pm$ 0.01 & 117.2 $\pm$ 0.3& & &\\ 
         &                &  &     I &0.51 $\pm$ 0.01 & -0.32 $\pm$ 0.01 & -0.40 $\pm$ 0.01 & 115.8 $\pm$ 0.5& & &\\ 
HD 96042& GG435 & 2015-01-02 07:29:02& &   &  &   &  &  5116 $\pm$ 242 & 0.59 $\pm$ 0.01 & 1.43 $\pm$ 0.30 \\ 
         &                &  &     B &0.59 $\pm$ 0.02 & -0.35 $\pm$ 0.02 & -0.48 $\pm$ 0.02 & 117.0 $\pm$ 1.0 & & &\\ 
         &                &  &     V &0.62 $\pm$ 0.01 & -0.39 $\pm$ 0.01 & -0.47 $\pm$ 0.01 & 115.1 $\pm$ 0.5& & &\\ 
         &                &  &     R &0.58 $\pm$ 0.01 & -0.36 $\pm$ 0.01 & -0.45 $\pm$ 0.01 & 115.4 $\pm$ 0.3& & &\\ 
         &                &  &     I &0.49 $\pm$ 0.01 & -0.31 $\pm$ 0.01 & -0.37 $\pm$ 0.01 & 115.0 $\pm$ 0.5& & &\\ 
HD 152853& free & 2015-02-03 07:45:45& &   &  &   &  &  5804 $\pm$ 75 & 1.78 $\pm$ 0.02 & 1.35 $\pm$ 0.14 \\ 
         &                &  &     B &1.57 $\pm$ 0.01 & 1.19 $\pm$ 0.01 & 1.02 $\pm$ 0.01 & 20.3 $\pm$ 0.2 & & &\\ 
         &                &  &     V &1.71 $\pm$ 0.01 & 1.29 $\pm$ 0.01 & 1.13 $\pm$ 0.01 & 20.6 $\pm$ 0.2& & &\\ 
         &                &  &     R &1.71 $\pm$ 0.01 & 1.26 $\pm$ 0.01 & 1.15 $\pm$ 0.01 & 21.1 $\pm$ 0.1& & &\\ 
         &                &  &     I &1.42 $\pm$ 0.01 & 1.04 $\pm$ 0.01 & 0.97 $\pm$ 0.01 & 21.5 $\pm$ 0.2& & &\\ 
HD 152853& GG435 & 2015-02-03 08:03:39& &   &  &   &  &  5619 $\pm$ 67 & 1.69 $\pm$ 0.01 & 1.11 $\pm$ 0.08 \\ 
         &                &  &     B &1.59 $\pm$ 0.02 & 1.18 $\pm$ 0.02 & 1.06 $\pm$ 0.02 & 21.0 $\pm$ 0.3 & & &\\ 
         &                &  &     V &1.64 $\pm$ 0.01 & 1.21 $\pm$ 0.01 & 1.10 $\pm$ 0.01 & 21.2 $\pm$ 0.2& & &\\ 
         &                &  &     R &1.60 $\pm$ 0.01 & 1.14 $\pm$ 0.01 & 1.12 $\pm$ 0.01 & 22.3 $\pm$ 0.1& & &\\ 
         &                &  &     I &1.41 $\pm$ 0.01 & 0.98 $\pm$ 0.01 & 1.00 $\pm$ 0.01 & 22.8 $\pm$ 0.1& & &\\ 
HD 137569\tablefootmark{a,b}& free & 2015-02-03 08:31:58& &   &  &   &  &  \dots& \dots & \dots \\ 
         &                &  &     B &0.11 $\pm$ 0.01 & -0.09 $\pm$ 0.01 & 0.05 $\pm$ 0.01 & \dots & & &\\ 
         &                &  &     V &0.13 $\pm$ 0.01 & -0.11 $\pm$ 0.01 & 0.07 $\pm$ 0.01 & \dots& & &\\ 
         &                &  &     R &0.11 $\pm$ 0.01 & -0.10 $\pm$ 0.01 & 0.06 $\pm$ 0.01 & \dots& & &\\ 
         &                &  &     I &0.09 $\pm$ 0.01 & -0.08 $\pm$ 0.01 & 0.05 $\pm$ 0.01 & \dots& & &\\ 
HD 137569\tablefootmark{a,b}& free & 2015-02-03T08:43:51& &   &  &   &  &  \dots & \dots & \dots \\ 
         &                &  &     B &0.11 $\pm$ 0.01 & -0.11 $\pm$ 0.01 & 0.04 $\pm$ 0.01 & \dots & & &\\ 
         &                &  &     V &0.12 $\pm$ 0.01 & -0.12 $\pm$ 0.01 & 0.04 $\pm$ 0.01 & \dots& & &\\ 
         &                &  &     R &0.11 $\pm$ 0.01 & -0.10 $\pm$ 0.01 & 0.05 $\pm$ 0.01 & \dots& & &\\ 
         &                &  &     I &0.09 $\pm$ 0.01 & -0.06 $\pm$ 0.01 & 0.06 $\pm$ 0.01 & \dots& & &\\ 
HD 137569\tablefootmark{a,b}& GG435 & 2015-02-06 07:58:28& &   &  &   &  &  \dots &\dots & \dots \\ 
         &                &  &     B &0.10 $\pm$ 0.03 & 0.08 $\pm$ 0.03 & 0.06 $\pm$ 0.03 & \dots & & &\\ 
         &                &  &     V &0.07 $\pm$ 0.02 & 0.07 $\pm$ 0.02 & 0.02 $\pm$ 0.02 & \dots& & &\\ 
         &                &  &     R &0.11 $\pm$ 0.01 & 0.10 $\pm$ 0.01 & 0.03 $\pm$ 0.01 & \dots& & &\\ 
         &                &  &     I &0.13 $\pm$ 0.01 & 0.11 $\pm$ 0.01 & 0.06 $\pm$ 0.01 & \dots& & &\\ 
HD 137569\tablefootmark{a,b}& GG435 & 2015-02-06 08:45:46& &   &  &   &  &  \dots & \dots & \dots \\ 
         &                &  &     B &0.47 $\pm$ 0.02 & -0.21 $\pm$ 0.02 & 0.42 $\pm$ 0.02 & \dots & & &\\ 
         &                &  &     V &0.39 $\pm$ 0.01 & -0.15 $\pm$ 0.01 & 0.36 $\pm$ 0.01 & \dots& & &\\ 
         &                &  &     R &0.31 $\pm$ 0.01 & -0.14 $\pm$ 0.01 & 0.27 $\pm$ 0.01 & \dots& & &\\ 
         &                &  &     I &0.19 $\pm$ 0.01 & -0.12 $\pm$ 0.01 & 0.15 $\pm$ 0.01 & \dots& & &\\ 
HD 137569\tablefootmark{a,b}& GG435 & 2015-02-06 08:49:35& &   &  &   &  &  \dots & \dots & \dots \\ 
         &                &  &     B &0.09 $\pm$ 0.02 & -0.09 $\pm$ 0.02 & 0.02 $\pm$ 0.02 & \dots & & &\\ 
         &                &  &     V &0.06 $\pm$ 0.01 & -0.04 $\pm$ 0.01 & 0.04 $\pm$ 0.01 & \dots& & &\\ 
         &                &  &     R &0.04 $\pm$ 0.01 & -0.02 $\pm$ 0.01 & 0.03 $\pm$ 0.01 & \dots& & &\\ 
         &                &  &     I &0.03 $\pm$ 0.01 &  0.02 $\pm$ 0.01 & 0.03 $\pm$ 0.01 & \dots& & &\\ 
HD 137569\tablefootmark{a,b}& GG435 & 2015-02-06 08:58:20& &   &  &   &  &  \dots & \dots & \dots \\ 
         &                &  &     B &0.47 $\pm$ 0.02 & -0.21 $\pm$ 0.02 & -0.42 $\pm$ 0.02 & \dots & & &\\ 
         &                &  &     V &0.40 $\pm$ 0.01 & -0.18 $\pm$ 0.01 & -0.36 $\pm$ 0.01 &\dots& & &\\ 
         &                &  &     R &0.34 $\pm$ 0.01 & -0.11 $\pm$ 0.01 & -0.32 $\pm$ 0.01 & \dots& & &\\ 
         &                &  &     I &0.30 $\pm$ 0.01 & -0.05 $\pm$ 0.01 & -0.29 $\pm$ 0.01 & \dots& & &\\ 
\hline
\end{longtable}
\tablefoot{The errors in this table are statistical only, while the root-mean-square of the Stokes Q and U is $\sim 0.05 \%$.\\
\tablefoottext{a}{The constant polarization curve of this star could not be fitted well with a Serkowski curve.}
\tablefoottext{b}{The polarization angle could not be determined due to low polarization degree.} }
}
}

{\tiny
\longtab{
\begin{longtable}{lllllllllll}
\caption{\label{tb_pol_individual_CAFOS} Individual epochs with CAFOS}\\
\hline\hline
&&&&&&&& \multicolumn{3}{c}{Serkowski curve} \\\cline{9-11}
Name & Filter & Epoch & Passband &  $P \hspace{0.3mm} (\%)$ &  $P_Q \hspace{0.3mm} (\%)$ &  $P_U \hspace{0.3mm} (\%)$ & $\theta \hspace{0.3mm} (^{\circ})$ & $\lambda_{\rm max} \hspace{0.3mm} (\AA)$ &  $P_{\rm max} \hspace{0.3mm} (\%)$ & $K$ \\
\hline
\endfirsthead
\caption{continued.}\\
\hline
&&&&&&&& \multicolumn{3}{c}{Serkowski curve} \\\cline{9-11}
Name & Filter & Epoch & Passband &  $P \hspace{0.3mm} (\%)$ &  $P_Q \hspace{0.3mm} (\%)$ &  $P_U \hspace{0.3mm} (\%)$ & $\theta \hspace{0.3mm} (^{\circ})$ & $\lambda_{\rm max} \hspace{0.3mm} (\AA)$ &  $P_{\rm max} \hspace{0.3mm} (\%)$ & $K$ \\
\hline
\endhead
\hline
\endfoot
BD+23d3762&  free & 2015-04-30 03:09:40& &   &  &   &  &  4964.8 $\pm$ 60.9 & 2.23 $\pm$ 0.01 & 0.92 $\pm$ 0.06 \\ 
         &                &   &   B &2.20 $\pm$ 0.01 & 1.83 $\pm$ 0.01 & 1.21 $\pm$ 0.02 & 16.7 $\pm$ 0.2& & & \\ 
         &                &   &   V &2.19 $\pm$ 0.01 & 1.76 $\pm$ 0.01 & 1.31 $\pm$ 0.02 & 18.3 $\pm$ 0.2& & &\\ 
         &                &   &   R &2.07 $\pm$ 0.01 & 1.62 $\pm$ 0.01 & 1.28 $\pm$ 0.01 & 19.2 $\pm$ 0.1& & &\\ 
BD+45d3341&  free & 2015-04-30 02:06:40& &   &  &   &  &  5046.5 $\pm$ 72.6 & 3.09 $\pm$ 0.01 & 0.83 $\pm$ 0.08 \\ 
         &                &   &   B &3.02 $\pm$ 0.02 & -0.19 $\pm$ 0.01 & 3.02 $\pm$ 0.02 & 46.8 $\pm$ 0.2& & & \\ 
         &                &   &   V &3.06 $\pm$ 0.02 & -0.31 $\pm$ 0.01 & 3.04 $\pm$ 0.02 & 47.9 $\pm$ 0.1& & &\\ 
         &                &   &   R &2.90 $\pm$ 0.01 & -0.31 $\pm$ 0.01 & 2.88 $\pm$ 0.01 & 48.0 $\pm$ 0.1& & &\\ 
BD+45d3341&  free & 2015-04-30 02:34:01& &   &  &   &  &  5192.5 $\pm$ 34.0 & 3.01 $\pm$ 0.01 & 1.09 $\pm$ 0.06 \\ 
         &                &   &   B &2.93 $\pm$ 0.01 & -0.20 $\pm$ 0.01 & 2.92 $\pm$ 0.01 & 47.0 $\pm$ 0.1& & & \\ 
         &                &   &   V &2.97 $\pm$ 0.01 & -0.30 $\pm$ 0.01 & 2.96 $\pm$ 0.01 & 47.9 $\pm$ 0.1& & &\\ 
         &                &   &   R &2.87 $\pm$ 0.01 & -0.30 $\pm$ 0.01 & 2.85 $\pm$ 0.01 & 48.0 $\pm$ 0.1& & &\\ 
HD1337\tablefootmark{a}&  free & 2015-04-30 03:47:04& &   &  &   &  & \dots & \dots & \dots \\ 
          &               &   &   B &0.56 $\pm$ 0.03 &  0.04 $\pm$ 0.02 & 0.56 $\pm$ 0.03 & 42.8 $\pm$ 1.7& & & \\ 
         &                &   &   V &0.54 $\pm$ 0.02 & -0.04 $\pm$ 0.02 & 0.54 $\pm$ 0.02 & 47.1 $\pm$ 1.3& & &\\ 
         &                &   &   R &0.55 $\pm$ 0.01 & -0.03 $\pm$ 0.01 & 0.55 $\pm$ 0.01 & 46.6 $\pm$ 0.7& & &\\ 
HD1337\tablefootmark{a}& free & 2015-04-30 03:55:09& &   &  &   &  & \dots & \dots & \dots \\ 
         &                &   &   B &0.54 $\pm$ 0.01 & -0.02 $\pm$ 0.01 & 0.54 $\pm$ 0.01 & 46.3 $\pm$ 0.5& & & \\ 
         &                &   &   V &0.56 $\pm$ 0.01 & -0.04 $\pm$ 0.01 & 0.56 $\pm$ 0.01 & 47.2 $\pm$ 0.5& & &\\ 
         &                &   &   R &0.55 $\pm$ 0.01 & -0.04 $\pm$ 0.01 & 0.55 $\pm$ 0.01 & 47.1 $\pm$ 0.3& & &\\ 
HD1337\tablefootmark{a}&  free & 2015-04-30 04:09:15& &   &  &   &  &  \dots & \dots & \dots \\ 
         &                &   &   B &0.58 $\pm$ 0.01 & -0.04 $\pm$ 0.01 & 0.57 $\pm$ 0.01 & 47.0 $\pm$ 0.6& & & \\ 
         &                &   &   V &0.58 $\pm$ 0.01 & -0.05 $\pm$ 0.01 & 0.57 $\pm$ 0.01 & 47.6 $\pm$ 0.6& & &\\ 
         &                &   &   R &0.56 $\pm$ 0.01 & -0.04 $\pm$ 0.01 & 0.56 $\pm$ 0.01 & 46.8 $\pm$ 0.4& & &\\ 
HD137569\tablefootmark{a,b}&  free & 2015-04-29T23:40:10& &   &  &   &  &   \dots & \dots & \dots \\ 
         &                &   &   B &0.19 $\pm$ 0.03 & -0.19 $\pm$ 0.03 & -0.04 $\pm$ 0.03 & \dots& & & \\ 
         &                &   &   V &0.34 $\pm$ 0.05 & -0.29 $\pm$ 0.05 & -0.18 $\pm$ 0.05 & \dots& & &\\ 
         &                &   &   R &0.21 $\pm$ 0.03 & -0.21 $\pm$ 0.03 & -0.04 $\pm$ 0.03 & \dots& & &\\ 
HD137569\tablefootmark{a,b}&  free & 2015-04-30T00:08:05& &   &  &   &  &   \dots & \dots & \dots \\ 
         &                &   &   B &0.20 $\pm$ 0.02 & -0.14 $\pm$ 0.03 & -0.15 $\pm$ 0.02 & \dots& & & \\ 
         &                &   &   V &0.18 $\pm$ 0.04 & -0.11 $\pm$ 0.04 & -0.14 $\pm$ 0.04 & \dots& & &\\ 
         &                &   &   R &0.16 $\pm$ 0.03 & -0.15 $\pm$ 0.03 & -0.06 $\pm$ 0.02 & \dots& & &\\ 
HD154445& free & 2015-04-30 00:39:26& &   &  &   &  &  5550.8 $\pm$ 13.8 & 3.67 $\pm$ 0.01 & 1.50 $\pm$ 0.04 \\ 
         &                &   &   B &3.37 $\pm$ 0.01 & -3.36 $\pm$ 0.01 & 0.19 $\pm$ 0.01 & 88.4 $\pm$ 0.1& & &\\ 
         &                &   &   V &3.63 $\pm$ 0.01 & -3.63 $\pm$ 0.01 & 0.07 $\pm$ 0.01 & 89.4 $\pm$ 0.1& & &\\ 
         &                &   &   R &3.53 $\pm$ 0.01 & -3.53 $\pm$ 0.01 & 0.02 $\pm$ 0.01 & 89.9 $\pm$ 0.1& & &\\ 
HD154445&  free & 2015-04-30 04:33:46& &   &  &   &  &  5640.8 $\pm$ 20.5 & 3.61 $\pm$ 0.01 & 1.85 $\pm$ 0.07 \\ 
         &                &   &   B &3.23 $\pm$ 0.01 & -3.22 $\pm$ 0.01 & 0.13 $\pm$ 0.01 & 88.9 $\pm$ 0.1& & &\\ 
         &                &   &   V &3.57 $\pm$ 0.01 & -3.57 $\pm$ 0.01 & 0.05 $\pm$ 0.01 & 89.6 $\pm$ 0.1& & &\\ 
         &                &   &   R &3.48 $\pm$ 0.01 & -3.48 $\pm$ 0.01 & -0.01 $\pm$ 0.01 & 90.1 $\pm$ 0.1& & &\\ 
HD194092&  free & 2015-04-30 01:21:46& &   &  &   &  &  5727.8 $\pm$ 235.1 & 0.64 $\pm$ 0.01 & 1.46 $\pm$ 0.47 \\ 
         &                &   &   B &0.51 $\pm$ 0.01 & -0.39 $\pm$ 0.01 & 0.32 $\pm$ 0.01 & 70.6 $\pm$ 0.5& & & \\ 
         &                &   &   V &0.61 $\pm$ 0.01 & -0.52 $\pm$ 0.01 & 0.31 $\pm$ 0.02 & 74.5 $\pm$ 0.6& & &\\ 
         &                &   &   R &0.61 $\pm$ 0.01 & -0.50 $\pm$ 0.01 & 0.34 $\pm$ 0.01 & 72.9 $\pm$ 0.4& & &\\ 
HD28446&  free & 2015-04-29 20:44:12& &   &  &   &  &  4835.7 $\pm$ 117.9 & 2.12 $\pm$ 0.01 & 1.08 $\pm$ 0.16 \\ 
         &                &   &   B &2.13 $\pm$ 0.02 & 1.16 $\pm$ 0.02 & -1.79 $\pm$ 0.02 & 151.4 $\pm$ 0.2& & & \\ 
         &                &   &   V &2.07 $\pm$ 0.01 & 1.13 $\pm$ 0.01 & -1.74 $\pm$ 0.01 & 151.6 $\pm$ 0.2& & &\\ 
         &                &   &   R &1.98 $\pm$ 0.01 & 1.12 $\pm$ 0.01 & -1.63 $\pm$ 0.01 & 152.3 $\pm$ 0.1& & &\\ 
HD28446&  free & 2015-04-29 20:52:17& &   &  &   &  &  4887.0 $\pm$ 100.4 & 2.07 $\pm$ 0.01 & 0.60 $\pm$ 0.08 \\ 
         &                &   &   B &2.03 $\pm$ 0.01 & 1.12 $\pm$ 0.01 & -1.69 $\pm$ 0.01 & 151.8 $\pm$ 0.1& & & \\ 
         &                &   &   V &2.03 $\pm$ 0.01 & 1.10 $\pm$ 0.01 & -1.70 $\pm$ 0.01 & 151.5 $\pm$ 0.1& & &\\ 
         &                &   &   R &1.96 $\pm$ 0.01 & 1.09 $\pm$ 0.01 & -1.63 $\pm$ 0.01 & 151.8 $\pm$ 0.1& & &\\ 
HD43384& free & 2015-04-29 20:14:24& &   &  &   &  &  5342.5 $\pm$ 42.8 & 2.96 $\pm$ 0.01 & 0.96 $\pm$ 0.06 \\ 
         &                &   &   B &2.88 $\pm$ 0.01 & 2.62 $\pm$ 0.01 & -1.19 $\pm$ 0.01 & 167.8 $\pm$ 0.1& & & \\ 
         &                &   &   V &2.94 $\pm$ 0.01 & 2.73 $\pm$ 0.01 & -1.08 $\pm$ 0.01 & 169.2 $\pm$ 0.1& & &\\ 
         &                &   &   R &2.85 $\pm$ 0.01 & 2.70 $\pm$ 0.01 & -0.92 $\pm$ 0.01 & 170.6 $\pm$ 0.1& & &\\ 
HD43384&  free & 2015-04-29 21:14:04& &   &  &   &  &  5215.5 $\pm$ 75.8 & 3.00 $\pm$ 0.01 & 0.80 $\pm$ 0.08 \\ 
         &                &   &   B &2.93 $\pm$ 0.02 & 2.72 $\pm$ 0.02 & -1.08 $\pm$ 0.03 & 169.2 $\pm$ 0.2& & & \\ 
         &                &   &   V &2.97 $\pm$ 0.02 & 2.79 $\pm$ 0.01 & -1.02 $\pm$ 0.02 & 170.0 $\pm$ 0.2& & &\\ 
         &                &   &   R &2.87 $\pm$ 0.01 & 2.72 $\pm$ 0.01 & -0.91 $\pm$ 0.01 & 170.8 $\pm$ 0.1& & &\\ 
\hline
\end{longtable}
\tablefoot{The errors in this table are statistical only, while the root-mean-square of the Stokes Q and U is $\sim 0.04 \%$.\\
\tablefoottext{a}{The constant polarization curve of this star could not be fitted well with a Serkowski curve.}
\tablefoottext{b}{The polarization angle could not be determined due to low polarization degree.}}
}
}

%

{\tiny
\longtab{
\begin{longtable}{lllllllllll}
\caption{\label{tb_pol_individual_AFOSC} Individual epochs with AFOSC}\\
\hline\hline
&&&&&&&& \multicolumn{3}{c}{Serkowski curve} \\\cline{9-11}
Name & Filter & Epoch & Passband &  $P \hspace{0.3mm} (\%)$ &  $P_Q \hspace{0.3mm} (\%)$ &  $P_U \hspace{0.3mm} (\%)$ & $\theta \hspace{0.3mm} (^{\circ})$ & $\lambda_{\rm max} \hspace{0.3mm} (\AA)$ &  $P_{\rm max} \hspace{0.3mm} (\%)$ & $K$ \\
\hline
\endfirsthead
\caption{continued.}\\
\hline
&&&&&&&& \multicolumn{3}{c}{Serkowski curve} \\\cline{9-11}
Name & Filter & Epoch & Passband &  $P \hspace{0.3mm} (\%)$ &  $P_Q \hspace{0.3mm} (\%)$ &  $P_U \hspace{0.3mm} (\%)$ & $\theta \hspace{0.3mm} (^{\circ})$ & $\lambda_{\rm max} \hspace{0.3mm} (\AA)$ &  $P_{\rm max} \hspace{0.3mm} (\%)$ & $K$ \\
\hline
\endhead
\hline
\endfoot
HD 28446& free & 2015-02-09 19:43:28& &   &  & &    &  4722 $\pm$ 49 & 2.00 $\pm$ 0.01 & 0.72 $\pm$ 0.04 \\
&                  &    &     B  & 1.99 $\pm$ 0.01  &  1.16 $\pm$ 0.01  &  -1.62 $\pm$ 0.01  & 152.8 $\pm$ 0.1 &   &   &   \\
&                  &     &    V  & 1.96 $\pm$ 0.01  &  1.12 $\pm$ 0.01  &  -1.61 $\pm$ 0.01  & 152.3 $\pm$ 0.1 &   &   &   \\
&                  &     &    R  & 1.85 $\pm$ 0.01  &  1.04 $\pm$ 0.01  &  -1.53 $\pm$ 0.01  & 152.1 $\pm$ 0.1 &   &   &   \\
HD 43384& free & 2015-02-09 20:03:30& &   &  & &    &  5201 $\pm$ 17 & 3.16 $\pm$ 0.01 & 1.03 $\pm$ 0.02 \\
&                  &    &     B  & 3.08 $\pm$ 0.01  &  2.92 $\pm$ 0.01  &  -0.96 $\pm$ 0.01  & 170.9 $\pm$ 0.1 &   &   &   \\
&                  &     &    V  & 3.14 $\pm$ 0.01  &  3.01 $\pm$ 0.01  &  -0.92 $\pm$ 0.01  & 171.5 $\pm$ 0.1 &   &   &   \\
&                  &     &    R  & 2.99 $\pm$ 0.01  &  2.87 $\pm$ 0.01  &  -0.84 $\pm$ 0.01  & 171.9 $\pm$ 0.1 &   &   &   \\
HD 43384& free & 2015-02-10 17:53:17& &   &  & &    &  5371 $\pm$ 14 & 3.19 $\pm$ 0.01 & 1.21 $\pm$ 0.03 \\
&                  &    &     B  & 3.04 $\pm$ 0.01  &  2.91 $\pm$ 0.01  &  -0.89 $\pm$ 0.01  & 171.5 $\pm$ 0.1 &   &   &   \\
&                  &     &    V  & 3.17 $\pm$ 0.01  &  3.04 $\pm$ 0.01  &  -0.92 $\pm$ 0.01  & 171.6 $\pm$ 0.1 &   &   &   \\
&                  &     &    R  & 3.03 $\pm$ 0.01  &  2.91 $\pm$ 0.01  &  -0.86 $\pm$ 0.01  & 171.8 $\pm$ 0.1 &   &   &   \\
HD 43384 & free & 2015-03-09 21:06:00& &   &  & &    &  5317 $\pm$ 20 & 2.92 $\pm$ 0.01 & 1.14 $\pm$ 0.03 \\
&                  &    &     B  & 2.82 $\pm$ 0.01  &  2.62 $\pm$ 0.01  &  -1.04 $\pm$ 0.01  & 169.2 $\pm$ 0.1 &   &   &   \\
&                  &     &    V  & 2.90 $\pm$ 0.01  &  2.71 $\pm$ 0.01  &  -1.02 $\pm$ 0.01  & 169.7 $\pm$ 0.1 &   &   &   \\
&                  &     &    R  & 2.76 $\pm$ 0.01  &  2.60 $\pm$ 0.01  &  -0.93 $\pm$ 0.01  & 170.2 $\pm$ 0.1 &   &   &   \\
HD 1337	\tablefootmark{a}& free & 2015-02-09 18:40:10& &   &  & &    &  \dots & \dots  & \dots  \\
&                  &    &     B  & 0.51 $\pm$ 0.01  &  -0.12 $\pm$ 0.01  &  0.49 $\pm$ 0.01  &  52.1 $\pm$ 0.2 &   &   &   \\
&                  &     &    V  & 0.50 $\pm$ 0.01  &  -0.14 $\pm$ 0.01  &  0.48 $\pm$ 0.01  &  53.2 $\pm$ 0.2 &   &   &   \\
&                  &     &    R  & 0.54 $\pm$ 0.01  &  -0.19 $\pm$ 0.01  &  0.51 $\pm$ 0.01  &  55.2 $\pm$ 0.1 &   &   &   \\
HD 1337	\tablefootmark{a}& free & 2015-03-10 18:06:42& &   &  & &    &   \dots & \dots  & \dots  \\
&                  &    &     B  & 0.65 $\pm$ 0.01  &  -0.07 $\pm$ 0.01  &  0.64 $\pm$ 0.01  &  48.0 $\pm$ 0.2 &   &   &   \\
&                  &     &    V  & 0.61 $\pm$ 0.01  &  -0.08 $\pm$ 0.01  &  0.60 $\pm$ 0.01  &  48.6 $\pm$ 0.2 &   &   &   \\
&                  &     &    R  & 0.61 $\pm$ 0.01  &  -0.11 $\pm$ 0.01  &  0.60 $\pm$ 0.01  &  49.9 $\pm$ 0.1 &   &   &   \\
HD 54439& free & 2015-02-10 20:48:43& &   &  & &    &  5138 $\pm$ 114 & 0.69 $\pm$ 0.01 & 1.28 $\pm$ 0.22 \\
&                  &    &     B  & 0.66 $\pm$ 0.01  & -0.03 $\pm$ 0.01  &  -0.66 $\pm$ 0.01  & 133.8 $\pm$ 0.2 &   &   &   \\
&                  &     &    V  & 0.69 $\pm$ 0.01  &  0.08 $\pm$ 0.01  &  -0.68 $\pm$ 0.01  & 138.4 $\pm$ 0.1 &   &   &   \\
&                  &     &    R  & 0.63 $\pm$ 0.01  &  0.03 $\pm$ 0.01  &  -0.63 $\pm$ 0.01  & 136.5 $\pm$ 0.1 &   &   &   \\
HD 21291& free & 2015-02-10 17:23:13& &   &  & &    &  5166 $\pm$ 27 & 2.95 $\pm$ 0.01 & 1.05 $\pm$ 0.04 \\
&                  &    &     B  & 2.87 $\pm$ 0.01  &  -1.73 $\pm$ 0.01  &  -2.29 $\pm$ 0.01  & 116.4 $\pm$ 0.1 &   &   &   \\
&                  &     &    V  & 2.92 $\pm$ 0.01  &  -1.81 $\pm$ 0.01  &  -2.29 $\pm$ 0.01  & 115.9 $\pm$ 0.1 &   &   &   \\
&                  &     &    R  & 2.77 $\pm$ 0.01  &  -1.74 $\pm$ 0.01  &  -2.16 $\pm$ 0.01  & 115.6 $\pm$ 0.1 &   &   &   \\
HD 137569	\tablefootmark{a,b}& free & 2015-02-11 03:51:49& &   &  & &    &  \dots & \dots  & \dots  \\
&                  &    &     B  & 0.22 $\pm$ 0.01  &  -0.22 $\pm$ 0.01  &  -0.02 $\pm$ 0.01  & \dots &   &   &  \\
&                  &     &    V  & 0.19 $\pm$ 0.01  &  -0.19 $\pm$ 0.01  &  -0.00 $\pm$ 0.01  & \dots &   &   &  \\
&                  &     &    R  & 0.19 $\pm$ 0.01  &  -0.19 $\pm$ 0.01  &   0.01 $\pm$ 0.01  &  \dots &   &   &  \\
HD 137569	\tablefootmark{a,b} & free & 2015-03-10 02:41:10& &   &  & &    &   \dots & \dots  & \dots  \\
&                  &    &     B  & 0.20 $\pm$ 0.01  &  -0.18 $\pm$ 0.01  &  0.10 $\pm$ 0.01  &  \dots &   &   &   \\
&                  &     &    V  & 0.18 $\pm$ 0.01  &  -0.17 $\pm$ 0.01  &  0.07 $\pm$ 0.01  &  \dots &   &   &  \\
&                  &     &    R  & 0.22 $\pm$ 0.01  &  -0.20 $\pm$ 0.01  &  0.10 $\pm$ 0.01  &  \dots &   &   &  \\
HD 198478& free & 2016-08-02 21:00:02& &   &  & &    &  5132 $\pm$ 41 & 2.77 $\pm$ 0.01 & 0.99 $\pm$ 0.06 \\
&                  &    &     B  & 2.70 $\pm$ 0.01  &  2.69 $\pm$ 0.01  &  0.25 $\pm$ 0.01  &  2.6 $\pm$ 0.1 &   &   &   \\
&                  &     &    V  & 2.73 $\pm$ 0.01  &  2.72 $\pm$ 0.01  &  0.30 $\pm$ 0.01  &  3.2 $\pm$ 0.1 &   &   &   \\
&                  &     &    R  & 2.61 $\pm$ 0.01  &  2.59 $\pm$ 0.01  &  0.30 $\pm$ 0.01  &  3.3 $\pm$ 0.1 &   &   &   \\
HD 194092& free & 2016-08-02 20:22:18& &   &  & &    &  5884 $\pm$ 107 & 0.74 $\pm$ 0.01 & 1.09 $\pm$ 0.18 \\
&                  &    &     B  & 0.69 $\pm$ 0.01  &  -0.67 $\pm$ 0.01  &  0.16 $\pm$ 0.01  &  83.1 $\pm$ 0.3 &   &   &   \\
&                  &     &    V  & 0.75 $\pm$ 0.01  &  -0.72 $\pm$ 0.01  &  0.20 $\pm$ 0.01  &  82.4 $\pm$ 0.2 &   &   &   \\
&                  &     &    R  & 0.73 $\pm$ 0.01  &  -0.70 $\pm$ 0.01  &  0.17 $\pm$ 0.01  &  83.1 $\pm$ 0.1 &   &   &   \\
\hline
\end{longtable}
\tablefoot{The errors in this table are statistical only, while the root-mean-square of the Stokes Q and U is $\sim 0.06 \%$.\\
\tablefoottext{a}{The constant polarization curve of this star could not be fitted well with a Serkowski curve.}
\tablefoottext{b}{The polarization angle could not be determined due to low polarization degree.}}
}
}
\end{landscape}

\end{appendix}

\end{document}